%% file: mnras_template.tex
\documentclass[fleqn,usenatbib]{mnras}

\usepackage{newtxtext,newtxmath}

\usepackage[T1]{fontenc}

\DeclareRobustCommand{\VAN}[3]{#2}
\let\VANthebibliography\thebibliography
\def\thebibliography{\DeclareRobustCommand{\VAN}[3]{##3}\VANthebibliography}

\usepackage{graphicx}	
\usepackage{amsmath}	
\usepackage{amssymb}	
\usepackage{booktabs}
\usepackage{float}

\newcommand{\ewlya}{\ifmmode W_{\rm Ly\alpha} \else $W_{\rm Ly\alpha}$\fi}
\newcommand{\ewlis}{\ifmmode W_{\rm LIS} \else $W_{\rm LIS}$\fi}
\newcommand{\rlis}{\ifmmode {\rm R(LIS)} \else ${\rm R(LIS)}$\fi}
\newcommand{\cflis}{\ifmmode C_f({\rm LIS}) \else $C_f({\rm LIS})$\fi}

\newcommand{\fescabs}{\ifmmode f_{\rm esc}^{\rm abs} \else $f_{\rm esc}^{\rm abs}$\fi}
\newcommand{\fesclis}{\ifmmode f\mathrm{_{esc}^{abs,~LIS}} \else $f\mathrm{_{esc}^{abs,~LIS}}$\fi}

\newcommand{\mobs}{\ifmmode M_{\rm 1500} \else $M_{\rm 1500}$\fi}
\newcommand{\mint}{\ifmmode M_{\rm 1500}^{\rm int} \else $M_{\rm 1500}^{\rm int}$\fi}
\newcommand{\bslope}{\ifmmode \beta _{\rm spec}^{1500} \else $\beta _{\rm spec}^{1500}$\fi}
\newcommand{\ebv}{\ifmmode E_{\rm B-V} \else $E_{\rm B-V}$\fi}
\newcommand{\Zsun}{\ifmmode {\rm Z_{\odot}} \else ${\rm Z_{\odot}}$\fi}
\newcommand{\xiion}{\ifmmode \xi_{\rm ion} \else $\xi_{\rm ion}$\fi}
\newcommand{\fxi}{\ifmmode f_{\rm esc}^{\rm abs} \times \xi_{\rm ion} \else $f_{\rm esc}^{\rm abs} \times \xi_{\rm ion}$\fi}
\newcommand{\Fion}{\ifmmode {\rm (F_{\lambda900}/F_{\lambda1500})_{int}} \else ${\rm (F_{\lambda900}/F_{\lambda1500})_{int}}$\fi}

\newcommand{\hi}{\ion{H}{i}}
\newcommand{\lya}{Ly$\alpha$}

\title[Ionizing properties of SFGs at $3 \leq z \leq 5$]{The VANDELS survey: the ionizing properties of star-forming galaxies at $3 \leq z \leq 5$ using deep rest-frame ultraviolet spectroscopy}

\author[A. Saldana-Lopez et al.]{
A. Saldana-Lopez,$^{1}$ \thanks{E-mail: alberto.saldanalopez@unige.ch}
D. Schaerer,$^{1,2}$
J. Chisholm,$^{3}$
A. Calabr\`o,$^{4}$
L. Pentericci,$^{4}$
\newauthor
F. Cullen,$^{5}$
A. Saxena,$^{6,7}$
R. Amor{\'i}n,$^{8}$
A. C. Carnall,$^{5}$
F. Fontanot,$^{9}$
J. P. U. Fynbo,$^{10,11}$
\newauthor
L. Guaita,$^{12}$
N. P. Hathi,$^{13}$
P. Hibon,$^{14}$
Z. Ji,$^{15}$
D. J. McLeod,$^{5}$
E. Pompei,$^{14}$
G. Zamorani $^{16}$
\\
$^{1}$ Department of Astronomy, University of Geneva, 51 Chemin Pegasi, 1290 Versoix, Switzerland\\
$^{2}$ CNRS, IRAP, 14 Avenue E. Belin, 31400 Toulouse, France\\
$^{3}$ Department of Astronomy, The University of Texas at Austin, 2515 Speedway, Stop C1400, Austin, TX 78712-1205, USA\\
$^{4}$ INAF - Osservatorio Astronomico di Roma, via Frascati 33, 00078, Monteporzio Catone, Italy\\
$^{5}$ Institute for Astronomy, University of Edinburgh, Royal Observatory, Edinburgh EH9 3HJ, UK\\
$^{6}$ Sub-department of Astrophysics, University of Oxford, Keble Road, Oxford OX1 3RH, UK\\
$^{7}$ Department of Physics and Astronomy, University College London, Gower Street, London WC1E 6BT, UK\\
$^{8}$ Instituto de Investigaci{\'o}n Multidisciplinar en Ciencia y Tecnolog{\'i}a, Universidad de La Serena, Ra{\'u}l Bitr{\'a}n, 1305, La Serena, Chile\\
$^{9}$ INAF - Trieste Observatory, via GB Tiepolo 11, 34143 Trieste, Italy\\
$^{10}$ Cosmic Dawn Center (DAWN), Jagtvej 128, DK2200 Copenhagen N, Denmark\\
$^{11}$ Niels Bohr Institute, University of Copenhagen, Blegdamsvej 17, DK2100 Copenhagen {\O}, Denmark\\
$^{12}$ Departamento de Ciencias Fisicas, Facultad de Ciencias Exactas, Universidad Andres Bello, Fernandez Concha 700, Las Condes, Santiago, Chile\\
$^{13}$ Space Telescope Science Institute, Baltimore, MD 21218, USA\\
$^{14}$ European Southern Observatory, Alonso de C{\'o}rdova 3107, Vitacura, Santiago de Chile, Chile\\
$^{15}$ Department of Astronomy, University of Massachusetts, Amherst, 710 North Pleasant Street, Amherst, MA 01003-9305, USA\\
$^{16}$ INAF - Astrophysics and Space Science Observatory, Via Piero Gobetti 93/3, 40129, Bologna, Italy
}

\date{Accepted 2023 April 24. Received 2023 April 24; in original form 2022 November 21}

\pubyear{2022}

\begin{document}
\label{firstpage}
\pagerange{\pageref{firstpage}--\pageref{lastpage}}
\maketitle

\begin{abstract}
The physical properties of Epoch of Reionization (EoR) galaxies are still poorly constrained by observations. To better understand the ionizing properties of galaxies in the EoR, we investigate deep, rest-frame ultraviolet (UV) spectra of $\simeq 500$ star-forming galaxies at $3 \leq z \leq 5$ selected from the public ESO-VANDELS spectroscopic survey. The absolute ionizing photon escape fraction (\fescabs, i.e., the ratio of \emph{leaking} against \emph{produced} ionizing photons) is derived by combining absorption line measurements with estimates of the UV attenuation. The ionizing production efficiency (\xiion, i.e., the number of ionizing photons produced per non-ionizing UV luminosity) is calculated by fitting the far-UV (FUV) stellar continuum of the VANDELS galaxies.
We find that the \fescabs\ and \xiion\ parameters increase towards low-mass, blue UV-continuum slopes and strong \lya\ emitting galaxies, and both are slightly higher-than-average for the UV-faintest galaxies in the sample. Potential Lyman Continuum Emitters (LCEs, $\fescabs \geq 5\%$) and selected Lyman Alpha Emitters (LAEs, $\ewlya \leq -20$\AA) show systematically higher \xiion\ ($\log \xiion ({\rm Hz/erg}) \approx 25.38, 25.41$) than non-LCEs and non-LAEs ($\log \xiion ({\rm Hz/erg}) \approx 25.18, 25.14$) at similar UV magnitudes. This indicates very young underlying stellar populations ($\approx 10~{\rm Myr}$) at relatively low metallicities ($\approx 0.2~{\rm Z_{\odot}}$). 
The FUV non-ionizing spectra of potential LCEs is characterized by blue UV slopes ($\leq -2$), enhanced \lya\ emission ($\leq -25$\AA), strong UV nebular lines (e.g., high \ion{C}{iv}1550/\ion{C}{iii}1908 $\geq 0.75$ ratios), and weak absorption lines ($\leq 1$\AA). The latter suggests the existence of low gas-column-density channels in the interstellar medium, which enables the escape of ionizing photons. 
By comparing our VANDELS results against other surveys in the literature, our findings imply that the ionizing budget in the EoR was likely dominated by UV-faint, low-mass and dustless galaxies.
\end{abstract}




\defcitealias{R16}{R16}
\defcitealias{SMC}{SMC}
\defcitealias{SL22}{SL22}
\defcitealias{C19}{C19}

\newpage
\input{1_intro}
\input{2_data}
\input{3_methods}
\input{4_results}
\input{5_discussion}
\input{6_conclusions}

\section*{Acknowledgements}
The authors thank the anonymous referee for providing useful comments, which have certainly improved the quality of this paper. The authors also thank Hakim Atek, Jorytt Matthee and Irene Shivaei for kindly providing tabulated data from \citet{Atek2022}, \citet{Matthee2017b} and \citet{Shivaei2018}. ASL and DS acknowledge support from Swiss National Science Foundation. The Cosmic Dawn Center is funded by the Danish National Research Foundation under grant No.140. RA acknowledges support from ANID Fondecyt Regular 1202007. JPUF is supported by the DFF grant 1026-00066. 




\section*{Data availability}
The VANDELS Data Release 4 (DR4) is now publicly available and can be accessed using the VANDELS database at \url{http://vandels. inaf.it/dr4.html}, or through the ESO archives. Any secondary product and/or data underlying this article will be shared on reasonable request to the corresponding author.

\bibliographystyle{mnras}
\bibliography{references} 




\appendix
\input{A_appendix}


\bsp	
\label{lastpage}
\end{document}

%% file: 1_intro.tex
\section{Introduction}\label{sec:intro}
Several data sets \citep[see][and references therein]{Goto2021} measuring the redshift ($z$) evolution of the volume-averaged neutral hydrogen fraction provide evidence for the last major phase change underwent by the Universe, the Cosmic Reionization. Between $z = 9$ and $z = 6$ \citep{Planck2016}, the number of ionizing photons emitted per unit time ($\dot{N}_{\rm ion}$) overcame the recombination rate ($\Gamma_{\rm HI}$) of hydrogen atoms, $\dot{N}_{\rm ion} \geq \Gamma_{\rm HI}$ \citep{Madau1999}, so that the bulk of \ion{H}{i} neutral gas within the Intergalactic Medium (IGM) \emph{progressively} transitioned to an ionized state \citep[see][for a review]{DayalFerrara2018}.

Yet, the sources mainly responsible for expelling such a vast amount of ionizing (or Lyman Continuum, LyC) photons to the IGM remain elusive \citep[see][for a review]{Robertson2022}. Overall consensus exists about the minor contribution of Active Galactic Nuclei (AGN) to the ionizing budget during the Epoch of Reionization \citep[EoR,][]{Hassan2018,Kulkarni2019,Dayal2020}, mainly because of their lower number density at early epochs \citep[e.g.,][but see \citet{Fontanot2012} and \citet{Cristiani2016} for a different viewpoint]{Matsuoka2018}. This said, AGNs have played a key role in keeping the IGM ionized after the EoR \citep[see e.g.,][]{Becker2013}, whereas stars seem to provide a small contribution to the ionizing radiation budget at $z < 5$ \citep{Tanvir2019}. Clearly, assuming that star-forming (SF) galaxies drove Cosmic Reionization shifts the current debate on whether more massive, UV-bright galaxies \citep{Madau2015} or in opposite, low-mass, UV-faint counterparts \citep{Robertson2013, Robertson2015} dominated the ionizing emissivity at the EoR.

On the one hand, the remarkable modeling efforts by \citet{Sharma2016} and \citet{Naidu2020}, based on the evolution of the star-formation rate surface-density of galaxies, and the works by \citet{Naidu2022} and \citet{Matthee2022}, based on the fraction of \lya\ emitters (LAEs) over time, support a late-completed and rapid Reionization purely dominated by more massive and moderately luminous sub$-M_{\rm UV}^{\star}$ systems. These results are compatible with the rapid reionization modeled by \citet{Mason2019}, in which the ionizing emissivity was constrained from CMB optical depth and \lya\ forest dark pixel fraction data. On the other hand, semi-empirical models like those in \citet{Finkelstein2019}, based on observational constraints on the UV luminosity function during the EoR, and independent cosmological hydrodynamical simulation such as \citet{Rosdahl2022}, suggest an early-completed Reionization conducted primarily by low-mass and fainter galaxies \citep[see also][]{Trebitsch2022}. The most recent measurements of the mean free path of ionizing photons by \citet{Becker2021} have shown a much shorter value than previously thought at $ 5 \leq z \leq 6$, supporting the rapid reionization scenario that, according to recent models, could still be conducted by the faintest galaxies \citep{Cain2021}.

In observations, the problem reduces to solving the equation for the ionizing emissivity of the average galaxy population ($\dot{n}_{\rm ion}$), i.e., the number of ionizing photons emitted per unit time and comoving volume \citep{Robertson2022}:
\begin{equation}
    \dot{n}_{\rm ion} = \fescabs \xiion \rho_{\rm UV}
    \label{eq:nion}
\end{equation}

\noindent where \fescabs\ stands for the absolute escape fraction of ionizing photons (i.e., the ratio between the number of \emph{escaping} versus \emph{produced} LyC photons by massive stars), and \xiion\ is the so-called ionizing photon production efficiency (ionizing photons generated per non-ionizing intrinsic UV luminosity). $\rho_{\rm UV}$ accounts for the non-ionizing UV luminosity density at the EoR, resulting from the integral of the UV luminosity function (UVLF, the number of galaxies per UV luminosity and comoving volume). 

The UVLF (or $\rho_{\rm UV}$, equivalently) of galaxies is relatively well-constrained up to the very high-redshift Universe \citep{Bouwens2015b, Davidzon2017, Bouwens2021, Donnan2023, Finkelstein2023}. At the EoR, some works observe a decrease in the UV luminosity density from $z = 8$ to 10 \citep{Oesch2018, Harikane2022}, compatible with a fast build-up of the dark matter halo mass function at those redshifts while, in contrast, some others do not find such a suppression at all \citep{McLeod2016, Livermore2017}. So far, UV-bright galaxies are found to be several orders of magnitude less numerous than the bulk of the UV-faint detected galaxies \citep{Atek2015}, and therefore thought to play a minor role during Reionization \citep[but see][]{MarquesChaves2022b}, although a possible excess in the number of sources at the bright-end of the UVLF \citep{RojasRuiz2020}, might make flip the argument towards an EoR governed by the ``oligarchs''. 

Our paradigm of the Early Universe is rapidly changing thanks to the \emph{James Webb Space Telescope} (\emph{JWST}). The JWST Early Release Science and Observations have just made possible the photometric selection \citep[e.g.,][]{Atek2023, Furtak2023, Labbe2023} and spectroscopic characterization \citep{ArellanoCordoba2022, Brinchmann2022, Schaerer2022b, Tacchella2022, Trussler2022, Carnall2023, Curti2023, Trump2023}  of galaxies at the EoR. Joint efforts combining both photometry and spectroscopy of some of the first JWST programmes \citep[see e.g.,][]{Harikane2023, Isobe2023, Nakajima2023} have shown as the best approach so far to study the properties of EoR galaxies in the context of galaxy evolution, with surveys such as GLASS \citep{Castellano2022b, Castellano2022a, Nanayakkara2022, Mascia2023, Santini2023}, UNCOVER \citep{Bezanson2022, Weaver2023}, CEERS \citep{Endsley2022, Topping2022, Finkelstein2023, Fujimoto2023, Tang2023} or JADES \citep{Cameron2023, Saxena2023, CurtisLake2023, Robertson2023}.

The ionizing production efficiency of galaxies (\xiion) is more uncertain. Surveys targeting H$\alpha$ emitters \citep[HAEs, see][]{Bouwens2016, Matthee2017b, Shivaei2018, Atek2022} and LAEs \citep{Harikane2018, Nakajima2018b} at intermediate redshifts show, in general, higher production efficiencies for the low-mass and UV-faint galaxies \citep{PrietoLyon2022}, although with a huge scatter within which \xiion\ spans a wide range of values depending on the galaxy type ($\log \xiion ({\rm Hz/erg}) = 24 - 26$). Reassuringly, the overall \xiion\ evolution with galaxy properties at these redshifts is in line with the usual formalism by which a $\log \xiion {\rm (Hz/erg)} \geq 25.2$ SFG population with constant SFH (over 100Myr) is able to fully reionize the Universe, assuming a fixed escape fraction of $\fescabs \geq 5\%$ \citep{Robertson2013}. Even so, given the stochastic nature of the LyC emission, linked to the predominantly bursty star-formation histories (SFHs) in low-mass SFGs \citep{Muratov2015}, these assumptions might not realistically apply anymore \citep[see discussion in][]{Atek2022}. 

Finally, the LyC absolute escape fraction (\fescabs) is by far the most unknown parameter in Eq.\ \ref{eq:nion} \citep{Faisst2016}. Starting a decade ago, the search for LyC emitters (LCEs), targeting Lyman Break Galaxies (LBGs) at $z = 0.7 - 4$ through expensive imaging \citep{Vanzella2010, Vanzella2012, Vanzella2015, Mostardi2015, deBarros2016, Grazian2016, Micheva2017, Japelj2017, Rutkowski2017, Grazian2017, Naidu2018, Alavi2020, Bian2020, Begley2022} and spectroscopic campaigns \citep{Steidel2001, Shapley2006, Bridge2010, Marchi2017, Mestric2021, Prichard2022} remained unsuccessful, where most of the estimates for global escape fraction of the SF galaxy-population relied on stacking and \fescabs\ upper limits. The systematic searches at $z \simeq 3$ by the Lyman Continuum Escape Survey \citep[LACES,][]{Fletcher2019, Nakajima2020} and other surveys by \citet{Saxena2022a} and \citet{RiveraThorsen2022} have provided the first statistical correlations between individual escape fraction measurements and diverse galaxy properties. Other remarkable individual detections are \citet{Vanzella2012, Vanzella2015, deBarros2016, Shapley2016, Vanzella2018, Saha2020, MarquesChaves2021, MarquesChaves2022b}.

However, our knowledge about LCEs and their physical properties is mainly due to the \emph{Keck Lyman Continuum Survey} \citep[KLCS,][]{Steidel2018, Pahl2021, Pahl2023} at $z \simeq 3$, and the pioneering work by \citet{Izotov16a, Izotov16b, Izotov18a, Izotov18b, Izotov2021, Izotov2022} and the recent \emph{Low-Redshift Lyman Continuum Survey} \citep[LzLCS,][]{Flury2022a} at $z \simeq 0.3$. Particularly, compactness, high ionization parameters (traced by optical line ratios), high star-formation rate (SFR) surface-densities, strong \lya\ emission, and low dust-attenuation (traced by the UV slope) seem to characterize the strongest LCEs at low-$z$ \citep[see also][]{Wang2019, Izotov2020}, with ionizing photon production efficiencies analogued to $z \simeq 6$ galaxies \citep{Schaerer2016, Chisholm2022}. Interestingly, and according to the analysis of new JWST observations by \citet{Mascia2023}, these characteristics resemble the properties of galaxies at the EoR \citep[see also][]{Endsley2022, Schaerer2022b, Cullen2023, Lin2023}.

Since the flux at LyC wavelengths will not be accessible at the EoR due to the increase of the IGM opacity \citep{Inoue2014}, indirect tracers of LyC radiation are needed. Thanks to the advent of the LzLCS \citep{Flury2022a}, in which both far-UV (FUV) and optical spectra are available for each galaxy, several \fescabs\ diagnostics have been statistically tested for the first time in \citet{Flury2022b}. Among them, those which properly account for the neutral gas and dust column densities as well as for geometrical effects \citep[see][]{Seive2022}, so that LyC radiation escapes only along favoured, cleared sight-lines in the interstellar medium (ISM), remain the most promising proxies. In particular, the peak separation of the \lya\ line \citep{Verhamme2017, Gazagnes2020, Naidu2022}, the depth of the low-ionization state (LIS) UV absorption lines \citep{Reddy2016b,Chisholm2018,Gazagnes2018,SL22} and the \ion{Mg}{ii} doublet ratio \citep{Henry2018, Chisholm2020, Xu2022} seem to closely probe the measured escape fraction of LzLCS galaxies \citep[but see cautionary theoretical work by][]{Mauerhofer2021, Katz2022}. 

In this work, we aim to indirectly study the ionizing properties of high-$z$ SFGs and their evolution along with the different galaxy-properties. For that, we make use of a sample of $\simeq 500$ deep, rest-frame ultraviolet spectra at $3 \leq z \leq 5$ drawn from the VANDELS survey \citep{McLure2018b, Pentericci2018, Garilli2021}.  In particular, LyC absolute escape fractions (\fescabs) are derived by measuring the depth of the absorption lines in combination with the UV attenuation, whilst ionizing production efficiencies (\xiion) are computed based on the best Spectral Energy Distribution (SED) fit to the FUV stellar continuum \citep[based on the work by][]{C19}. Our study complements ongoing efforts to understand the properties of LyC emitting galaxies at low \citep[LzLCS,][]{Flury2022a} and high redshifts \citep[KLCS,][]{Steidel2018}, thereby setting the pathway to interpret the ionizing signatures of EoR-galaxies, whose number of detections has been dramatically boosted thanks to the high-quality performance of the first \emph{JWST} observations.

The layout of this article is as follows. The VANDELS survey and sample definition are described in Sect.\ \ref{sec:data}. The code for fitting the stellar SED of VANDELS spectra, and the methods for predicting individual ionizing efficiencies and escape fractions are outlined in Sect\ \ref{sec:method}. The main results of this research, looking for correlations between \fescabs\ and \xiion\ with the different galaxy properties are summarized in Sect.\ \ref{sec:results}. Our results on the ionizing efficiency of high-$z$ galaxies are compared with different estimates in the literature in Sect.\ \ref{sec:discussion_xiion}, and finally the possible redshift evolution of the $\fescabs \times \xiion$ product is discussed in Sect.\ \ref{sec:discussion_lowzhighz}, by comparing our values against state-of-the-art low- and high-$z$ surveys. We summarize our findings in Sect.\ \ref{sec:conclusions}. 

Throughout this paper, a standard flat $\Lambda$CDM cosmology is used, with a matter density parameter $\Omega_{\rm M}$ = 0.3, a vacuum energy density parameter $\Omega_{\Lambda}$ = 0.7, and a Hubble constant of $H_0$ = 70~km s$^{-1}$ Mpc$^{-1}$. All magnitudes are in AB system \citep{Oke1983}, and we adopt a solar metallicity value of $12 + \log({\rm O/H})_{\odot}=8.69$. All the stellar metallicities are quoted relative to the solar abundance (${\rm Z_{\odot}}$) from \citet{Asplund2009}, which has a composition by mass of ${\rm Z_{\odot}} = 0.014$. Emission and absorption line equivalent widths (EWs) are given in rest-frame (unless stated otherwise), with positive (negative) EWs meaning lines seen in absorption (emission).

%% file: 2_data.tex
\begin{figure}
    \includegraphics[width=0.99\columnwidth, page=1]{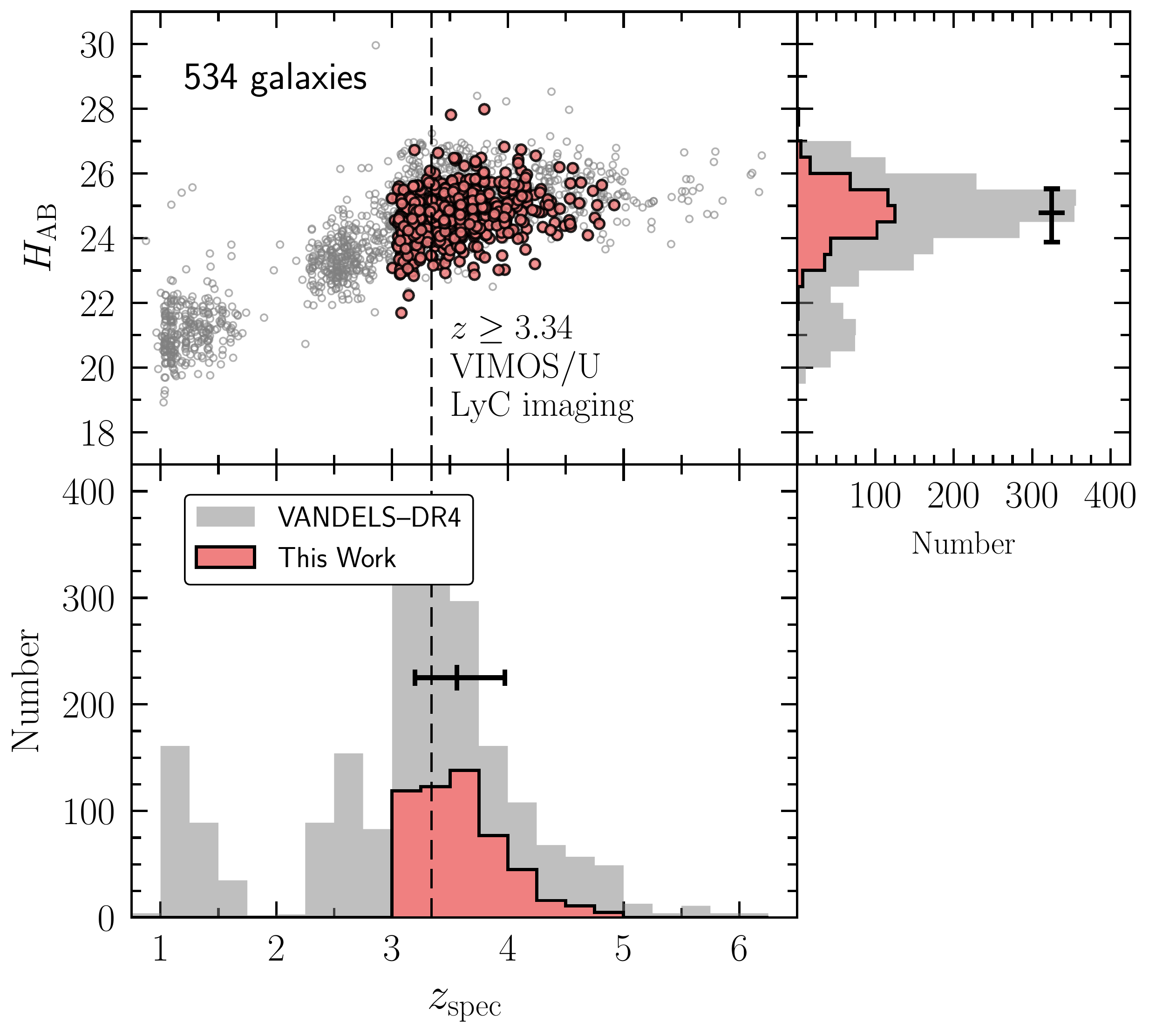}
\caption{$H-$band AB magnitude versus spectroscopic redshift ($z_{\rm spec}$) for the VANDELS-DR4 (gray open circles, 2087 galaxies). Our working sample of 534 galaxies is highlighted with red filled symbols, and the vertical dashed line indicates the minimum redshift at which LyC imaging is available through the VIMOS/U--band. The \emph{top-right and bottom} histograms show the $H_{\rm AB}$ and $z_{\rm spec}$ distributions of both the original VANDELS-DR4 and working samples (shaded gray and red, respectively). Errorbars comprise the 16$^{th}$, 50$^{th}$ and 84$^{th}$ percentiles of the distributions.}
\label{fig:sample}
\end{figure}

\section{Sample}\label{sec:data}
We rely on rest-frame FUV spectroscopy in order to estimate the ionizing properties of high-$z$ SFGs. At $3 \leq z \leq 5$, the rest-frame UV spectrum of galaxies ($1200-2000$\AA) is accessible from the ground through optical spectroscopy. The criteria by which we build our sample of high-$z$ rest-UV spectra (\ref{sub:vandels}), and the estimation of the main SED parameters (\ref{sub:seds}) and other spectroscopic features (\ref{sub:lya}) are described in detail in the following sections. Additionally, the survey properties of the two main comparison samples of LCEs in the literature are summarized (\ref{sub:literature}).

\subsection{Deep rest-frame UV spectra: the VANDELS survey}\label{sub:vandels}
The VANDELS survey \citep[final Data Release 4 in][]{Garilli2021} is an ESO Public Spectroscopic Survey composed of around $2100$ optical, high signal-to-noise ratio (S/N) and medium-resolution (R) spectra of galaxies at redshifts $z = 1 - 6.5$. The VANDELS footprints are centered on two of the HST \emph{Cosmic Assembly Near-Infrared Deep Extragalactic Legacy Survey} \citep[CANDELS,][]{Grogin2011,Koekmoer2011} fields, in particular the CDFS \citep[\emph{Chandra Deep Field South}, see][]{Guo2013} and the UDS \citep[UKIDSS \emph{Ultra Deep Survey}, see][]{Galametz2013}, but covering a wider area. The primary targets were selected from the parent CDFS and UDS photometric catalogs attending to the quality of their photometric redshifts \citep[see][for details]{McLure2018b, Pentericci2018}.

Ultra deep, multi-object optical spectroscopy (at $4800-10000$\AA\ observed-frame) was conducted with the VIMOS instrument \citep{LeFevre2003}, at the \emph{Very Large Telescope} (VLT), resulting in 1019 and 1068 sources observed in the CDFS and UDS, respectively, over 2087 measured spectroscopic reshifts in total. The final VANDELS catalog reaches a target selection completeness of $40\%$ at $i_{\rm AB} = 25$. VANDELS spectra have an unprecedented average S/N $\geq 7$ per resolution element over 80$\%$ of the spectra, thanks to exposure times spanning from 20 up to 80 hours on source. The VIMOS --Full-With at Half Maximum-- spectral resolution is $R \equiv \lambda/ \Delta \lambda \approx 600$ at 5500\AA, with a wavelength dispersion of 2.5\AA/pixel. For additional information about the survey design and data reduction, we encourage the reader to check \citet{McLure2018b, Pentericci2018} and \citet{Garilli2021} papers.

According to the selection criteria, the galaxies in VANDELS can be classified into passive/quiescent ($H_{\rm AB} \leq 22.5$, $1 \leq z_{\rm phot} < 2.5$), bright SFGs ($i_{\rm AB} \leq 25 $, $2.4 \leq z_{\rm phot} < 5.5$), LBGs ($25 \leq H_{\rm AB} < 27, i_{\rm AB} \leq 27.5$, $3 \leq z_{\rm phot} < 6.5$), and a smaller subset of AGN candidates \citep{Garilli2021}. For non-AGN type galaxies, the spectroscopic redshifts are ``flagged'' as a function of the reliability of  the measurement as marginal ($\simeq40\%$), fair ($\simeq80\%$) or robust ($\geq95\%$) reliability, corresponding to \texttt{flag = 1, 2, 3/4}, respectively \citep[see][for more details]{LeFevre2015, Garilli2021}. Based on a subset of VANDELS galaxies with \ion{C}{iii]}1908 detection, \citet{Llerena2022} found that the the systemic redshifts are slightly larger than the spectroscopic redshifts, with a mean difference of 0.002.

Our final sample comprises VANDELS-DR4 galaxies under the SFGs and LBGs categories only, whose spectroscopic redshifts fall in the $3 \leq z_{\rm spec} \leq 5$ range and have the highest reliability in the redshift measurement (i.e., \texttt{flag=3/4}). Given the VIMOS instrumental sensitivity, the redshift constraint ensures complete wavelength coverage from 1200 to 1600\AA\ for all galaxies in the sample (in practice, from \lya1216 to \ion{C}{iv}1550). We then compute the mean S/N per pixel over two regions free of absorption features, specifically $1350-1375$\AA\ and $1450-1475$\AA\ in the rest-frame (25\AA). Then, we select the galaxies with ${\rm S/N } \geq 2$ in the two spectral windows simultaneously ($\simeq 50\%$ of the remaining sample). Worth noticing is that the uncertainties of the DR4 spectra were systematically underestimated \citep[see][]{Garilli2021}. For this reason, we applied individual correction factors to the error spectra provided by the VANDELS collaboration (Talia et al., in preparation). The average correction factor is $\times 1.4$.

In summary, our VANDELS working sample includes 534 galaxies at $3 \leq z_{\rm spec} \leq 5$ with a median ${\rm S/N} \approx 5$ {in the 1400\AA\ continuum range}, where 297 sources were observed in the CDFS versus 237 in the UDS field. Fig. \ref{fig:sample} shows the observed $H-$band (either HST/F160W or VISTA/H) apparent magnitude distribution as a function of the spectroscopic redshift for the selected sources, together with the entire VANDELS-DR4 sample. The median (16$^{th}$ and 84$^{th}$ percentiles) F160W and $z_{\rm spec}$ of the working sample correspond to $\langle H_{\rm AB} \rangle = 24.8_{-0.9}^{+0.8}$ AB and $\langle z_{\rm spec} \rangle = 3.56_{-0.36}^{+0.41}$, respectively.

\begin{figure}
    \includegraphics[width=0.9\columnwidth, page=2]{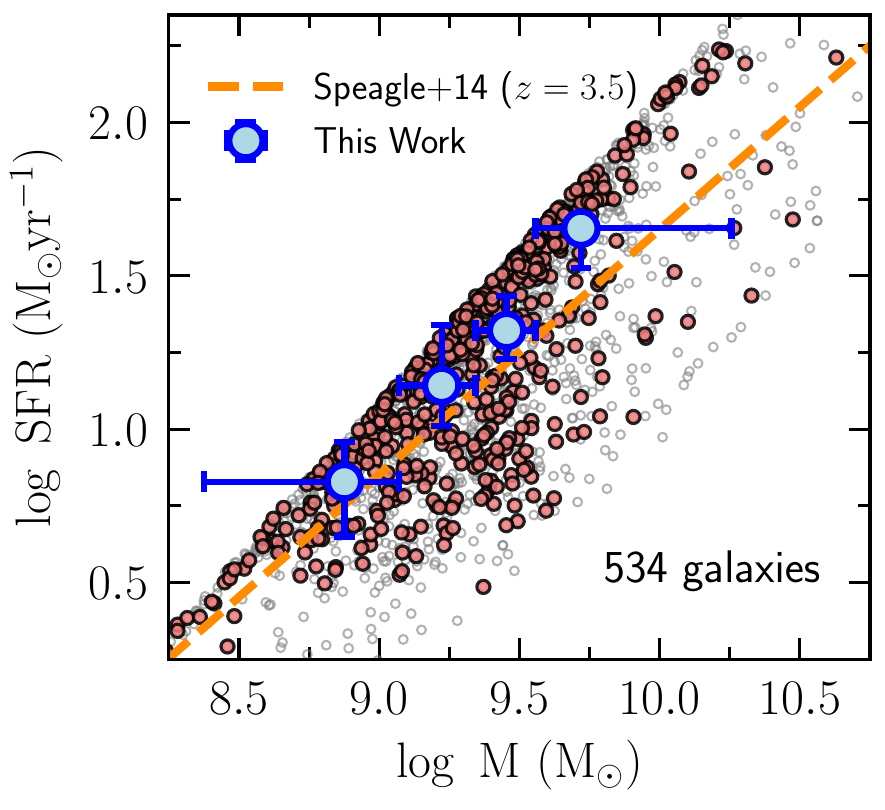}
\caption{Star-forming Main Sequence (MS) diagram i.e., ${\rm \log SFR - \log M_{\star}}$, for the VANDELS-DR4 (gray open circles). Our working sample is indicated through red filled symbols, and the large blue circles display the SFR running medians at each inter-quartile range of stellar mass. The golden dashed line follows the MS relation by \citet{Speagle2014}, for comparison. Our sample fall $\sim 0.1$dex. systematically above the MS at $z = 3.5$, but it still probes the upper bound of the SFG population at these redshifts.}
\label{fig:ms}
\end{figure}

\subsection{Photometric properties of the VANDELS galaxies: stellar masses}\label{sub:seds}
The wide imaging coverage of VANDELS allow a robust characterization of the SED of every galaxy \citep{Garilli2021}. The physical integrated properties of the VANDELS-DR4 galaxies were obtained by SED fitting using the Bayesian Analysis of Galaxies for Physical Inference and Parameter EStimation (\textsc{Bagpipes}) code\footnote{\textsc{Bagpipes} \citep{Carnall2018} is a state-of-the-art \textsc{Python} code for modeling galaxy spectra and fitting spectroscopic and photometric data, for details go to \url{https://bagpipes.readthedocs.io/en/latest/}.}. \textsc{Bagpipes} \citep[see][]{Carnall2018} inputs the spectroscopic redshifts measured by the VANDELS team \citep{Pentericci2018} and all the available CANDELS \citep{Galametz2013,Guo2013} plus ground-based photometry \citep{McLure2018b} to run a Bayesian algorithm which samples the posterior of the SED parameters. It makes use of the 2016 updated version of the \citet{BC03} models, including the MILES stellar spectral library \citep{MILES2011} and the stellar evolutionary tracks of \citet{Bressan2012} and \citet{Marigo2013}. An exponentially declining SFH is assumed, with a minimum timescale of 10 Myr and a minimum age of 10 Myr. The stellar metallicity was fixed to 0.2$Z_{\odot}$, coinciding with the average stellar metallicity found by \citet{Cullen2019} and \citet{Calabro2021}, probing a similar sample of galaxies in VANDELS. The dust attenuation was modeled using the \citet{Salim2018} prescription, and nebular emission was considered by adopting a constant ionization parameter of $\log(U)=-3$ in the fits.

For our VANDELS-DR4 working sample, the median stellar mass and SFR are $\langle \log (M_{\star}/\rm M_{\odot}) \rangle = 9.34_{-0.39}^{+0.31}$ and ${\rm \langle SFR/({\rm M_{\odot}~yr^{-1}}) \rangle} = 14.46_{~-8.47}^{+25.30}$, respectively, with $0.1 {\rm M_{\odot}}$ and $2.4 {\rm M_{\odot}~yr^{-1}}$ typical uncertainties. The ${\rm \log SFR - \log M_{\star}}$ distribution for the selected sample lays 0.1 to 0.2~dex above the SF main-sequence (MS) relation of \citet{Speagle2014} at similar redshifts, with a scatter of $\pm1$~dex in SFR, approximately \citep[see][]{McLure2018b}. In Fig.\ \ref{fig:ms}, we explicitly show how the SFR running medians at the 25$^{th}$, 50$^{th}$ and 75$^{th}$ stellar mass inter-percentile ranges fall above the \citet{Speagle2014} MS relation at $z = 3.5$. This means that, at comparable stellar masses (i.e., $\log(M_{\star}/{\rm M_{\odot})} = 8.5 - 10.5$), our sample is probing a slightly higher SFR regime than the bulk of SFGs, although it is still representative of the whole population given the intrinsic scatter of the MS at high-$z$ \citep[see the discussion in][]{Cullen2021}. A similar behavior was found in \citet{Calabro2022}. Finally, our sample also suffers from luminosity (or stellar mass) incompleteness i.e., the highest redshifts objects ($z \gtrsim 4$) are biased towards the brightest AB absolute magnitudes because of the current flux-limiting nature of the VANDELS survey.

\subsection{Spectral properties of the VANDELS galaxies: \lya\ equivalent widths}\label{sub:lya}
In this work, we make use of the methods and measurements provided by the VANDELS collaboration (Talia et al., in preparation) to compute the \lya\ equivalent width (\ewlya) of galaxies, either in emission or absorption.

The \lya\ (\ion{H}{i}1216) flux and equivalent width were measured by fitting a single-Gaussian profile to the \lya\ line using the dedicated \textsc{slinefit} code\footnote{\textsc{slinefit} is a \texttt{C++} simple software that can be used to derive spectroscopic redshifts from 1D spectra and measure line properties (fluxes, velocity widths, offsets, etc.). More in \url{https://github.com/cschreib/slinefit}, by Corentin Schreiber.}. {If the ${\rm S/N}$ of the line reach an input threshold,} the script allows a $\pm$1000~kms$^{-1}$ offset of the line center respect to rest-frame value given by the spectroscopic redshift. The continuum level was defined by the \citet{BC03} best-fit template to the entire spectrum (which gives a first order correction for underlying stellar absorption), and the flux and equivalent width was measured over $\pm$8000~kms$^{-1}$ each side of the line peak (see Sect.\ \ref{sub:fesc_abs} for the actual definition of the equivalent width). By convention, the fits report negative equivalent widths when the \lya\ line appears in emission, whereas a positive value is reported when the line is absorption dominated. Given the limited variety of \lya\ profiles observed in the VANDELS sample due to a medium $R \approx 600$ spectral resolution \citep[see][and methods therein]{Kornei2010, Cullen2020}, this approach only constitutes a first order estimation of \ewlya. We also note that, according to Talia et al. (in preparation), for 198 object ($\simeq 37\%$ of the sample) a single Gaussian fit did not provide a good fit to the \lya\ line. Rather, two Gaussian components were needed. However, a more detailed calculation is out of the scope of this paper, and we finally adopt single Gaussian fits as our fiducial estimates of the \lya\ flux. Doing so, the median \lya\ equivalent width is $\langle \ewlya~$(\AA)$\rangle = 4.58^{+19.01}_{-9.43}$ for our selected sample. According to the definition by \citet{Pentericci2009}, 104 out of 534 of galaxies ($\simeq 20\%$ of the sample) can be classified as LAEs with $\ewlya \geq -20$\AA\ \citep[although other authors may adopt different criteria, e.g., see][]{Stark2011, Nakajima2018b, Kusakabe2020}.

\subsection{Comparison samples of LCEs}\label{sub:literature}
Asides from our VANDELS sample, primarily composed by SFGs at $3 \leq z \leq 5$, we will consider as benchmarks other samples of confirmed LCEs from the literature (see Sect.\ \ref{sec:discussion_lowzhighz}). The LCEs comparison samples are the following:

\begin{description}
    \item[\textbf{The Low-$z$ Lyman Continuum Survey, or LzLCS,}] is a large HST programme \citep[][PI: Jaskot, HST Project ID: 15626]{Flury2022a} targeting 66 SFGs at $0.22 \leq z \leq 0.43$, selected from SDSS and GALEX observations. Each galaxy was observed with the low-resolution COS/G140L grating, covering the LyC and the non-ionizing rest-frame FUV continuum ($600-1450$\AA), with a spectral resolution of $R \approx 1000$ at 1100\AA. In combination with these 66 galaxies, the LzLCS also includes a compilation of 23 archival sources drawn from the literature \citep{Izotov16a,Izotov16b,Izotov18a,Izotov18b,Wang2019,Izotov2021}. Out of 89 galaxies, 50 were detected in the LyC, with a median absolute escape fraction of $\fescabs = 0.04_{-0.02}^{+0.16}$, while the remaining 39 galaxies show strong \fescabs\ upper limits, typically below $1\%$.
    \item[\textbf{The Keck Lyman Continuum Survey, or KLCS,}] is an extensive Keck/LRIS spectroscopic campaign \citep{Steidel2018} to obtain deep rest-UV spectra of selected LBGs at $2.8 \leq z \leq3.4$. The LRIS spectra of KLCS galaxies cover the LyC region and a wide FUV window longwards of the \lya\ line ($880-1650$\AA), at a spectral resolution of $R \approx 800$ for LRIS-B and $R \approx 1400$ for LRIS-R. In the most recent HST imaging follow-up by \citet{Pahl2021,Pahl2023}, they re-analyze a sample of 124 KLCS galaxies and look for foreground and low-redshift contamination that could potentially lie in projection within the angular extent of each LyC detected galaxy. In total, 13 individually-detected and 107 LyC undetected sources resulted in a sample-averaged escape fraction of $\fescabs = 0.06 \pm 0.01$. 
\end{description}

%% file: 3_methods.tex
\section{Methods}\label{sec:method}
In this section, we first describe the \textsc{FiCUS} code (Sect.\ \ref{sub:fits}), a customized \textsc{Python} script to fit the stellar continuum of extragalactic UV spectra. Then, we present the methodology to indirectly infer the ionizing absolute escape fraction of galaxies (\fescabs), by using the information provided by the UV absorption lines and the global SED fit to the non-ionizing UV continuum (Sect.\ \ref{sub:fesc_abs}). Finally, we discuss about the procedure for stacking individual VANDELS spectra, and the systematic effects included in the measurements due to the instrumental resolution and stacking (Sect.\ \ref{sub:stacks}).

\subsection{The \textsc{FiCUS} code: \textsc{Fi}tting the stellar Continuum of \textsc{Uv} Spectra}\label{sub:fits}
With the purpose of quantifying the stellar continuum properties underlying the VANDELS spectra, we have created \textsc{FiCUS} \footnote{The \textsc{FiCUS} code is publicly available and can be cloned from the author's \textsc{GitHub} repository: \url{https://github.com/asalda/FiCUS.git}.}. \textsc{FiCUS} is a customized \textsc{Python} script that stands for \emph{FItting the stellar Continuum of Uv Spectra} and, in short, it returns an estimation of the galaxy light-weighted stellar age, metallicity and dust extinction as well as other secondary SED parameters by using a combination of best-fit stellar population templates. This methodology was widely described and previously tested in \citet[][hereafter \citetalias{C19}]{C19}, and has been used in other papers such as \citet{Gazagnes2018, Gazagnes2020} and \citet[][hereafter \citetalias{SL22}]{SL22}. We refer the reader to those papers for similar approaches but slightly different applications of the current method. 

\begin{figure*}
    \includegraphics[width=0.9\textwidth]{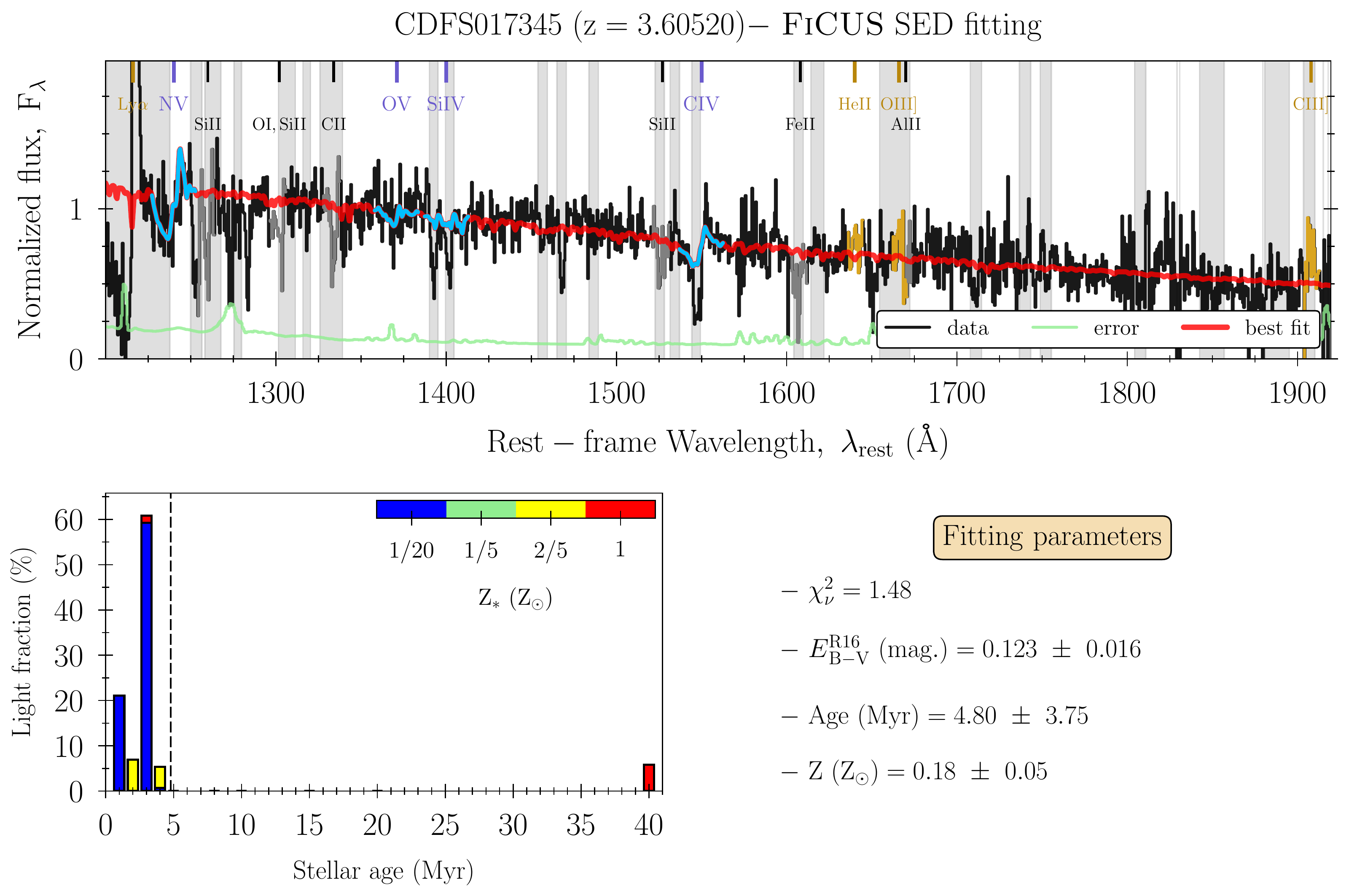}
\caption{{\bf Top:} \textsc{FiCUS} SED fitting results for the CDFS017345 VANDELS galaxy. The observed and error spectra are displayed in black and light-green. The best-fit stellar continuum is overplotted in red, while the spectral regions masked during the fit are shown in gray. The spectrum, shown in ${\rm F_{\lambda}}$ units, has been normalized over $1350-1370$\AA. The most prominent stellar features, nebular and ISM absorption lines are indicated with blue, gold and black vertical lines at the top part of the figure. {\bf Bottom left:} light-fractions as a function of age and metallicity for the best-fit composite stellar population model. {\bf Bottom right:} summary chart with the best-fit SED fitting parameters, i.e., reduced-$\chi^2$, \ebv\ (in mag., \citetalias{R16}), light-weighted stellar age (in Myr) and metallicity (in \Zsun).}
\label{fig:SEDexample}
\end{figure*}

\subsubsection{Stellar population models}
As described in \citetalias{SL22}, the stellar continuum modeling was achieved by fitting every observed spectrum with a linear combination of multiple bursts of single-age and single-metallicity stellar population models. By default, \textsc{FiCUS} inputs the fully theoretical \textsc{starburst99} single-star models without stellar rotation \citep[S99,][]{Leitherer2011, Leitherer2014} using the Geneva evolution models \citep{Meynet1994}, and computed with the \textsc{WM-Basic} method \citep{Pauldrach2001, Leitherer2010}. The S99 models assume a \citet{Kroupa2001} initial mass function (IMF) with a high-(low-)mass exponent of 2.3 (1.3), and a high-mass cutoff at 100~M$_{\odot}$. The spectral resolution of the S99 models is $R \equiv \lambda / \Delta \lambda \simeq 2500$, and it remains approximately constant at FUV wavelengths. 

Four different metallicities (0.05, 0.2, 0.4 and 1 $Z_{\odot}$) and ten ages for each metallicity (1, 2, 3, 4, 5, 8, 10, 15, 20 and 40 Myr) were chosen as a representative set of $\times$40 models for our high-$z$ UV spectra. In detail, ages were chosen to densely sample the stellar ages where the stellar continuum features appreciable change \citepalias[see][]{C19}. For example, at older ages than 40~Myr, the FUV stellar continuum does not appreciably change such that a 40~Myr model looks very similar to a 100~Myr population. We caution the reader that the S99 high-resolution models do not densely sample the HR diagram at effective temperatures below 15,000K \citep{Leitherer2010}. The cooler models have artificially bluer continuum slopes, which makes older stellar populations appear slightly bluer than expected for their temperatures \citep[for details, see][]{Chisholm2022}.

Finally, a nebular continuum was generated by self-consistently processing the stellar population synthesis models through the \textsc{cloudy v17.0} code\footnote{\url{https://trac.nublado.org/}} \citep{Ferland2017}, assuming similar gas-phase and stellar metallicities, an ionization parameter of $\log(U)=-2.5$, and a volume hydrogen density of n$_{H} = 100~{\rm cm}^{-3}$. The output nebular continua for each stellar population were added to the stellar models. The inclusion of the nebular continuum is only appreciable at the youngest ages of the bursts ($\leq10$Myr) at wavelengths $\geq 1200$\AA, and produces redder stellar models than before, which has a pronounced effect on the fitted \ebv\ of the young stellar populations. \citetalias{C19} tested the effect of different ionization parameters on the fitted stellar ages and metallicities, and found that the \emph{goodness} of the fits does not significantly change for different $\log(U)$ values.

\subsubsection{Fitting method}
The observed spectra were first manually placed into the rest-frame by multiplying by the corresponding $1/(1+z)$ factor. Both the spectra and the models were then normalized by the median flux within a wavelength interval free of stellar and ISM features ($1350-1370$\AA), and all the fits were performed in the same rest-wavelength range ($1200-1925$\AA\ in the case of VANDELS spectra). Finally, the models were convolved by a Gaussian kernel to the instrumental resolution ($R{\rm (VIMOS)} \approx 600$).

First, the non-stellar features and the spectral regions that are affected by host-galaxy ISM absorptions and sky subtraction residuals had to be masked out. Bad pixels with zero flux were neither considered in the fit. When ran with \textsc{FiCUS}, the UV stellar continuum ($F^{\star}(\lambda)$) is fitted with a linear combination of multiple S99 models plus the nebular continuum. $F^{S99}_{ij}$ single stellar population is usually not able to fully reproduce both the stellar features and the shape of the continuum of extragalactic spectra, such that a combination of instantaneous burst models is actually needed \citepalias{C19}.

Adopting a simple geometry where the dust is located as a uniform foreground screen surrounding the galaxy\footnote{The effect of a clumpy gas-to-dust geometry has been discussed in \citet{Gazagnes2018} and \citetalias{SL22}. Regrettably, the current resolution and S/N of the VANDELS spectra is not enough to disentangle between the two geometries, so we assume the most simple scenario of a uniform screen of dust as a our fiducial case.}, this results in: 
\begin{equation}
    \begin{aligned}
    F^{\star}(\lambda) &= 10^{-0.4 k_{\lambda} \ebv} \sum_{i,j} X_{ij} F^{S99}_{ij}, \\
    &i \equiv 1, 2, 3, 4, 5, 8, 10, 15, 20, 40 ~Myr \\
    &j \equiv 0.05, 0.2, 0.4, 1 ~Z_{\odot}\\
    \end{aligned}
    \label{eq:s99_fit}
\end{equation}

\noindent where $10^{-0.4 A_{\lambda}}$ is the UV attenuation term, $A_{\lambda} = k_{\lambda} \ebv$, and $k_{\lambda}$ is given by the adopted dust-attenuation law. The $X_{ij}$ linear coefficients determine the weight of single-stellar population within the fit, and the best fit is chosen through a non-linear $\chi^2$ minimization algorithm with respect to the observed data \citep[\textsc{lmfit}\footnote{A \textsc{python} version of the \textsc{lmfit} package can be found at \url{https://lmfit.github.io/lmfit-py/}.} package, ][]{lmfit}. Errors were derived in a Monte-Carlo way, varying observed pixel fluxes by a Gaussian distribution whose mean is zero and standard deviation is the 1$\sigma-$error of the flux in the same pixel, and re-fitting the continuum over $N$ iterations per spectrum (we chose $N = 100$, enough realizations to sample the posterior continuum so that it approaches ``Gaussianity'' on each pixel).

Fig. \ref{fig:SEDexample} shows the \textsc{FiCUS} output for one of the galaxies in VANDELS (\texttt{objID:CDFS017345}, at $z = 3.61$), with the observed (in black) and fitted stellar continuum (in red). The bottom panel of this figure additionally shows the distribution of the light-fractions at a given burst-age and metallicity for the CDFS017345 best-fit. This galaxy is dominated by a young and low-metallicity population ($\simeq 60\%$ of the total light at 3Myr) with 4.80 Myr average light-weighted age and 0.18 \Zsun\ light-weighted metallicity (see the following section). CDFS017345 is moderately attenuated, with a UV dust-attenuation parameter of $\ebv \simeq 0.12$~mag. In Fig. \ref{fig:SEDexample}, and together with other nebular (gold) and ISM lines (black) usually present in VANDELS spectra, we also highlight (in blue) the main stellar features that helped our algorithm to estimate the age and the metallicity of the stellar population: the \ion{N}{v}$\lambda$1240 and \ion{C}{iv}$\lambda$1550 P-Cygni stellar wind profiles (see also \ion{O}{v}$\lambda$1371 and \ion{Si}{iv}$\lambda$1402). Note that the inclusion of \ion{C}{iv}1550 requires a careful masking of the strong ISM component of this line, where only the blue wing of the P-Cygni was used in the fit. While \ion{N}{v}$\lambda$1240 is mostly sensitive to the age of the burst --with the P-Cygni profile being more prominent at younger ages--, the \ion{C}{iv}$\lambda$1550 feature, on the other hand, changes accordingly to the metallicity \citepalias{C19} of the stellar population --whose asymmetric profile is stronger at higher stellar metallicities--. Therefore, these two distinct spectral features partially break the age-$Z_{\star}$ degeneracy in the FUV, although the method is still affected by age-attenuation degeneracy effects.

\subsubsection{Results}
Given the best-fit SED for the stellar continuum and the best-fit $X_i$ coefficients in Eq.\ \ref{eq:s99_fit}, \textsc{FiCUS} also provides a handful of other secondary SED parameters, that we described here. 

\begin{figure}
    \includegraphics[width=0.99\columnwidth, page=3]{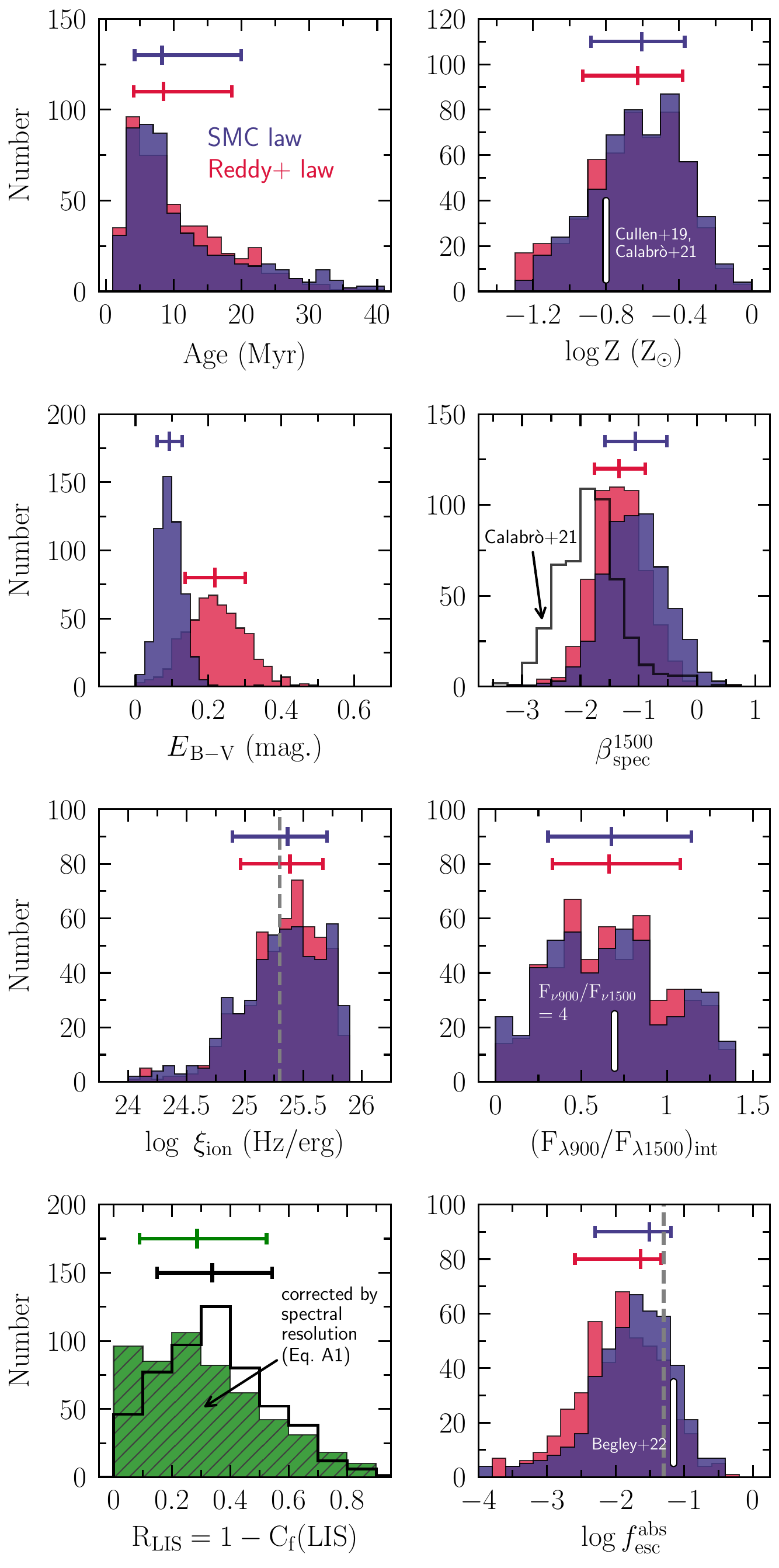}
\caption{Histograms showing the distribution of diverse secondary products resulting from our \textsc{FiCUS} SED fitting to the VANDELS spectra. {\bf From top-left to bottom-right:} light-weighted stellar Age, $\log Z_{\star}$, \ebv, \bslope, $\log \xiion$, \Fion, measured \rlis\ and predicted $\log \fescabs$. Results derived using \citetalias{R16} or \citetalias{SMC} attenuation laws are plotted in red and dark-blue, respectively. The error bars on top of each histogram encompass the 16$^{th}$, 50$^{th}$ and 84$^{th}$ percentiles. The distributions of the mean residual flux of the LIS lines are shown in different colors because, by definition, this quantity does not depends on the attenuation law. Dashed vertical lines indicate $\log \xiion ({\rm Hz/erg}) = 25.2$ and $\fescabs = 5\%$.}
\label{fig:fit_histos}
\end{figure}

\subparagraph{{\bf Stellar age and metallicity}}
The light-weighted stellar age and metallicity ($Z_{\star}$) of the best-fit stellar population can be easily obtained using the $X_k$ weights as:
\begin{equation}
    {\rm Age} = \dfrac{\sum_k X_k {\rm age}_k}{\sum_k X_k}, ~Z_{\star} = \dfrac{\sum_k X_k z_k}{\sum_k X_k}
    \label{eq:avg_age_Z}
\end{equation}

\noindent where ${\rm age}_k, z_k$ are the age and metallicity of $k$-th synthetic stellar model. For instance, in the example of Fig. \ref{fig:SEDexample}, the light-weighted age is 4.80 Myr and the light-weighted metallicity is 0.18 \Zsun. The median stellar age and metallicity of the VANDELS sample are ${\rm \langle Age \rangle} = 8.5_{-4.4}^{+10.1}$~Myr and ${\rm \langle Z_{\star}/\Zsun \rangle} = 0.23_{-0.12}^{+0.18}$, respectively. \citet{Cullen2019} and \citet{Calabro2021} estimate $\langle \log (Z_\star/\Zsun)\rangle \simeq -0.8$ from independent methods (see white vertical line in Fig.\ \ref{fig:fit_histos}). Compare to these works, our average metallicity is slightly higher: $\langle \log ( Z_\star/\Zsun)\rangle \simeq -0.63$, although still compatible within the spread of the distributions. 

Our shift in $\log Z_{\star}$ with respect to the previous works must be driven by use of instantaneous burst of star-formation instead of a continuous SFH in the theoretical stellar models. Our election of single-burst models is motivated by the fact that mixed-age populations usually do a better job when reproducing age-sensitive tracers (such us \ion{N}{v}1240) of individual FUV galaxy-spectra \citepalias[see Sect. 5.3 in][]{C19}. Moreover, by accounting for the dominance of young stellar populations, bursty star formation histories, which are expected to hold for intermediate-to-low-mass SFGs at high-redshift \citep[e.g.,][]{Trebitsch2017}, produce significantly more ionizing photons, at higher energies, than continuous star formation histories. 

The effect that the choice of either constant SFRs or instantaneous bursts of star-formation have in the stellar age, metallicity and ionizing properties of the mean SFG galaxy population at $z \simeq 3$ will be discussed in a future publication. We refer the reader to \citet{Cullen2019} and \citet{Calabro2021} for more details on the stellar metallicity of VANDELS galaxies.

\subparagraph{{\bf Dust-attenuation parameter: \ebv}}
The amount of dust-attenuation is given by the $10^{-0.4 A_{\lambda}}$ term in Eq.\ \ref{eq:s99_fit}, where $A_{\lambda} = k_{\lambda} \ebv$ is defined as the ``UV attenuation'', and the specific functional form for $k_{\lambda}$ is determined by the dust-attenuation law. Therefore, the resulting best-fit values for \ebv\ will differ depending on the assumed dust-law, and this will have important consequences for other related quantities which explicitly depend on \ebv, like the absolute escape fraction (\fescabs) as we will discuss later on (see Sect.\ \ref{sec:discussion_lowzhighz}). 

With the goal of testing the influence of the dust-attenuation law in our results, we use two extreme cases for $k_{\lambda}$ in SF galaxies \citep{Shivaei2020}: the \citet[][hereafter \citetalias{R16}]{R16} and Small Magellanic Cloud \citep[][hereafter \citetalias{SMC}]{SMC84, SMC85, SMC} attenuation curves. As widely discussed in the dedicated sections in \citetalias[][and references therein]{SL22}, the use of \citetalias{R16} is, on the one hand, motivated by the fact that this law is one of the only properly defined below 1000\AA\ by a significant number of galaxies. On the other hand, the use of steeper SMC-like curves has appeared to be more suitable for high-$z$ SFGs \citep{Salim2018}, low-metallicity starburst \citep{Shivaei2020}, and LCEs \citep{Izotov16b}. 

After performing the SED fits with \textsc{FiCUS} using both the \citetalias{R16} and \citetalias{SMC} prescriptions for $k_{\lambda}$, we find an average attenuation parameter of $\langle \ebv \rangle = 0.22_{-0.08}^{+0.09}$~mag. for \citetalias{R16} and $\langle \ebv \rangle = 0.10_{-0.03}^{+0.04}$~mag. when using \citetalias{SMC} (see Fig. \ref{fig:fit_histos}). This corresponds to UV attenuations of 1.82 and 2.83~mag. at 1600\AA\ and 912\AA, using \citetalias{R16} (1.25, 2.63~mag. for \citetalias{SMC}). Although the resulting light-weighted stellar ages and metallicities are similar in both sets of fits (similar light-fractions), the UV attenuation term $10^{-0.4 A_{\lambda}}$ is slightly higher for \citetalias{SMC} at all wavelengths, meaning that the fitted $X_i$ coefficients are slightly lower for \citetalias{SMC} than \citetalias{R16} law, so that both fits can match the observations similarly (reduced-$\chi^2$ distributions are similar). Lower $X_i$ coefficients for \citetalias{SMC} means that all absolute quantities derived from the intrinsic SEDs will also change accordingly.

\subparagraph{{\bf UV-continuum slope (at 1500\AA): \bslope}}
The UV spectroscopic continuum slope at 1500\AA, also called $\beta-$slope, was computed by fitting a power-law of the form $F_{\lambda} \propto \lambda^{\beta}$ to every individual spectrum \citep{Meurer1999}. To do so, we take the average flux density in seven 15\AA-width spectral windows, between $1275-1825$\AA\ \citep[similar to the ones in][]{Calzetti1994}, and fit a linear relation to the $\log F_{\lambda} - \log \lambda$ values using \textsc{Curve-Fit}\footnote{\textsc{Curve-Fit} is a \textsc{python}-based optimization algorithm included in the \textsc{SciPy} package for scientific analysis: in \url{https://docs.scipy.org/doc/scipy/reference/generated/scipy.optimize.curve_fit.html}.}. For consistency, and in order to avoid inhomogeneous wavelength sampling for every source due to the different redshifts, $F_{\lambda}$ corresponds to the continuum flux obtained from the best-fit SED model of the galaxy ($F^{\star}(\lambda)$), which was described in the previous Sect.\ \ref{sub:fits}. This assumption imprints an intrinsic dependency of the $\beta-$slope on the chosen attenuation law, since it modifies the shape of the dust-attenuated spectrum through the $k_{\lambda}$ term. Thus, median of the $\beta-$slope distribution for the VANDELS sample is $\langle \bslope \rangle = -1.34_{-0.42}^{+0.45}$ for \citetalias{R16} and $\langle \bslope \rangle = -1.06_{-0.51}^{+0.54}$ for \citetalias{SMC}. When compared to the slopes obtained in \citet{Calabro2021} for the same objects ($\langle \beta_{1500} \rangle \simeq -1.80$), our spectroscopic $\beta-$slopes are overall redder (lower) than the former by $\simeq 0.4$, approximately (see distribution in Fig.\ \ref{fig:fit_histos}). In \citet{Calabro2021}, the UV slopes were derived by fitting a power-law to the available multi-band VANDELS photometry, taking the photometric bands whose bandwidths lie inside the 1230–2750\AA\ rest-frame wavelength range. We investigated the possible reasons that may reach to this discrepancy and conclude that it is mainly caused by the use of a wider range in wavelength with respect to the spectroscopic slopes. Although more effects will be discussed later in Sect.\ \ref{sec:discussion_lowzhighz}, there is always the possibility that the SED of galaxies in the rest-UV is not a perfect power law \citep{Bouwens2012}, a behavior that was previously observed in the VANDELS galaxies by \citet{Calabro2021}. Differential shifts between UV slopes derived from different methods (either photometry or spectroscopy) have been already reported in the literature \citep[e.g., see][]{Hathi2016}.

\subparagraph{{\bf Ionizing photon production efficiency: \xiion}}
The ionizing photon production efficiency is defined as the total number of ionizing photons per unit time produced by a radiation field ($Q$) normalized by its intrinsic UV luminosity ($L_{\rm UV}$). In practice \citepalias[see][]{C19}:

\begin{equation}
    \xiion~[{\rm Hz/erg}] = \dfrac{Q~[{\rm s^{-1}}]}{L_{\nu 1500}~[{\rm erg~s^{-1}~Hz^{-1}]}}
    \label{eq:xiion}
\end{equation}

\noindent where $L_{\nu 1500}$ is the intrinsic (dust-free) SED luminosity at 1500\AA\ (in $F_{\nu}$ units) and $Q$ is the number of ionizing photons produced by the best-fit stellar population. $Q$ is calculated as the integral of the intrinsic SED below the Lyman limit ($\lambda < \lambda {\rm (H^0)} = 912$ \AA, or equivalently $ E > E {\rm (H^0)} = 13.6$ eV): $Q = \int_{<\lambda {\rm (H^0)}} F^{\star}(\lambda) d\lambda$. When averaged over the whole galaxy population and integrated over the full range of UV luminosities, \xiion\ can be used to compute the total emissivity of ionizing photons ($\dot{n}_{\rm ion}$) at any redshift (see Eq.\ \ref{eq:nion}).

For typical SF galaxies that can be described through a single-burst of star-formation, S99 templates predict an exponentially declining \xiion\ with age, whose $\log \xiion$ values span over $26-24$~Hz/erg for $1-10$Myr stellar populations at any metallicity, dropping dramatically towards ages older than 10~Myr \citepalias{C19}. In contrast, mixed-age models can provide systematically higher \xiion\ for a given mean age than instantaneous burst models at the same evolutionary stage. Accordingly, our mixed-age fits give a median \xiion\ of $\langle \log \xiion \rangle = 25.30_{-0.42}^{+0.28}~{\rm Hz/erg}$ using the \citetalias{R16} law. \xiion\ does not strongly depend on the assumed attenuation law because it has been calculated as the ratio between two absolute quantities ($Q$ and $L_{\rm UV}$). Since the fitted stellar ages and metallicities are similar for both \citetalias{R16} and \citetalias{SMC} (as they are primarily fixed by single spectral features), the shape of the intrinsic SED is also similar for both laws and therefore \xiion\ stays unaltered.

\subparagraph{{\bf Intrinsic 900-to-1500\AA\ flux ratio: \Fion}}
The ionizing-to-nonionizing flux ratio, namely at 900-to-1500\AA, depends on the physical properties of galaxies like the mean stellar age, metallicity, IMF and star formation history. Following \citetalias{C19}, S99 single-burst bases set a limit of $\Fion \leq 2$, exponentially declining with age down to negligible values at ages older than 20Myr, where $F_{\lambda 900}^{\rm int} \approx 0$. Again, a mixed-age population could in principle increase the flux ratio at intermediate ages with respect to a single-burst of star-formation. The median of the 900-to-1500\AA\ flux ratio for the VANDELS sample results in $\langle F_{\rm \lambda 900} / F_{\rm \lambda 1500} \rangle_{\rm int} = 0.66_{-0.32}^{+0.41}$. This quantity does not show any dependency on the attenuation law for the same reasons as \xiion.

In photometric campaigns, when searching for LCEs at high-$z$ \citep[e.g.,][]{Grazian2016, Grazian2017}, authors usually relate the observed to the intrinsic 900-to-1500\AA\ flux ratio in order to infer the relative LyC escape fraction of galaxies. Usually, these works only have access to the observed but not the intrinsic ratio, and they end up by assuming a constant value typically set by stellar population models predictions \citep[e.g,][]{Steidel2001}. In the same works, a more common way to look at this quantity is indeed the 1500-to-900\AA\ flux ratio but in $F_{\nu}$ units. Typically assumed values in the literature are $\langle F_{\rm \nu 1500} / F_{\rm \nu 900} \rangle_{\rm int} = 3-5$ \citep[see e.g.,][]{Guaita2016}. We obtain a median value of $\langle F_{\rm \nu 1500} / F_{\rm \nu 900} \rangle_{\rm int} \approx 4.2$ for the VANDELS sample at $3 \leq z \leq 5$ (Fig.\ \ref{fig:fit_histos}), in agreement with the previous studies.\\

Fig. \ref{fig:fit_histos} offers a summary of the above-mentioned distributions: Age, $Z_{\star}$, \ebv, \bslope, \xiion, and \Fion; where the values resulting from our SED fits using either \citetalias{R16} or \citetalias{SMC} law are shown in red and dark-blue, respectively. Error bars cover the 16$^{th}$ to 84$^{th}$ percentiles of each distribution, whose values have been quoted with respect to the median in the current section. 

\subsection{Predicting ionizing escape fractions using UV absorption lines}\label{sub:fesc_abs}

\subsubsection{The picket-fence model}
The observation that neutral lines of hydrogen and other strong low-ionization state (LIS) lines do not become black at minimum depth (or maximum optical depth) suggests a partial covering of the stellar continuum sources by the same cold, neutral and low-ionized gas \citep{Heckman2001, Heckman2011}. In this scenario, it is expected that the residual flux of the absorption lines correlates with the fraction of LyC photons that escape the galaxy via uncovered channels.

The commonly used `picket-fence' model \citep{Reddy2016b,Vasei2016} connecting the UV absorption features of a galaxy with the escaping ionizing radiation assumes a galaxy described by a patchy, ionization-bounded ISM \citep{Zackrisson2013}, where both the neutral and low-ionized enriched gas are distributed in high-column-density regions surrounding the ionizing radiation field (clumps). The fraction of sight-lines which are optically thick to the transition along these dense clouds is usually parametrized by the so-called covering fraction, $C_f({\lambda_i})$. Optically thick gas absorbs all of the continuum light at a given velocity whereas optically thin gas transmits all of the continuum. If the dust is homogeneously distributed as a foreground screen on top of the stars, the residual flux of the lines can be simply related to the gas covering fraction as follows:
\begin{equation}
    R_{\lambda_i} = 1 - C_f(\lambda_i),
    \label{eq:cfA}
\end{equation}

\noindent This simple model also assumes that the lines are described by a single gas component or, in other words, that all velocity components of the gas have the same covering fraction. Additionally, if one accounts for the dust attenuation within the galaxy (\ebv), the escape fraction of ionizing photons (\fesclis) can be predicted from the depth of the neutral and other LIS absorption lines using the following formulae \citep{Reddy2016b,Gazagnes2018,Steidel2018}:
\begin{equation}
    \fesclis = 10^{-0.4k_{912}\ebv} \times (a \times R_{\lambda{_i}} + b),
    \label{eq:fesc_LIS}
\end{equation}

\noindent where the \ebv\ is the UV dust-attenuation parameter measured according to the methods described in the previous Sect.\ \ref{sub:fits}, using a uniform screen geometry (see Eq.\ \ref{eq:s99_fit}), and $R_{\lambda{_i}}$ is the measured residual flux of the $\lambda_i$ transition. The $[a, b]$ coefficients correspond to the LIS to \hi\ lines residual flux conversion \citepalias[see \cite{Gazagnes2018}, \cite{Reddy2022} and][]{SL22}. 

This methodology has been successfully validated against direct measurements of the absolute escape fraction of low-$z$ LCEs in \citet{Chisholm2018} and \citet{Gazagnes2020} and recently thanks to the advent of the LzLCS by \citetalias{SL22}, where we pointed to this method as a good predictor for the real \fescabs\ of galaxies at any redshift. The depth of the LIS lines have also been applied to predict the escape fraction of high-$z$ galaxy composites in \citet{Steidel2018} and \citet{Pahl2021}, and some individual high-$z$ spectra in \citet{Chisholm2018}, with reasonable agreement with the observed \fescabs\ values. 

However, this approach is not without of caveats \citepalias[see Sect.\ 7.4 in][]{SL22}: the choice of the dust-attenuation law and the assumptions on the gas and dust geometry are some of the limitations of the picket-fence model. Indeed, as suggested by \citet{Mauerhofer2021}, a `picket-fence' gas/dust distribution is shown to be a very simplistic approximation to the real ISM geometry in state-of-the-art galaxy simulations. In principle, all these assumptions and model dependencies can contribute to explain the scatter seen in the observed $\rlis - \fescabs$ relation \citepalias[see][]{SL22}, but this model could still be a good approximation for unresolved high-$z$ studies, where the LyC emission coming from a \emph{single} sight-line usually dominates.

\subsubsection{Absorption line measurements}
Our goal is to apply Eq.\ \ref{eq:fesc_LIS} to VANDELS spectra in order to indirectly estimate their ionizing escape fractions. To do so, we first measured the equivalent widths ($W_{\lambda}$) and residual fluxes ($R_{\lambda}$) for a set of $\times 4$ UV LIS lines, namely \ion{Si}{ii}1260, \ion{O}{i}+\ion{Si}{ii}1302, \ion{C}{ii}1334 and \ion{Si}{ii}1527, which had simultaneous wavelength coverage in all the spectra at $3 \leq z \leq 5$. The equivalent width ($W_{\lambda_i}$) was then computed individually for every absorption line following \citet{Trainor2019}:
\begin{equation}
    W_{\lambda_i} = \int _{\Delta \lambda_i} \left(1 - f_\lambda/F^\star(\lambda)\right) d\lambda
    \label{eq:ew}
\end{equation}

\noindent where $f_{\lambda}$ is the observed spectral flux density and $F^{\star}(\lambda)$ is the modeled stellar continuum. The integration window ($\Delta \lambda_i$) was defined as $\pm$1000kms$^{-1}$ from the wavelength of minimum depth for the line in question. Then, the residual flux was measured as the median over a narrrow velocity interval of $\pm 150$kms$^{-1}$ around the minimum flux of the line, or equivalently: 
\begin{equation}
    R_{\lambda_i} = \langle f_{\lambda_i}/F^{\star}\rangle
    \label{eq:r}
\end{equation}

\noindent One of the conditions of applicability of the picket-fence model requires that the column density of gas is large enough that the absorption lines are saturated in the curve of growth (i.e., optically thick limit). In order to test the condition of saturation, we performed a similar analysis to the one in \citetalias{SL22} \citep[see also][]{Trainor2015, Calabro2022}, comparing equivalent-width ratios for transitions of the same ion at different wavelengths. Most of the galaxies have a \ion{Si}{ii}1260-to-\ion{Si}{ii}1527 equivalent-width ratio which is compatible within the optically thick limits ($W_{1527}/W_{1260} \leq 2.55$) given by the theoretical curve-of-growth for these transitions \citep{Draine}. We conclude that the ISM conditions are such that \ion{Si}{ii} is optically thick (saturated), and we assume that \ion{C}{ii} is also saturated for typical Si/C abundances \citep[see discussion in][]{Mauerhofer2021}. \citet{Calabro2022} independently found optically thick \ion{Si}{ii} ($\log (N_{\rm SiII} / {\rm ~cm^{-2})} > 12.7$) along a sample of VANDELS \ion{C}{iii} emitters at similar redshifts, whereas higher ionization lines like \ion{Si}{iv}1400 were closer to the optically thin limit as they probe more rarefied gas.

\subparagraph{The effect of resolution on the absorption lines}
When measuring absorption line properties from observed spectra, the low spectral resolution ($R$) tends to make the lines broader and less deep, and the actual residual flux ($R_{\lambda_i}$) can be overestimated. To account for these effects we performed mock absorption line simulations. Considering the picket-fence model with a uniform dust-screen, we simulated \ion{Si}{ii}1260 absorption Voigt profiles assuming Gaussian distributions for the column density of \ion{Si}{ii} ($N_{\rm SiII}, {\rm cm^{-2}}$), the Doppler broadening parameter ($b, {\rm kms^{-1}}$) and the line velocity shift ($v, {\rm kms^{-1}}$). The simulated spectra were then degraded to $R{\rm (VIMOS)} \approx 600$ spectral resolution. Finally, the impact of the S/N was folded onto the simulations by assuming a constant ${\rm S/N} = 5$ that matches the median of the S/N distribution of the VANDELS spectra. A set of $\times$100 simulations were run for every of the input covering fractions in the $C_f(\lambda_i)= 0-1$ range, and the median measured residual flux of each distribution was then compared to the theoretical value of $1-C_f$. See App.\ \ref{app:resolution} for a more comprehensive view of our simulations.

As a result, a theoretical residual flux of $(1-C_f) = 0.1, 0.3$ and 0.5 would be observed as $\rlis = 0.25, 0.4$ and 0.55 due to the \emph{smearing} effect of the VIMOS instrumental resolution (i.e., $60\%, 25\%$ and $10\%$ relative error), while a negligible correction will be applied at $(1-C_f) \geq 0.6$ because this $C_f$ regime is dominated by the ${\rm S/N}$ of the spectra. The resulting calibration (Eq.\ \ref{eq:cal_R}) was applied as a correction factor to the individually measured residual fluxes.

\subsubsection{Indirect \fescabs\ estimations}
The mean residual flux of the LIS lines, \rlis, is calculated by taking the average of the individual \ion{Si}{ii}1260, \ion{O}{i}+\ion{Si}{ii}1302, \ion{C}{ii}1334 and \ion{Si}{ii}1527) line's depths. If these lines were all saturated, their residual fluxes should be very similar and the mean is representative, allowing to gain ${\rm S/N}$ over the measurement. E.g, the average LIS residual flux of CDFS01735 (Fig.\ \ref{fig:SEDexample}) is $\rlis \approx 0.69$ (0.75 previous to the correction by resolution). 

Fig. \ref{fig:fit_histos} shows the histogram of the average \rlis\ distribution, characterized by $\langle \rlis \rangle = 0.31_{-0.23}^{+0.22}$ ($0.35_{-0.20}^{+0.19}$ without resolution correction). Then, using the dust-attenuation parameter (\ebv) provided by the \textsc{FiCUS} SED fits, we applied Eq.\ \ref{eq:fesc_LIS} to every single galaxy in our sample. For the $[a, b]$ coefficients in this equation, we chose the calibration by \citet{Gazagnes2018} i.e., [$a, b$] $=$ [$(0.63 \pm 0.12), (0.44 \pm 0.07)$]. We remark that this calibration was obtained over a sample of emission line galaxies but, as stated in \citetalias{SL22}, we do not expect this relation to change with galaxy type. As an example, the LyC escape fraction for CDFS017345 results in $\fescabs \approx 16\%$, adopting the \citetalias{R16} attenuation law.

The predicted \fescabs\ distribution following this method is also presented in Fig. \ref{fig:fit_histos}. As expected, the inferred \fescabs\ values show a clear dependency on the dust-attenuation law, with the median of the \citetalias{R16} distribution $\times 1.5$ lower than the \citetalias{SMC} law \fescabs\ distribution. When using the \citetalias{R16} dust-attenuation law, the median ionizing absolute escape fraction, $\langle \fescabs \rangle$, for our VANDELS sample of 534 SFGs at $3 \leq z \leq 5$ is:\\

$\langle \fescabs \rangle = 0.02 \pm 0.01 $ ($0.03 \pm 0.02$),\\

\noindent while, when using the \citetalias{SMC} attenuation law, it results in:\\

$\langle \fescabs \rangle = 0.03 \pm 0.02$ ($0.04 \pm 0.03$),\\

\noindent where the numbers in brackets correspond to the median \fescabs\ previous to any resolution correction on the residual flux of the lines (see Eq.\ \ref{eq:fesc_LIS}). As a comparison, \citet{Begley2022} obtained $\langle \fescabs \rangle = 0.07 \pm 0.02$ for a similar sample of $z \simeq 3.5$ VANDELS galaxies with deep VIMOS/U$-$band observations (see Fig.\ \ref{fig:fit_histos}), that is, a factor of $\times 2$ higher than our median value using the \citetalias{SMC} law, although compatible within 1$\sigma$ uncertainty. Moreover, our average \fescabs\ is in agreement with the extrapolations of the recent \citet{Trebitsch2022} model at $z \simeq 4$. In Sect.\ \ref{sec:discussion_lowzhighz}, we will compare our results with the escape fraction derived from other surveys targeting LCEs in the literature, at lower and higher redshifts.

\subsection{Composite spectra}\label{sub:stacks}
Stacked spectra were built with the goal of increasing the S/N with respect to individual galaxy data, allowing us to clear out some of the underlying physical correlations between the different parameters in this study. 

Following \citet{Cullen2019, Calabro2021, Llerena2022}, all the individual spectra in the sample were first shifted into the rest-frame using the VANDELS spectroscopic redshift and then resampled onto a common wavelength range, specifically $1200-1925$\AA. According to the median redshift of the sample i.e., $\langle z_{\rm spec} \rangle = 3.56$, the resulting spectral binning was chosen to be 2.5\AA$/( 1+ \langle z_{\rm spec} \rangle ) = 0.55$\AA\ (2.5 \AA\ equals the VIMOS wavelength dispersion per resolution element). Before co-adding the spectra, they were normalized to the mean flux in the $1350-1370$\AA\ rest-frame interval. The final (normalized) flux at each wavelength was taken as the unweighted median of all the individual flux values after a regular 3$\sigma$ clipping in order to reject outliers and bad pixels. The uncertainty on the stacked spectrum was calculated via bootstrap resampling of the spectra included in the composite.

Guided by this scheme, we performed stacks in bins of UV magnitude (\mobs), UV intrinsic luminosity (\mint), UV-continuum slope (\bslope) and \lya\ equivalent width (\ewlya). Concretely, we divided the sample according to the 25$^{th}$, 50$^{th}$ and 75$^{th}$ percentiles (quartiles) of every property, resulting in four sub-samples for each quantity sorted as Q1, Q2, Q3 and Q4. Then, the resulting stacks were processed through the \textsc{FiCUS} code and all the secondary SED parameters described in Sect.\ \ref{sub:fits} were obtained. App.\ \ref{app:stack_results}, Table \ref{tab:stack_table} contains the main properties and inferred SED parameters (\fescabs\ and \xiion) of the different composites. 

\subparagraph{The effect of stacking on the absorption lines}
Even though the equivalent widths of the \hi\ and metal lines do not change with the gas column density in an optically thick medium, the lines can be slightly broader or narrower depending on the gas thermal (Doppler) and turbulence velocities. More importantly, different galaxies may have different gas flow velocities which translates into red- or blue-shifted line centers relative to the systemic velocity. Moreover, the use of the spectroscopic instead of the systemic redshift introduces an additional source of uncertainty in the position of minimum depth of the lines \citep[see][]{Llerena2022}. All these effects contribute so that the UV-lines residual flux of the resulting galaxy composite can potentially be overestimated.

To correct for this \emph{smearing} effect and making use of the simulations described in App.\ \ref{app:resolution}, we randomly generated a set of $N = 100$ simulated \ion{Si}{ii}1260 line profiles with different intrinsic gas properties (column densities, Doppler broadening, inflow/outflow velocities, etc.), but fixing the covering fraction (i.e., equivalent to one minus the residual flux). We then stacked the line profiles following the same methods and assumptions described in this section, where the effects of instrumental resolution ($R{\rm (VIMOS)} \approx 600$) and a constant ${\rm S/N} = 5$ were also incorporated in the mock realizations. After that, we measured the depth of the composite \ion{Si}{ii}1260 line profile and compared it to the input value given by the covering fraction. According to the our simulations (Eq.\ \ref{eq:cal_stack}), an input $(1-C_f) = 0.1, 0.3, 0.5$ and 0.8 would require corrections factors as large as $70\%, 40\%, 25\%$ and $10\%$ due to the effect of stacking. 

\citet{Calabro2022} showed that, although the bulk ISM velocity is globally in outflow, the average shift of the LIS lines is very close to the systemic velocity ($-60 ~{\rm km/s}$), i.e., within the actual spectral resolution. Therefore, our simulations -- which assume a single VIMOS resolution element ($\pm 150~{\rm km/s}$) as the standard deviation of the distribution for the velocity shift of the lines -- might actually over-predict this effect. For that reason, we prefer not to apply such stacking corrections to our line measurements, but we encourage the reader to check App.\ \ref{app:resolution} for more details.  

It is also worth mentioning that \citet{Calabro2022} do not find any correlation between the stellar mass or SFRs and the velocity shift of the ISM lines. Thence, we will not expect differential corrections on the residual flux when combining galaxies with very different masses or SFRs, a conclusion that can be extrapolated to other galaxy properties (e.g., UV magnitudes).

%% file: 4_results.tex
\section{Results}\label{sec:results}
The following paragraphs describe on the main results of this paper. In Sect.\ \ref{sub:fescxiion_props}, the global relations between the ionizing escape fractions and production efficiencies with different galaxy properties are shown on an individual galaxy-basis. In Sect.\ \ref{sub:restUV_props}, the LCEs and non-LCEs composites are presented, and the differences in their non-ionizing rest-UV spectra are placed in the context of the physical ISM conditions which enable the ionizing radiation to escape. In Sect.\ \ref{sub:ion_properties}, the global ionizing properties of the LCEs versus non-LCEs samples are discussed. Lastly, Sect.\ \ref{sub:confirmed_LCEs} summarizes the properties of previously known LCEs reported in the literature which were included in our sample.

\begin{figure*}
    \includegraphics[width=0.9\textwidth, page=5]{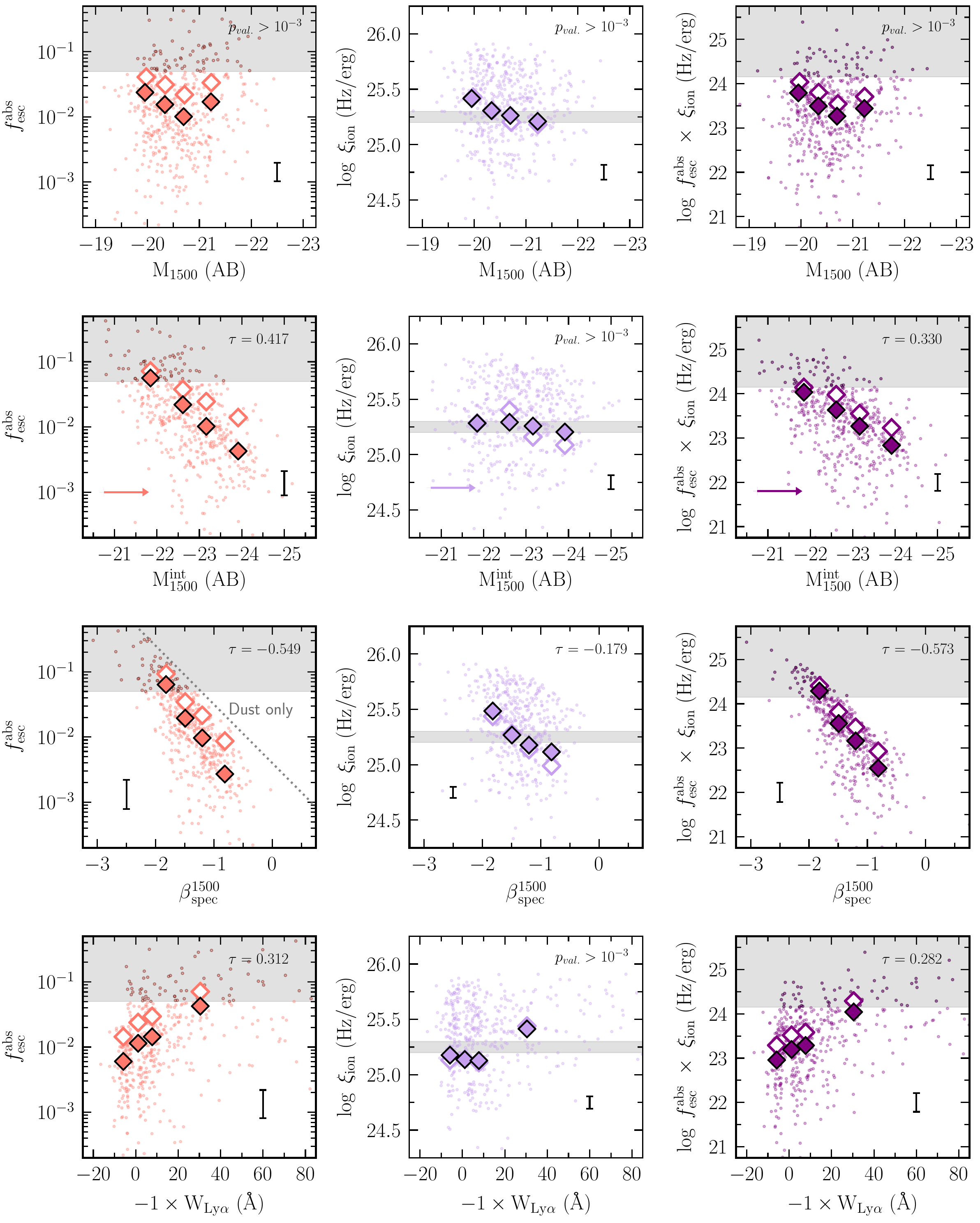}
\caption{Scatter plots searching for correlations between the predicted ionizing photon escape fraction (\fescabs, {\bf first column}), the ionizing production efficiency ($\log \xiion$, {\bf second column}) and the product of the two ($\log \fescabs \times \xiion$, {\bf third column}, see Eq.\ \ref{eq:nion}), versus different galaxy properties: observed and intrinsic UV absolute magnitudes (\mobs, \mint), UV-continuum slope at 1500\AA\ from the best-fit SED (\bslope), and \lya\ equivalent width ($-1 \times \ewlya$). The \emph{Kendall} ($\tau$) correlation coefficients for individual \citetalias{R16} measurements (coloured dots in the background) are shown at the top but only for significant correlations (thick-framed panels). Results from our stacking analysis are displayed with filled \citepalias{R16} and open \citepalias{SMC} thick diamonds. Typical error bars for individual sources are plotted at the bottom part of each panel, and the arrows along the \mint\ panels measure the shift in \mint due to the use of the \citetalias{SMC} dust-attenuation law. Grey-shaded regions mark assumed canonical values of $\fescabs \geq 5\%$ and $\log \xiion = 25.2-25.3$ in classical Reionization models \citep{Robertson2013}.}
\label{fig:fescXiion_phot}
\end{figure*}

\subsection{The ionizing escape fraction and production efficiency dependence with observed galaxy properties}\label{sub:fescxiion_props}
Fig.\ \ref{fig:fescXiion_phot} investigates how our predicted ionizing absolute escape fraction (\fescabs) and the ionizing photon production efficiency (\xiion) depends on galaxy properties. The individual VANDELS measurements are shown in the background together with the results from our stacking analysis on top (large symbols). Systematic differences in \fescabs\ and \xiion\ due to the use of a shallower \citepalias[][filled diamonds]{R16} or steeper \citepalias[][open diamonds]{SMC} attenuation law are also explored. In general, the escape fraction values derived using the \citetalias{SMC} law shift to slightly systematically higher escapes (by a factor of $\times 1.5$) compared to the \citetalias{R16} ones.

A \emph{Kendall}$-\tau$ statistics was applied to each pair of variables in order to figure out the strength ($\tau$) and significance ($p-$value) of the correlation. For a sample of $\simeq 500$ objects, we consider correlations to be significant  if $p_{val.} \lesssim 10^{-3}$ (3$\sigma$) and strong if $\left| \tau \right| \gtrsim 0.1$. Coincidentally, all significant correlations studied in this work showed up to be strong correlations, so that in Fig.\ \ref{fig:fescXiion_phot} we only indicate the $\tau$ coefficients for significant correlations (thick-framed panels), otherwise we write $p_{\rm val.} > 10^{-3}$.

\subsubsection{Observed absolute UV magnitude}
The predicted \fescabs\ and the observed UV absolute magnitude (\mobs) for the VANDELS sample at $3 \leq z \leq 5$ do not show any clear correlation when considering the whole range of UV magnitudes. \citet{Tanvir2019} found no significant correlation of $N_{\rm HI}$ with galaxy UV luminosity across a wide range of redshifts. Assuming that the escape fraction is primarily regulated by the same optically thick neutral gas along the line-of-sight, as suggested by Eq.\ \ref{eq:fesc_LIS}, a lack of correlation between \fescabs\ and \mobs\ would therefore be expected, supporting our global trend.

However, fainter (low-mass) galaxies usually host lower gas and dust fractions compared to brighter systems \citep{Trebitsch2017}. Together with their low-gravitational potential and often bursty SFHs \citep{Muratov2015}, the expelling of gas and dust in a turbulent ISM \citep{Kakiichi2021} favours the creation of holes through which the LyC photons can freely escape. In this scenario, a higher \fescabs\ is naturally expected for the faintest galaxies \citepalias[see Sect. 7.2 in][]{SL22}.

Even though the correlation is tentative, $\mobs \gtrsim -20$ and fainter VANDELS galaxies tend to have higher escape fractions (Fig.\ \ref{fig:fescXiion_phot}). This mild tendency has been reported through the study of galaxy composites at high-$z$ by the KLCS \citep{Steidel2018, Pahl2021, Pahl2023} as well as observed at low-$z$ by the LzLCS \citep[][and \citetalias{SL22}]{Flury2022a, Chisholm2022}. Contrarily, models such as \citet{Sharma2016} and \citet{Naidu2020} give an increasing escape fraction towards bright $M_{\rm UV}$ systems, in disagreement with our picket-fence formulation \footnote{This said, the hypothesis proposed in \citet{Naidu2020} of a reionization driven by $-20 \leq \mobs \leq -18$ galaxies (what they called the ‘oligarchs’), cannot necessarily be ruled out by our observations because our galaxies do not span beyond $M_{\rm UV} \geq -19$ AB.}.

The ionizing production efficiency shows a very smooth (although non-significant) dependence on \mobs, with the stacked measurement at the faintest UV magnitude bin ($\mobs \gtrsim -20$) being higher than the typically assumed canonical value for cosmic reionization of $\log \xiion ({\rm Hz/erg}) = 25.2$ \citep{Robertson2013}. Our results are in agreement with other studies in the literature, for example with the results by \citet{Bouwens2016} at similar UV magnitudes but, contrary to the former, our \mobs\ range is actually not wide enough to show a clear $\xiion - \mobs$ trend. Even though, it has been shown that, at similar redshifts, the \xiion\ evolution with UV luminosity is a very smooth correlation \citep[see][]{Lam2019, Emami2020, PrietoLyon2022}. A more complete picture on the $\xiion - \mobs$ relationship will be given in Sect.\ \ref{sec:discussion_xiion}.

In general, both \fescabs\ and \xiion\ distributions with \mobs\ are characterized by a large scatter ($\simeq 1$dex, e.g., 0.1$\%$ to 10$\%$ in \fescabs) and a weak evolution with \mobs, where only the faintest galaxies in the sample have tentatively higher values in \fescabs\ ($\geq 5\%$) and $\log \xiion$ ($\geq 25.2{~\rm Hz/erg}$). Finally, the $\log \fxi$ product preserves the overall evolution of \fescabs\ with the UV magnitude.

\subsubsection{Intrinsic (dust-free) UV luminosity}
The intrinsic UV absolute magnitude for each galaxy (\mint) was calculated from the best-fit SED by taking the flux at 1500\AA\ previous to attenuation by dust (i.e., the dust-free SED), and computing the AB magnitude via the usual distance modulus formulae. 

On the one hand, \fescabs\ versus \mint\ shows one of the strongest correlations of the present study ($\tau \approx 0.4$), where the less intrinsically luminous galaxies have the highest \fescabs. From our \fescabs\ prescription (Eq.\ \ref{eq:fesc_LIS}), this correlation is expected since the dust-attenuation is by definition directly related to the escape fraction of galaxies, where the most UV-bright galaxies are attenuated \citep{Finkelstein2012, Bouwens2014} and host substantially higher, more extended gas and dust reservoirs at the same time \citep[see e.g.,][from the perspective of the FIRE-2 simulation]{Ma2020}. This behavior was previously reported in \citet{Begley2022}, where they demonstrate how intrinsically UV-faint galaxies at $z \simeq 3.5$ would require statistically higher escape fractions in order to reproduce the observed distribution of the ionizing-to-nonionizing flux ratio. 

On the other hand, the \xiion\ versus \mint\ distribution appears completely flat irrespective of the \mint\ bin and the dust-attenuation law. According to Eq.\ \ref{eq:xiion}, the dust correction factor applied to any \xiion\ measurement would be: ${\rm A_{<912}/A_{1500}}$. This ratio does not strongly depend on the dust-law, thus no \xiion\ dependence on the attenuation will be expected neither. Therefore, the \fxi\ product inherits the same dependence on the UV luminosity as \fescabs\ does. When comparing the \citetalias{R16} and \citetalias{SMC} stacks predictions (see Fig.\ \ref{fig:fescXiion_phot}), their \fxi\ show the largest differences at the bright \mint\ end i.e., the more deviation appears for the more attenuated, redder galaxies.

\subsubsection{UV-continuum slope at 1500\AA}
Standing out as the strongest correlation of this study ($\tau \approx -0.6$, see Fig.\ \ref{fig:fescXiion_phot}), the \fescabs\ inversely scales with the best-fit UV-continuum slope at 1500\AA\ (\bslope) so that the bluest galaxies in the sample emit a significantly larger fraction of ionizing photons than their redder counterparts. The ``Dust Only'' case (dotted line) follows the resulting Eq.\ \ref{eq:fesc_LIS} assuming there is no gas along the line of sight, so the dust is the only source of attenuation for LyC photons. This scenario, although provides a physical upper limit to the previous equation, does not expect to hold observationally. Because the dust and cold gas within the ISM are spatially correlated, galaxies which are more dust-attenuated will also have lower residual fluxes of the LIS lines, and they will deviate from the ``Dust only'' case towards low \fescabs\ values, as we see in Fig. \ref{fig:fescXiion_phot}.

This trend has been directly observed by \citet{Flury2022b} in the LzLCS survey, also lately investigated by \citet{Chisholm2022}. Additionally, a decrease in \fescabs\ with increasing UV colors was shown in \citet{Pahl2021} using KLCS galaxy composites at high-$z$. Also interestingly, \citet{Begley2022} found that UV blue galaxies at $z \simeq 3.5$ would require statistically higher escape fractions in order to reproduce the observed ionizing-to-nonionizing flux ratio distribution.

Regarding our indirect approach to predict \fescabs\ (Eq.\ \ref{eq:fesc_LIS}), a negative correlation between the escape fraction and \bslope\ must be expected by definition, since the UV slope is inherently linked to the dust-attenuation (\ebv), so that redder SEDs usually means more attenuated galaxies \citep{Meurer1999}. For example, we obtain $\fescabs \geq 5\%$ for galaxies whose $\bslope \leq -1.5$, decreasing to $\fescabs \approx 0.1\%$ at $\bslope \approx -0.5$. As stated in \citet{Chisholm2022}, this steep relation between the escape fraction and the UV slope is particularly important at the highest redshifts because (1) it can be easily used as an observational proxy to indirectly infer \fescabs\ at the EoR \citep[see][]{Trebitsch2022}, and (2) primordial/fainter galaxies are thought to host less dust than actual SF galaxies \citep{Finkelstein2012, Bouwens2014, Cullen2023}, which naturally explains why higher redshift galaxies emit more ionizing photons than their lower redshift counterparts \citep{Finkelstein2019}. As we will discuss in Sect. \ref{sec:discussion_lowzhighz}, our $\fescabs-\bslope$ results agree with the relation $z = 0.3$ published by \citet{Chisholm2022} but extrapolated to redder UV slopes and lower escape fractions. This similarity suggests that the $\fescabs-\bslope$ does not strongly change with redshift.

The \xiion\ parameter is also strongly correlated with \bslope, so that \xiion\ rapidly increases as the UV slope decreases, only giving $\log \xiion ({\rm Hz/erg}) \geq 25.2$ for galaxies whose slopes are $\bslope \leq -1.5$. Our results are in agreement with other studies in the literature, for example with the results by \citet{Bouwens2016} and \citet{Matthee2017b}, but shifted to redder $\beta_{1500}$ values. For more details on the $\xiion - \beta_{1500}$ relationship and comparison with the literature, we refer to Sect. \ref{sec:discussion_xiion}.

Qualitatively, the largest differences between the \citetalias{R16} and \citetalias{SMC} derived \fescabs\ values are found among the redder objects i.e., the most attenuated galaxies in the sample. However, both \citetalias{R16} and \citetalias{SMC} derived \xiion\ values are similar irrespective of the dust-attenuation law.

\subsubsection{\lya\ equivalent width}
The relation between the properties of the \lya1216 line and the escape and production efficiency of ionizing photons is also striking. This can be seen in Fig.\ \ref{fig:fescXiion_phot}, where the predicted \fescabs\ monotonically raises as the \lya\ equivalent width also increases (\ewlya). For typically assumed values of $\ewlya \leq -20$\AA\ in LAEs \citep[e.g.,][but see \citet{Stark2011,Kusakabe2020}]{Pentericci2009}, we obtain $\fescabs \geq 5\%$. This result is in concordance with previous results in the literature. As demonstrated by \citetalias{SL22} \cite[but see also][]{Gazagnes2020, Izotov2021}, the strongest LCEs are usually among the strongest LAEs i.e., having the highest \lya\ equivalent widths.

With the exception of photon scattering and the effect of gas kinematics, the same mechanisms that regulates the escape of ionizing photons also regulate the leakage of \lya\ photons and therefore the shape and strength of the \lya\ line \citep[e.g.][]{Henry2015, Verhamme2015}. Considering the aforementioned picket-fence model with a very simplistic assumption on the geometry of galaxies, these main mechanisms are (1) the neutral gas column density -- whose optically-thick spatial distribution is parametrized in Eq.\ \ref{eq:fesc_LIS} via the covering fraction -- and (2) the dust-attenuation term (\ebv). In this way the \lya\ emission (\ewlya) and the escape of LyC photons (\fescabs) would remain physically related \citep[see][]{Verhamme2017, Izotov2020, Flury2022b, Maji2022}. In fact, studying the afterglow gas of a significant sample of Gamma Ray Burst (GRB) host galaxies, \citet{Vielfaure2021} found that the gas columns density of such gas was indeed optically thick, so that ``the bulk of \lya\ photons produced by massive stars in the star-forming region hosting the GRB will be surrounded by these opaque lines of sights''.

\begin{figure}
    \includegraphics[width=0.99\columnwidth, page=6]{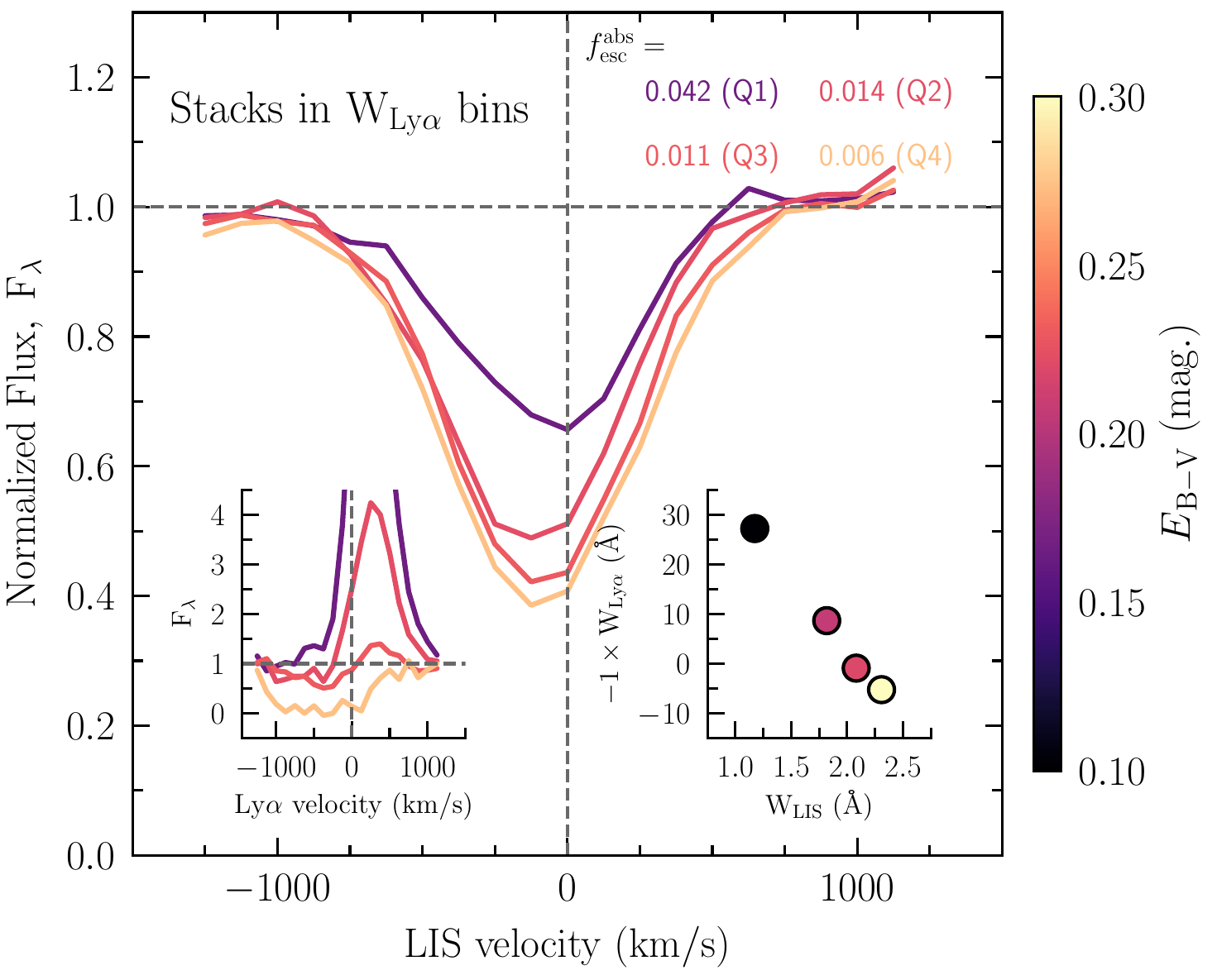}
\caption{Combined low-ionization-state (LIS) absorption profile for the stacks in bins of \lya\ equivalent width (\ewlya, see values in App.\ \ref{app:stack_results}), normalized by the local continuum and constructed by averaging the profiles of individual LIS absorption lines (\ion{Si}{ii}1260, \ion{O}{i}+\ion{Si}{ii}1302, \ion{C}{ii}1334, \ion{Si}{ii}1527). The composites are color coded by the dust-attenuation parameter (\ebv) derived from the \textsc{FiCUS} SED fits, and the predicted escape fraction for each composite (\fescabs) is indicated in the legend. The insets show the resulting continuum-normalized \lya\ profile for the stacked spectra ({\em left}), and the evolution of the \lya\ equivalent width as function of the LIS equivalent width in absorption ({\em right}).}
\label{fig:lya_lis}
\end{figure}

In order to illustrate the role of dust attenuation and the gas covering fraction in the escape of \lya\ and LyC photons, in Fig.\ \ref{fig:lya_lis} we plot the average LIS absorption profile for different stacks in bins of \lya\ equivalent width \citep[see also][]{Trainor2019}. This plot shows the combined profile of several ISM LIS lines (\ion{Si}{ii}1260, \ion{O}{i}+\ion{Si}{ii}1302, \ion{C}{ii}1334, \ion{Si}{ii}1527), color-coded by the inferred UV dust-attenuation (\ebv) and the predicted escape fraction for each composite (\fescabs) is indicated in the legend. The resulting \lya\ profiles can also be seen in the inset, and finally \ewlya\ (in \AA) is plotted against the LIS equivalent width of the combined LIS absorption profiles ($W_{\rm LIS}$, in \AA). 

As a result, as long as the dust-attenuation increases and the gas covering decreases (increasing residual fluxes), the resulting escape fractions and the \lya\ equivalent widths also decrease, from \ewlya\ values typically found in LAEs towards absorption, even damped, \lya\ profiles. Relatedly, the relation between the \lya\ and the strength of the LIS lines has been widely studied in individual and stacked rest-UV spectra of LBGs at $z = 2 - 5$ \citep{Shapley2003, Jones2012, Henry2015, Trainor2015, Du2018, Pahl2020}, and our results agree with the overall picture described in these papers. On the one hand, the observed connection between the LIS equivalent width and the reddening can only be explained if the dust resides within the same clouds as the neutral gas in a clumpy ISM geometry \citepalias{SL22}. On the other hand, the dwindling of the \lya\ emission with the dust-attenuation is a a natural consequence of the Mass Metallicity Relation \citep[MZR, see][for a review]{Maiolino2019}, and the empirical relation between the \lya\ strength and the stellar mass \citep{Cullen2020}.

A large \ewlya\ can also be ascribed to a boost in the production of ionizing photons \citep{Nakajima2018b}. The VANDELS high-$z$ $\ewlya-\xiion$ relation can be found in the bottom-mid panel of Fig.\ \ref{fig:fescXiion_phot}, where a significantly higher $\log \xiion {\rm (Hz/erg)} \geq 25.2$ is found for the strongest LAEs only, with a $\ewlya \approx -30$\AA. Thanks to JWST data in combination with HST photometry, an increasing $\log \xiion$ with \lya\ strength has also been shown for individual galaxies at the late-edge of the EoR \citep{Ning2022}. However, in the works by \citet{Cullen2020} and \citet{Reddy2022}, \xiion\ has been demonstrated to be insufficient to solely account for the whole variation in \ewlya, rather ``the covering fraction of optically-thick \ion{H}{i} (gas) appears to be the principal factor modulating the escape of \lya, with most of the \lya\ photons in down-the-barrel observations of galaxies escaping through low-column-density or ionized channels in the ISM'', as stated in \citet{Reddy2022}.

Similarly, our $\fxi - \ewlya$ relationship results from the convolution of the \fescabs\ and \xiion\ dependence on the \lya\ strength, where the $\fescabs - \ewlya$ behavior plays the major role. The higher \xiion\ values for the objects with the highest \lya\ equivalent width in the sample (LAEs) naturally increase the \fxi\ product \citep{Matthee2022}.

\subsubsection{Stellar mass}
\begin{figure}
    \includegraphics[width=0.85\columnwidth, page=7]{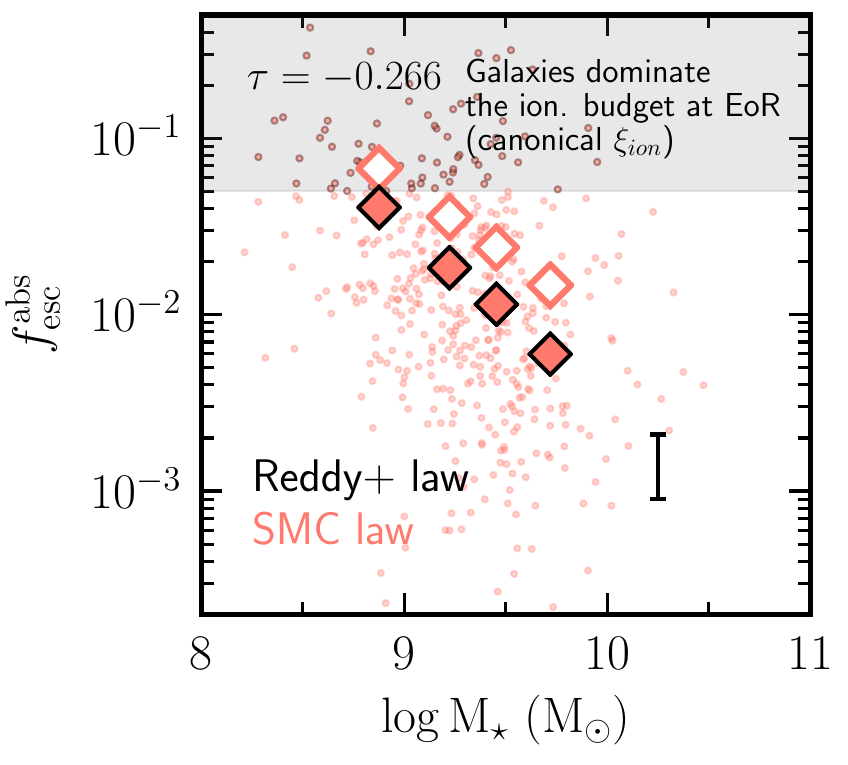}
\caption{Relation between our derived \fescabs\ and the stellar mass ($M_{\star}$, in $M_{\odot}$). The $\fescabs - M_{\star}$ correlation is strong and significant for both \citetalias{R16} and \citetalias{SMC} dust-laws, although systematically higher escapes are reported when using steeper, SMC-like attenuation laws. The layout is the same as in Fig.\ref{fig:fescXiion_phot}.}
\label{fig:fesc_mass}
\end{figure}

The predicted \fescabs\ strongly scales with the stellar mass ($M_{\star}$) of the VANDELS galaxies so that low-mass galaxies have statistically higher escape fractions than more massive systems (Fig.\ \ref{fig:fesc_mass}). For instance, only $\log M_{\star}(M_{\odot}) < 9$ VANDELS galaxies show cosmologically relevant escape fractions of $\fescabs \geq 5\%$ \citep[][see grey-shaded area in Fig.\ \ref{fig:fescXiion_phot}]{Robertson2013}, and these are actually $\simeq 1$dex higher than the average escape fraction of the most massive galaxies in the sample ($\log M_{\star}(M_{\odot}) \geq 10$).

According to our picket-fence model, a physical $\fescabs - M_{\star}$ connection is expected (Eq.\ \ref{eq:fesc_LIS}) because more massive galaxies are intrinsically more attenuated than low-mass systems \citep[see e.g.,][]{Finkelstein2012, McLure2018a, Fudamoto2020}. 
As $\log M_{\star}$ increases, the dust-attenuation increases and the line-depth decreases accordingly, and both yield progressively decreasing escape fractions (\fescabs). This is consistent with the results by \citet{Reddy2022} where, using a joint modeling of the composite FUV and optical spectra of high-$z$ SFGs, they demonstrate how the \ion{H}{i} and the gas-enriched covering fraction decreases with the galaxy stellar mass.

So far, there is no consensus in simulations on whether more massive galaxies should emit a higher \citep[e.g.,][]{Naidu2020} or lower \citep{Rosdahl2022} fraction of their produced ionizing photons to the IGM compared to their less massive counterparts, with some of them even suggesting a turnover at intermediate masses \citep[see][]{Ma2020}. In observations, while at low-$z$ there does not seem to be any clear relation between these two quantities \citep{Izotov2021, Flury2022b}, even when spanning a wide range of stellar masses, a gradual increase in \fescabs\ towards lower masses have been recently reported at high-$z$ in \citet{Fletcher2019,Saxena2022a}, for individual detections. 

Lately, \citet{Pahl2023} have also shown a negative and significant $\fescabs - \log M_{\star}$ trend in the $9 \leq \log (M_{\star}/M_{\odot}) \leq 10$ stellar mass range, using $z \simeq 3$ KLCS stacks. However, at similar redshifts, \citet{Begley2022} did not find any statistical distinction when splitting their sample into lower and higher masses with respect to the median, suggesting that $M_{\star}$ is at best a secondary indicator of the average \fescabs. In the $8 \leq \log (M_{\star}/M_{\odot}) \leq 10$ mass interval, our stacking analysis with VANDELS support the former scenario by which low-mass galaxies have higher escape fractions \citep[similar to][]{Ma2020, Pahl2023}. In this vein, our picket-fence formalism put the physical picture suggested by models like \citet{Naidu2020} up against the ropes, since the latter yields a monotonic increase of the escape fraction with stellar mass, behaviour also disfavoured by other works in the recent literature \citep[see compilation by ][]{Pahl2023}. 

Finally, whilst AGNs may not contribute directly to the ionizing photon budget at the EoR \citep[e.g.,][]{Hassan2018}, semi-analytical simulations by \citet{Seiler2018} show that they could indirectly influence reionization by clearing out holes in the ionized regions, so that \fescabs\ can be boosted after quasar wind events. In such scenario, the mean escape fraction peaks for intermediate-mass galaxies, around $\log (M_{\star}/M_{\odot}) \approx 8$. Regrettably, our sample is not able to probe this hypothesis, since it does not extend to lower stellar masses.

\begin{figure*}
    \includegraphics[width=0.95\textwidth, page=8]{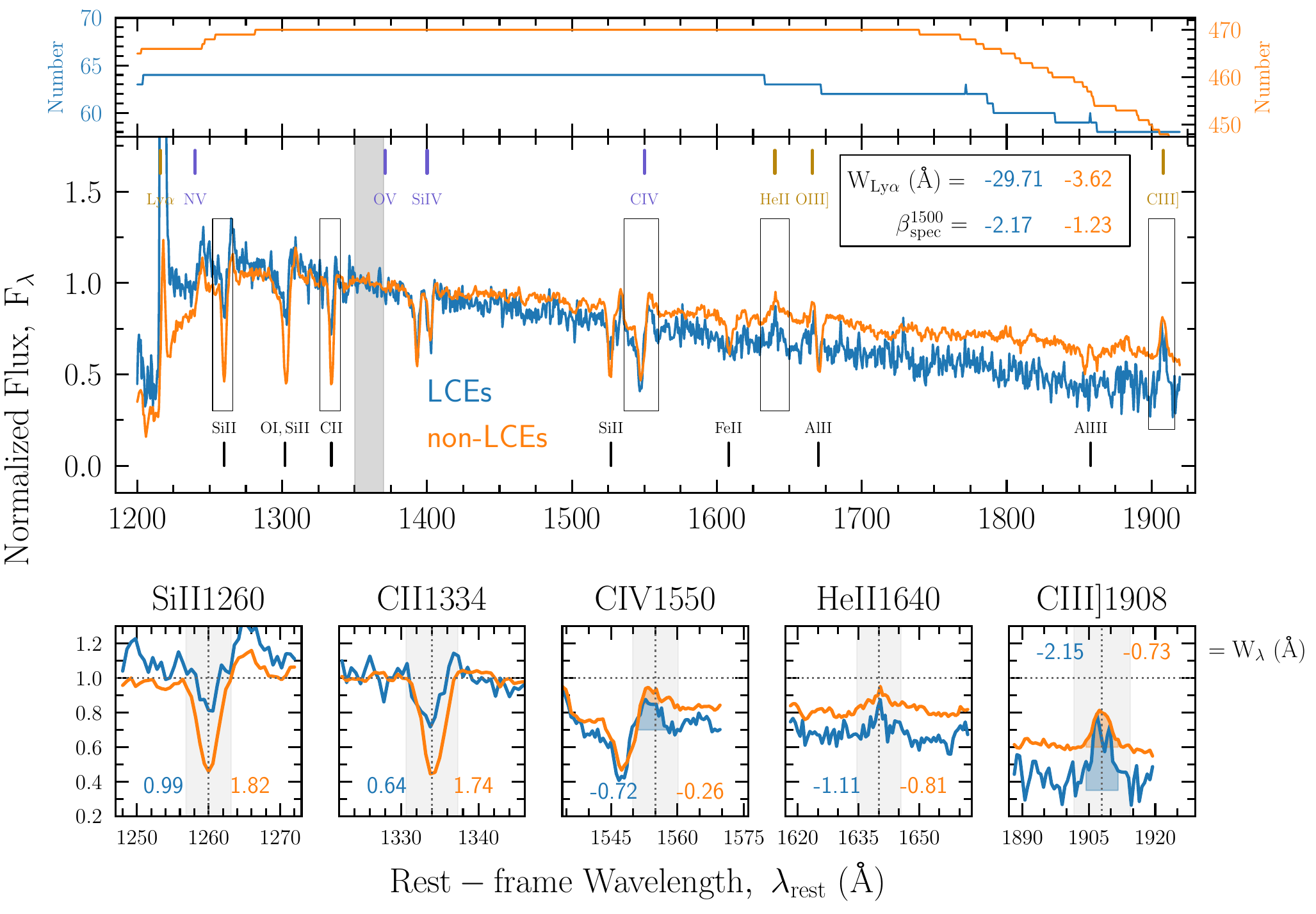}
\caption{Composite spectra (main plot) for LCEs (blue) and non-LCEs samples (orange) normalized at 1360\AA\ (grey band). The position of the main stellar-wind lines (\ion{N}{v}1240, \ion{O}{v}1371, \ion{Si}{iv}1400 doublet, \ion{C}{iv}1550), nebular lines (\lya1216, \ion{He}{ii}{1640}, \ion{O}{iii}]1666, \ion{C}{iii}]1908) and ISM lines (\ion{Si}{ii}1260, \ion{O}{i}+\ion{Si}{ii}1302, \ion{C}{ii}1334, \ion{Si}{ii}1527, \ion{Fe}{ii}1608, \ion{Al}{ii}1670 and \ion{Al}{iii}1858 doublet) are marked in dark-blue, gold and black labels, respectively. The \ewlya\ (in \AA) and \bslope\ values are also indicated in the inset. The {\bf top} panel show the number of objects included in each composite as a function of wavelength, and the {\bf bottom} panel zooms in some of the lines, highlighting fundamental differences in their non-ionizing spectra ({equivalent width for each line are listed, see text}).}
\label{fig:composite_spectra}
\end{figure*}

\begin{figure}
    \includegraphics[width=0.99\columnwidth, page=9]{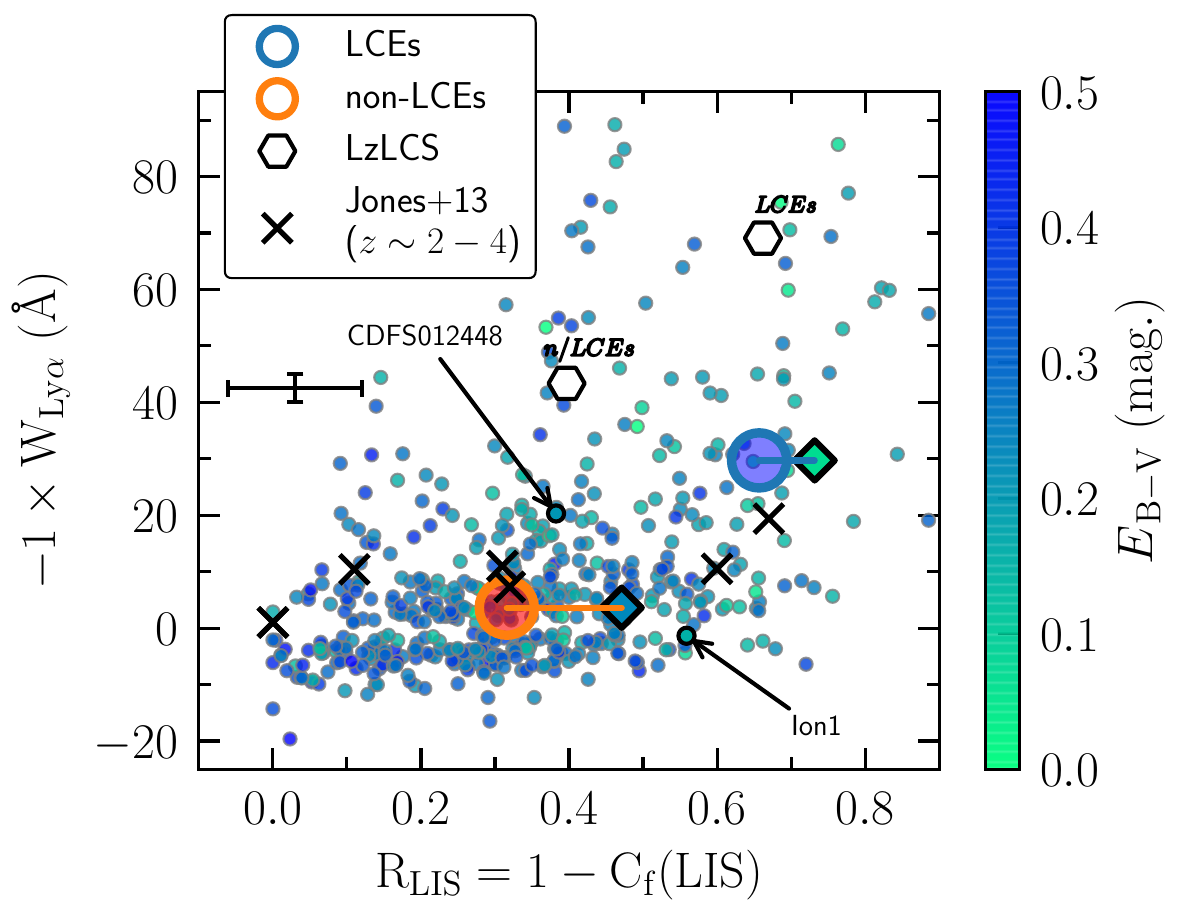}
\caption{The relation between the \lya\ equivalent width ($-1 \times \ewlya$, in \AA) and the measured mean residual flux of the LIS lines (\rlis). The points are color-coded by the dust-attenuation parameter (\ebv, in mag.). Blue and red circles indicate the position of the LCEs and non-LCEs composite spectra, once corrected by the effect of resolution on the depth of the lines (see App. \ref{app:resolution}). For comparison, the empty hexagons represent the measurements of LCEs and non-LCEs stacks from the LzLCS \citep[][$z = 0.2-0.4$]{Flury2022a}, and the black crosses show the individual measurements by \citet[][$z = 2-4$]{Jones2013}. Arrows point to already confirmed LCEs in our sample: \emph{Ion1} by \citet{Vanzella2012} and CDFS012448 by \citet{Saxena2022a}.}
\label{fig:Lya_Rlis}
\end{figure}

\subsection{Non-ionizing rest-UV properties of potential LCEs at $3 \leq z \leq 5$}\label{sub:restUV_props}
Having predicted the ionizing escape fraction (\fescabs) for every galaxy allows us to split the sample into potential LCEs, with $\fescabs \geq 0.05$ (64 out of 534 galaxies, i.e., $\approx 10\%$ of LyC detection rate), and non-LCEs with $\fescabs < 0.05$ (470 galaxies, assuming \citetalias{R16}). We then build composite spectra for LCEs and non-LCEs candidates, following the same method as described in Sect.\ \ref{sub:stacks}. The main properties, fitted SED parameters and UV lines measurements are summarized in  App.\ \ref{app:stack_results}, Table \ref{tab:lces_table}. Both LCEs and non-LCEs stacks are presented Fig.\ \ref{fig:composite_spectra}, together with a handful of insets and labels which highlight the differences in their non-ionizing FUV spectra.

\begin{figure}
    \includegraphics[width=0.95\columnwidth, page=10]{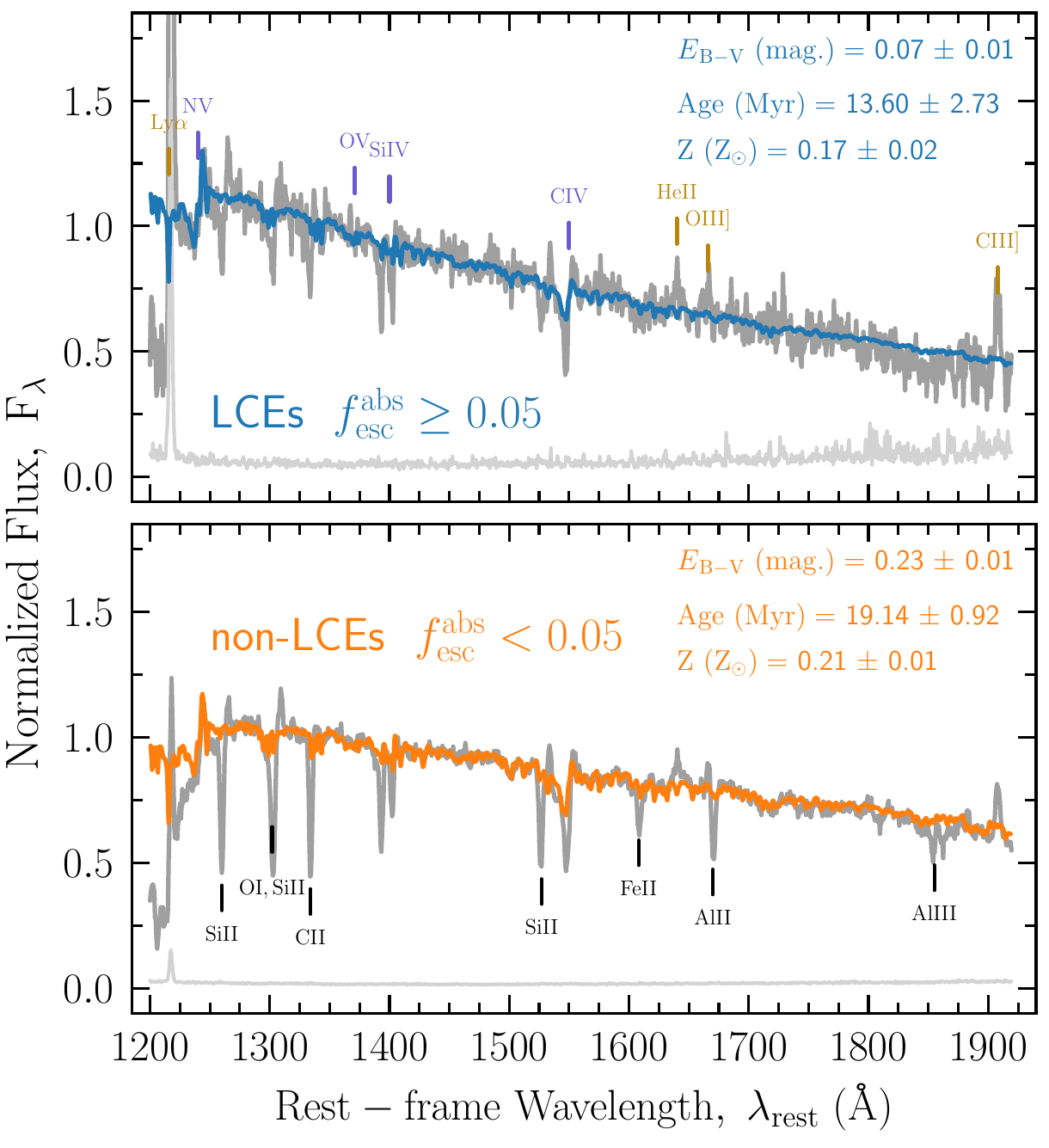}\caption{Main \textsc{FiCUS} SED fitting results for the LCEs {\bf (top)} and non-LCEs {\bf (bottom)} composite spectra: \ebv, stellar age and metallicity (see legend). The best-fit SED models for LCEs (blue) and non-LCEs (orange) are plotted on top of each composite (in grey), which is normalized at 1360\AA. The main stellar, nebular and ISM lines are labeled through dark-blue, gold and black labels, as in Fig. \ref{fig:composite_spectra}.}
\label{fig:composite_fits}
\end{figure}

\begin{figure*}
    \includegraphics[width=0.90\textwidth, page=11]{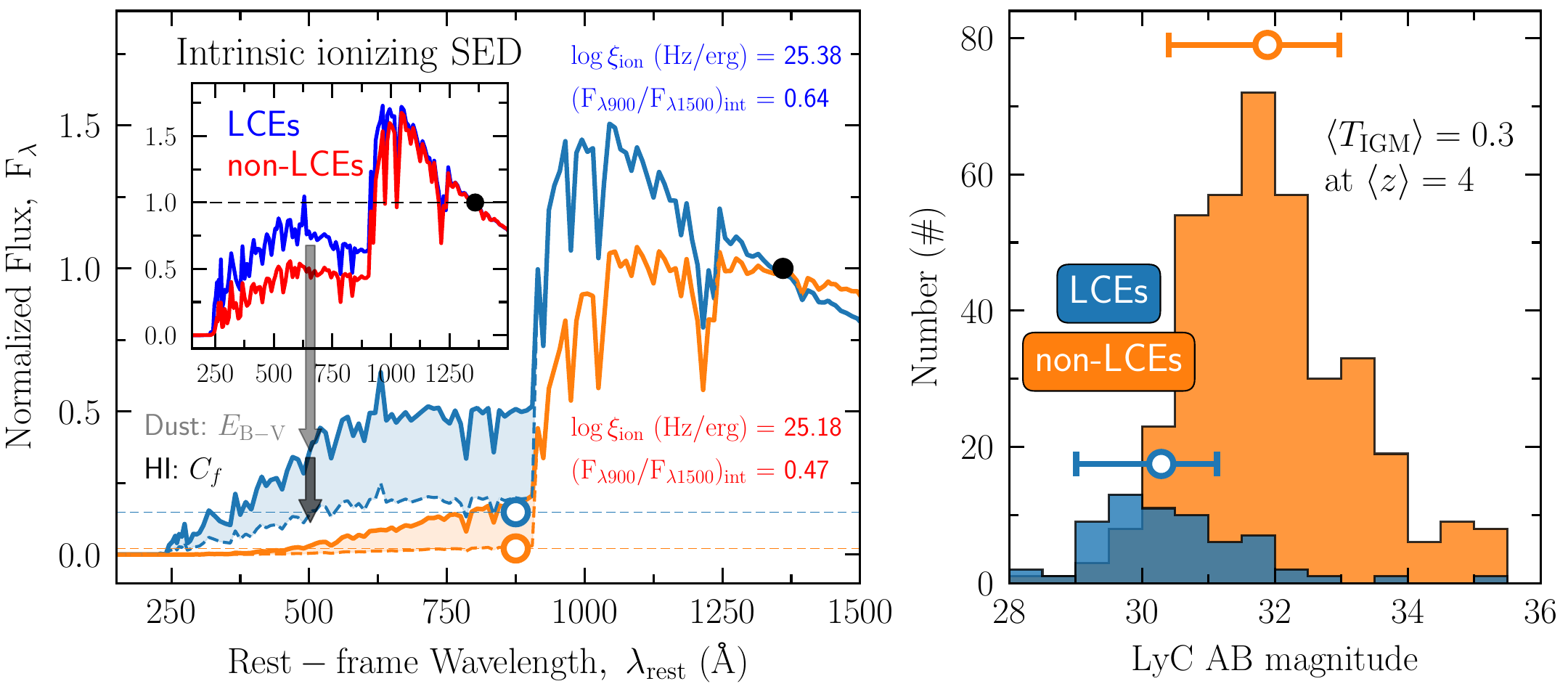}
\caption{The ionizing continuum of potential LCEs at $3 \leq z \leq 5$. {\bf Left:} a comparison between the dust-attenuated SED for LCEs (blue) versus non-LCEs (orange) down to ionizing wavelengths, normalized at 1360\AA. The inset shows the intrinsic ionizing spectrum (i.e., dust-free) of LCEs and non-LCEs (in blue and red, respectively). As written in the legend, the intrinsic 900-to-1500 flux ratio (in F$_{\lambda}$ units) and the ionizing photon production efficiency (\xiion) are higher for LCEs. The effect of dust-attenuation and neutral hydrogen absorption in both ionizing spectra is sketched through the grey and black arrows, where the blue and orange points represent the corresponding observed LyC flux once both absorptions have been accounted (a constant \ion{H}{i} cross-section with wavelength has been considered). {\bf Right:} the AB magnitude distribution of the predicted observed flux in the LyC for the LCEs (blue) and non-LCEs (orange) samples, once multiplied by the average IGM transmission at $\langle z \rangle = 4$ of $\langle T_{\rm IGM} \rangle = 0.3$.}
\label{fig:composite_ion}
\end{figure*}

\subsubsection{UV-continuum slope at 1500\AA}
The UV slope for LCEs and non-LCEs are remarkably different: $\bslope = -2.17 \pm 0.03$ for LCEs versus $-1.23 \pm 0.01$ for non-LCEs composites (see Fig.\ \ref{fig:composite_spectra}). As seen in the previous section, the UV slope is by definition related to \fescabs\ in the sense that it constitutes a proxy for dust attenuation at UV wavelengths \citep{Chisholm2022}, therefore favouring a picture in which bluer galaxies with lower levels of dust attenuation (\ebv) display higher values of escape fraction (by construction, see Eq.\ \ref{eq:fesc_LIS}). Having different levels of dust attenuation in LCEs and non-LCEs will partially influence the strength of other nebular emission lines in their UV spectra.

The intrinsically dustier nature of non-LCEs in opposite to the LCEs population has been explicitly reported in \citetalias{SL22} and further investigated in \citet{Chisholm2022} using LzLCS data at $z \simeq 0.3$. Also recently, \citet{Pahl2023} showed a monotonic increase of \fescabs\ with \ebv\ from FUV SED fitting in KLCS composite spectra. Likewise, \citet{Pahl2021} and \citet{Begley2022} observationally demonstrated a decrease in \fescabs\ with increasing UV colors at $z \simeq 3$, using HST (KLCS) and ground-based (VANDELS) LyC imaging, respectively. All these different LyC data sets at low- and high-$z$ will be put together in Sect.\ \ref{sec:discussion_lowzhighz}.

\subsubsection{\lya\ emission}
LCEs clearly show a stronger \lya\ emission than non-LCEs (Fig.\ \ref{fig:composite_spectra}): $\ewlya = -29.71 \pm 2.46$\AA\ against $-3.62 \pm 0.41$\AA\ for LCEs and non-LCEs stacks, respectively. Worth mentioning is that the \ewlya\ for the LCEs composite is compatible with the typical definition of LAEs in the literature \citep[i.e., $\ewlya \leq -20$\AA\ see][]{Pentericci2009}, so that the LCEs composite is mostly composed of the \lya\ emitting galaxies in the VANDELS sample. The non-LCEs composite, however, shows little \lya\ emission compared to LCEs, with a damped \lya\ red-wing extending up to 1240\AA. This suggests a higher column density of \ion{H}{i} gas beneath the bulk of the non-LCEs compared to the LCEs population, an unequivocal necessary condition for preventing the leak of \lya\ and LyC photons \citep[see][]{Henry2015}. This is also compatible with the strong decrease of the \lya\ equivalent width (by a factor $\simeq 9$) between LCEs and non-LCE, which is stronger than the decrease of \xiion\ (a factor $\simeq 4$), and which is naturally explained by the impact of a higher column density and higher dust content on the \lya\ escape fraction \citep[see e.g.][]{Atek2014}.


Observational evidences for an increase in \ewlya\ with increasing \fescabs\ has been reported by a few recent studies: \citet{Flury2022b} through FUV spectroscopy of LzLCS sources at $z \simeq 0.3$; \citet{Fletcher2019} and \citet{Pahl2021, Pahl2023} through LyC imaging of LCEs at $z \simeq 3$ (the LACES and KLCS surveys, respectively); and by \citet{Begley2022}, through the statistical measurement of the average \fescabs\ at $z \simeq 3.5$ (VANDELS).

\subsubsection{Other nebular emission lines}
In Fig.\ \ref{fig:composite_spectra}, the \ion{C}{iv}1550, \ion{He}{ii}{1640} and \ion{C}{iii}]1908 nebular line-profiles are displayed for the LCEs (in blue) and non-LCEs composites (in red). The \ion{C}{iv}1550 and \ion{He}{ii}{1640} lines for LCEs show slightly stronger emission and narrower profiles, while the equivalent width of the \ion{C}{iii}]1908 line is remarkably higher in the LCEs stacked spectra compared to the non-LCEs one. We obtain $W_{\rm HeII} = -1.11 \pm 0.25$\AA\ ($-0.81 \pm 0.06$\AA) and $W_{\rm CIII]} = -2.15 \pm 0.67$\AA\ ($-0.73 \pm 0.13$\AA) for LCEs (non-LCEs).

Similarly, \citet{Naidu2022} stacked the UV spectra of potential LyC emitting candidates according to their \lya\ line properties, showing a very different \ion{He}{ii}{1640} profile respect to the non-emitting candidates (probably attributed to sample selection). Contrarily, in the recent work by \citet{MarquesChaves2022} based on direct LyC measurements of low-$z$ LzLCS galaxies, the authors do not find any significant correlation between the LyC escape fraction and the spectral hardness, where the \ion{He}{ii}{1640} intensity is fundamentally driven by changes in the metallicity rather than by \fescabs. Therefore, the slightly higher observed equivalent widths in \ion{He}{ii}{1640} and \ion{O}{iii]}{1666} for the LCEs composite \citep[see][]{Llerena2022} may indicate a much dissimilar gas-phase metallicity between the LCEs and non-LCEs stacks than the stellar metallicities that are actually derived from our stellar population modeling (see Sect.\ \ref{sub:ion_properties}).

Regarding the UV carbon nebular lines and based on the work by \citet{Schaerer2022, Saxena2022b} and \citet{Mascia2023b}, we compute the \ion{C}{iv}1550 over \ion{C}{iii}]1908 flux ratio, and we obtain $\ion{C}{iv}/\ion{C}{iii]} = 0.79 \pm 0.21$ for LCEs while $\ion{C}{iv}/\ion{C}{iii]} = 0.42 \pm 0.20$ for non-LCEs (the \ion{C}{iv}1550 doublet is blended at the current resolution). 
\citet{Schaerer2022} empirically demonstrated that both \ion{C}{iv}1550 and \ion{C}{iii}]1908 lines tend to be stronger in the LzLCS galaxies with the highest \fescabs, and proposed a $\ion{C}{iv}/\ion{C}{iii]} \geq 0.75$ threshold for significant LyC leakage which, once again, is in agreement with our stacking analysis. As we will see in the next section, a combined effect of escaping ionizing radiation (favoured neutral gas and dust geometry) with a higher \xiion\ parameter is responsible for enhancing the emission of high-ionization-state lines such as \ion{C}{iv}1550 and \ion{C}{iii}]1908 in LCEs.

\subsubsection{Absorption lines and ISM properties}
Several stellar-wind (\ion{N}{v}1240, \ion{Si}{iv}1400, \ion{C}{iv}1550) and other ISM UV absorption lines (\ion{Si}{ii}1260, \ion{O}{i}+\ion{Si}{ii}1302, \ion{C}{ii}1334, \ion{Si}{ii}1527, \ion{Fe}{ii}1608, \ion{Al}{ii}1670 nd \ion{Al}{iii}1858 doublet) are detected in our VANDELS composite spectra (Fig.\ \ref{fig:composite_spectra}). 

The \ion{N}{v}1240 stellar-wind line in LCEs shows a characteristic P-Cygni profile that can be fully reproduced by stellar templates (excluding potential contamination by AGNs), and it indicates the presence of a young stellar population ($\leq 5~{\rm Myr}$). Conversely, \ion{N}{v}1240 is suppressed by the absorbing part of the damped \lya\ red-wing in non-LCEs, indicating a high \ion{H}{i} column density. Going to redder wavelengths, the \ion{Si}{iv}1393,1402 high-ionization state absorption lines, which share both a stellar and ISM origin, have the same equivalent width in the LCEs and non-LCEs composites, suggesting a homogeneous distribution of the diffuse high-ionized species in the ISM of LBGs. 

The remaining set of LIS absorption lines in LCEs (\ion{Si}{ii}1260, \ion{O}{i}+\ion{Si}{ii}1302, \ion{C}{ii}1334, \ion{Si}{ii}1527, \ion{Fe}{ii}1608, \ion{Al}{ii}1670 and \ion{Al}{iii}1858 doublet) clearly show lower equivalent widths than in the non-LCEs stack. For example, the \ion{S}{ii}1260 and \ion{C}{ii}1334 in Fig.\ \ref{fig:composite_spectra}, $W_{\rm SiII} = 0.99 \pm 0.13$ ($1.82 \pm 0.05$) for LCEs (non-LCEs), while $W_{\rm CII} = 0.64 \pm 0.13$ ($1.74 \pm 0.04$) in the LCEs (non-LCEs) composite. Following the picket-fence model, the fact that strong LCEs have significantly weaker absorption lines, i.e., lower equivalent widths with higher residual fluxes, can be attributed to a lower covering fraction of the spatially co-existing neutral and low-ionized gas, so that the photons from a given transition escape through low column-density channels in the ISM, as well as LyC photons do \citep[][\citetalias{SL22}]{Gazagnes2018,Gazagnes2020}. 

The observed connection between LIS absorption lines and the strength of the \lya\ line is also observed in our galaxy composites, and it can be explained by the picket-fence formulation, in which the gas covering fraction along the line-of-sight primarily governs the escape of the absorbed LIS, \lya\ and LyC photons. This behavior is well summarized in Fig.\ \ref{fig:Lya_Rlis}, where the \ewlya\ is plotted against the measured (averaged) residual flux of the LIS lines, \rlis, and the points are color-coded by \ebv. Blue and red symbols correspond to our LCEs and non-LCEs galaxy composites, respectively. 

As long as the line-of-sight covering fraction of gas decreases and the dust-attenuation increases (traced by higher \rlis\ and lower \ebv\ values), the \lya\ equivalent width dramatically increases towards values typically found in LAEs. For being the case, the stronger LAEs, which has been previously suggested in this work to be a proxy for the bulk of the LCEs population, fall in the top-right part of the plot.

This trend matches the results by \citet{Steidel2018} in the framework of the KLCS, and echoes the main conclusions by \citet{Gazagnes2020}, using individual LyC detected galaxies at low-$z$. In the same Fig.\ \ref{fig:Lya_Rlis}, the median results for LCEs and non-LCEs in the LzLCS \citep{Flury2022a}, are indicated by white open hexagons \citepalias[see also][]{SL22}. As we can see, the LzLCS results clearly agree with our VANDELS stacks, which reassure the use of the picket-fence approach for describing the ISM of LCEs. Additionally, the results by \citet{Jones2013} for individual LAEs at $z = 2-4$ are shown on top of the VANDELS results, following the general trend.

\subsection{Ionizing properties of potential LCEs at $3 \leq z \leq 5$}\label{sub:ion_properties}
The ionizing properties of LCEs and non-LCEs can be accessed by extrapolating the non-ionizing best-fit SED to ionizing wavelengths. In particular, it is of our most interest to check whether potential differences in the ionizing-to-nonionizing flux ratio, \Fion, and production efficiency, \xiion, exist.

First, we input the LCEs and non-LCEs stacked spectra into the \textsc{FiCUS} code in order to unveil the average properties of the underlying stellar population of each galaxy sample. Then, we use the best-fit stellar continua to define different galaxy observables (see Sect.\ \ref{sub:fits}). As we can see, the fit to LCEs provides a lower stellar age ($13 \pm 3$~Myr) than for non-LCEs ($19 \pm 1$~Myr), tentatively suggesting younger stellar populations for LCEs. On the other side, stellar metallicities are very similar for both (${\rm Z_{\star}} \approx 0.2 ~{\rm Z_{\odot}}$), and compatible with the median of the total sample. These values can be seen in Fig.\ \ref{fig:composite_fits}, where the best-fit \textsc{FiCUS} SEDs are superimposed to the LCEs and non-LCEs spectra. 

The inferred dust-attenuation parameter is remarkably dissimilar for LCEs and non-LCEs, being $\ebv \approx 0.07$ and $0.23~{\rm mag.}$, respectively. This is by construction, since the parent sample has been splitted based on the \fescabs\ of individual VANDELS galaxies (Eq. \ref{eq:fesc_LIS}), which is explicitly related to \ebv. It reflects the expected less dusty nature of LCEs compared to the bulk of the LBG population, as we explained in the previous section.

The intrinsic (i.e., dust-free) ionizing spectra of the resulting best-fit SED for LCEs and non-LCEs is presented in the inset of Fig.\ \ref{fig:composite_ion} (\emph{left}) through blue and red solid lines, correspondingly. The slight decrease in age and metallicity makes the intrinsic ionizing-to-nonionizing flux ratio \citepalias[900-to-1500\AA, see][]{C19} to grow up to $\Fion = 0.64$ in LCEs from $0.47$ for non-LCEs. A different (intrinsic) 900-to-1500\AA\ ratio due to an decrease in age also fosters a rise in the ionizing efficiency to $\log \xiion~{\rm (Hz/erg)} = 25.38$ for LCEs, while it is $25.18$ for the non-LCEs composites. In combination with the covering fraction of neutral gas, the \xiion\ parameter is partially responsible for shaping the strength of high-ionization-state lines such as \ion{C}{iv}1550 and \ion{C}{iii}]1908, and other nebular resonant lines such \lya.

Ultimately, we infer what the observed AB magnitude at LyC wavelengths would look like for the LCEs and non-LCEs populations, as a guidance for upcoming surveys targeting LyC emitting galaxies at similar FUV magnitudes. According to the fundamental definition of the absolute ionizing escape fraction of galaxies \citep[see][\citetalias{SL22}]{Izotov16b}, the observed flux close to the Lyman edge ($F_{\lambda 900}$) can be simply predicted by doing: $F_{\lambda 900} = \fescabs \times F_{\lambda 900}^{\rm int}$, where $F_{\lambda 900}^{\rm int}$ is the intrinsic LyC flux or, in other words, the ionizing flux previous to any gas or dust absorption (in $F_{\lambda}$ units). 

Since we have access to the intrinsic ionizing SED through the \textsc{FiCUS} fits, and the \fescabs\ has been predicted following Eq. \ref{eq:fesc_LIS}, we compute $F_{\lambda 900}$ for every galaxy in the VANDELS sample. We then derive the AB apparent magnitude from the $F_{\lambda 900}$ synthetic flux measurements ($m_{\rm LyC}, {\rm AB}$). The resulting distribution can be seen in Fig.\ \ref{fig:composite_ion} (\emph{right}), where the blue (orange) histogram represent the LyC AB distribution for LCEs (non-LCEs), once we manually apply a $\langle T_{\rm IGM} \rangle = 0.3$ mean IGM transmission factor, i.e., the mean $T_{\rm IGM}$ at $z = 4$ according to \citet{Inoue2014}.

The consecutive effects of neutral \hi\ gas absorption and dust attenuation in the LCEs and non-LCEs ionizing spectra are represented in Fig.\ \ref{fig:composite_ion} (\emph{left}) through the black and grey arrows (assuming a constant \ion{H}{i} cross-section with wavelength), whilst the blue and orange open circles denote the LyC flux measured at $870-910$\AA. After applying a $\langle T_{\rm IGM} \rangle = 0.3$ factor and translating into AB magnitudes, these two points would equal the median of the individual LyC AB magnitude distribution shown in the right panel of the same figure. 

Clearly, the combined effects of a lower ionizing-to-nonionizing flux ratio plus lower escape fractions for non-LCEs makes their median output LyC flux dwindle $\simeq 2$mag. down with respect to a hypothetical LCEs population drawn from the same LBG sample. Therefore, a dedicated survey which attempts to detect the bulk of LCEs at $3 \leq z \leq 5$ would required a sensitivity such that it reaches, at least, $m_{\rm LyC} \simeq 31$~AB at LyC wavelengths.

Using very deep VIMOS/U--band imaging of the CDFS field ($1\sigma$ limiting magnitude of 30.4~AB), \citet{Begley2022} estimated 10 to 15 LyC detections at $2\sigma$ for sources whose redshift is $3.35 \leq z \leq 3.95$, assuming an average $\fescabs \approx 0.07$. Restricting to the CDFS only, we found 7 sources with predicted $m_{\rm LyC} \leq 30$~AB in the same redshift range. The differences can be attributed to our lower average escape fraction compared to the former work. This said, only two robust LyC emitters have been detected so far in the CDFS field (see following section). The stochasticity of the IGM opacity might reduce our expectations.

\begin{figure*}
    \includegraphics[width=0.90\textwidth, page=12]{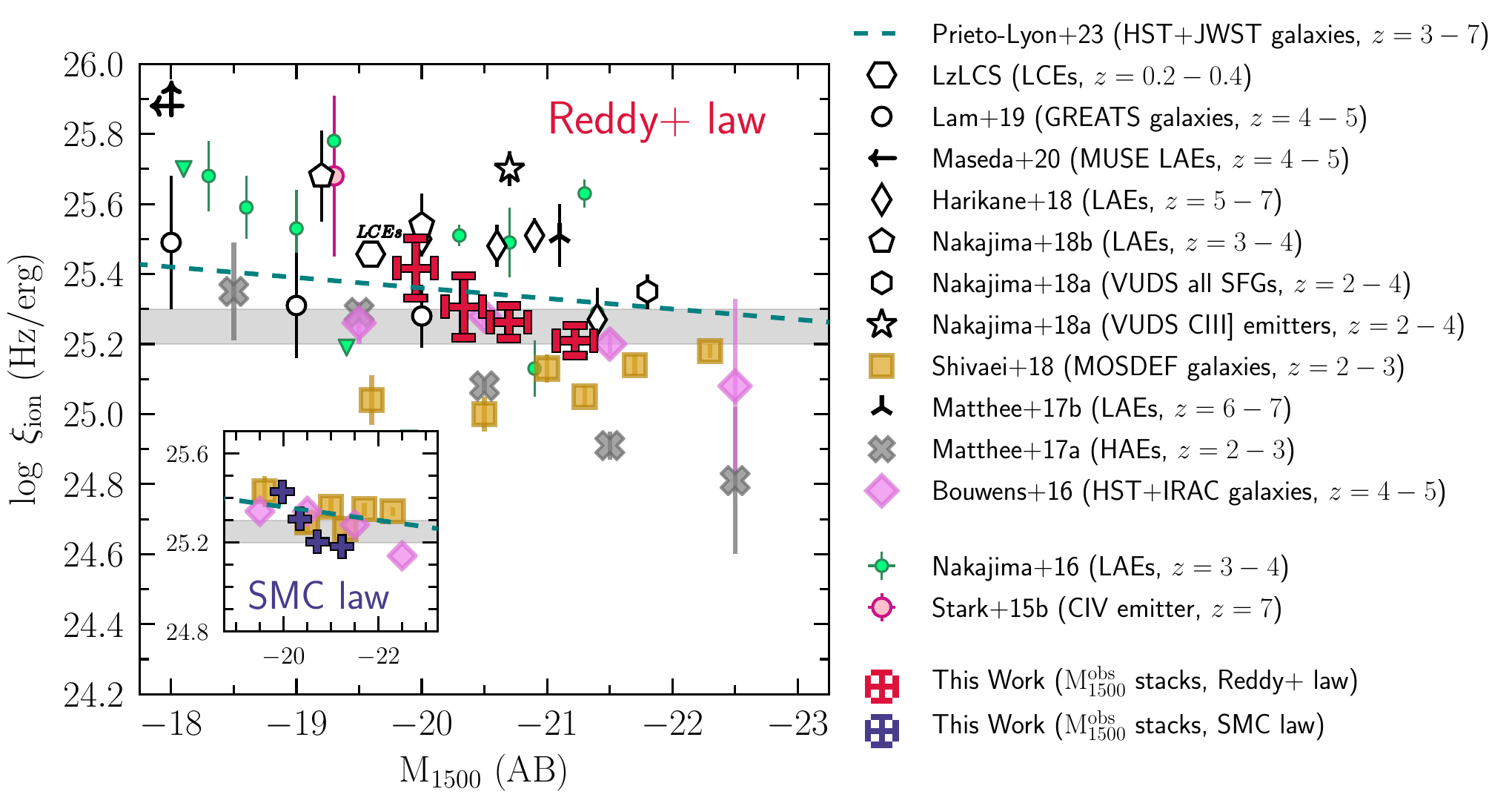}
\caption{$\log \xiion$ as a function of the observed absolute UV magnitude. The red thick symbols show the results of our $3 \leq z \leq 5$ VANDELS composite spectra in bins of UV magnitude when using the \citetalias{R16} attenuation law (\citetalias{SMC} dust-law results are shown in the inset). Stacked results from other literature samples at various redshifts are represented via open white symbols (see legend), in particular for low-$z$ LCEs \citep{Flury2022a}, high-$z$ LAEs \citep{Maseda2020, Nakajima2018b, Harikane2018, Matthee2017d}, and normal high-$z$ SFGs \citep{Nakajima2018a, Lam2019}. Our main comparison sources (studies targeting LBGs at UV magnitudes similar to the VANDELS galaxies) are the MOSDEF galaxies at $z = 2-3$ \citep[][orange squares]{Shivaei2018}, the extensive HST+IRAC campaigns at $z = 4-5$ by\citet[][blue diamonds]{Bouwens2016}, and the HAEs sample at $z = 2-3$ by \citet[][grey crosses]{Matthee2017b}. The individual LAEs and CIV emitters measurements from \citet{Nakajima2016} at $z = 3-4$ and \citet{Stark2015} at $z=7$ are plotted in coloured circles. The teal dashed line follows the extrapolation to brighter magnitudes of the recent HST+JWST results at $z  = 3-7$ \citep{PrietoLyon2022}. The grey-shaded area marks the canonical $\log \xiion = 25.2-25.3$ value given by a simple stellar population at constant SFR over 100Myr \citep{Robertson2013}. For reference, $M_{\rm UV}^{\star} \approx -21$ at $z = 6$ \citep{Bouwens2021}. Overall, $\log \xiion$ monotonically increases with \mobs.}
\label{fig:XiionMuv}
\end{figure*}

\subsection{Confirmed LCEs in the VANDELS sample}\label{sub:confirmed_LCEs}
In Fig.\ \ref{fig:Lya_Rlis}, we highlighted two LyC emitting galaxies included in our VANDELS sample which have been already published in the literature, in particular within the CDFS field \citep[see][]{Begley2022}: CDFS012448 \citep{Saxena2022a} and \emph{Ion1} \citep{Vanzella2012, Ji2020}. 

While both galaxies share a similar SED dust-attenuation ($\ebv \simeq 0.2$~mag.), we measure a much lower residual flux in CDFS012448 ($\rlis \simeq 0.4$) than in \emph{Ion1} ($\simeq 0.55$), yielding to a predicted escape fraction of $\fescabs \simeq 3\%$ versus $\simeq 6\%$. It is worth stressing that our inferred \fescabs\ value for \emph{Ion1} equals the reported value by \citet{Ji2020}. In the case of CDFS012448, our \fescabs\ is remarkably different to the one quoted in \citet{Saxena2022a} of $\approx 20\%$, possibly because the different approaches between papers. Our method was able to identify both galaxies as potential LCEs. 

Additionally, these galaxies exhibit very different \lya\ properties between each other: while CDSFS012448 shows a relatively strong \lya\ in emission ($\ewlya \simeq -20$\AA), the \emph{Ion1} \lya\ profile appears in absorption ($\ewlya \simeq 1$\AA). In our VANDELS sample, $\simeq 10\%$ of the LCEs ($\fescabs \geq 5\%$) appear with \lya\ in absorption. This manifests, once again, the variety of \lya\ profiles that can be found among the LCEs galaxy population \citep{Izotov2020, Flury2022b, Naidu2022} where, given the fact that \lya\ can be resonantly scattered whereas LyC cannot be scattered, both \lya\ and the ionizing emission do not necessarily come from the same spatial location within the host galaxy \citep[e.g.,][]{Vanzella2012}. A low IGM transmission could also, in principle, have killed out most of the \lya\ photons of \emph{Ion1}.

%% file: 5_discussion.tex
\begin{figure}
    \includegraphics[width=0.925\columnwidth, page=13]{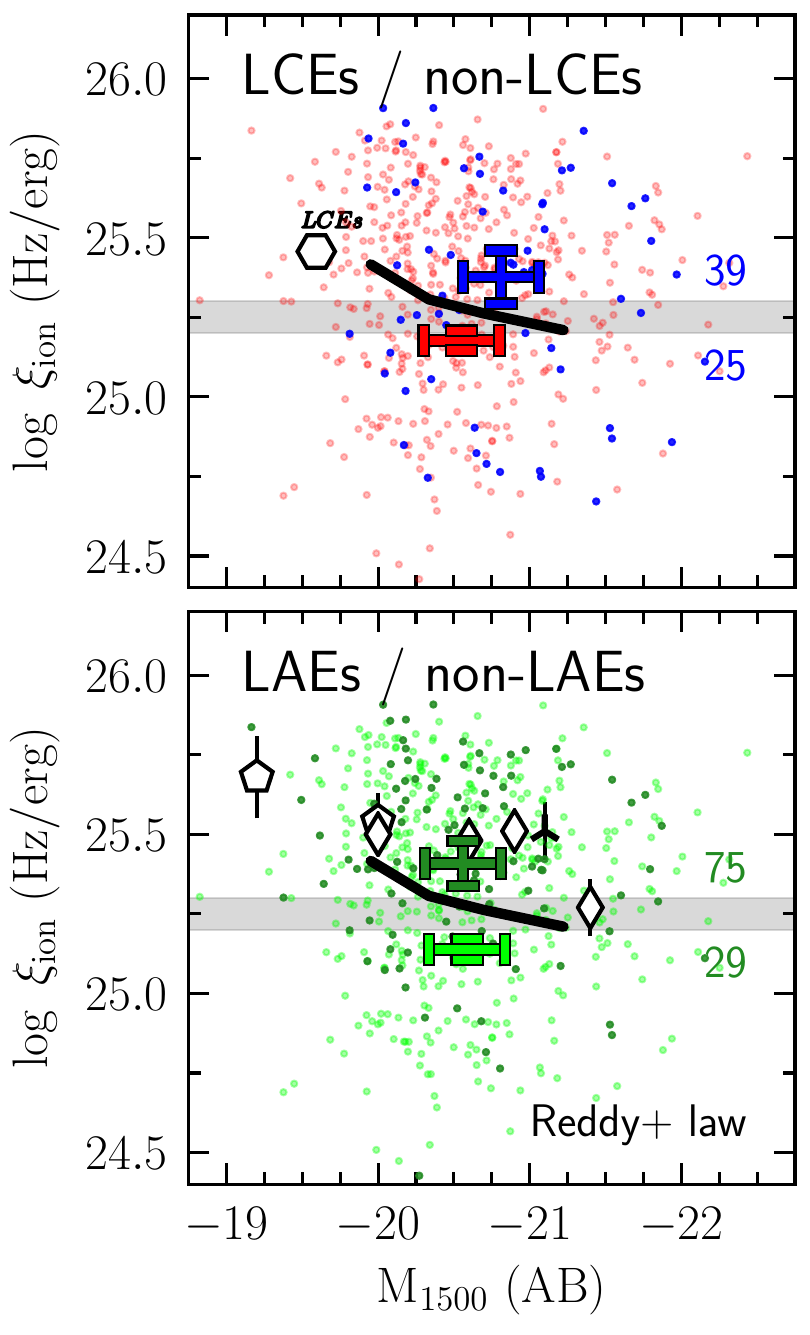}
\caption{$\log \xiion$ versus \mobs\ color-coded for different galaxy samples. {\bf Top:} potential LCEs versus non-LCEs (in blue and red, respectively). {\bf Bottom:} selected LAEs and non-LAEs (in green and light-green). The corresponding measurements for LCEs/non-LCEs and LAEs/non-LAEs composite spectra are represented through thick filled symbols, and the thick black line represents our stacked measurements in bins of UV magnitude. Stacked results from other literature samples are represented via open white symbols, in particular for low-$z$ LCEs \citep{Flury2022a} and high-$z$ LAEs \citep[][see Fig.\ \ref{fig:XiionMuv} for symbols]{Matthee2017d,Harikane2018,Nakajima2018b}. The grey-shaded area marks the canonical $\log \xiion = 25.2-25.3$ value given by a simple stellar population at constant SFR over 100Myr \citep{Robertson2013}, and the number of LCEs and LAEs above and below this limit is indicated in the right side of the plot. In conclusion, LCEs and LAEs have systematically higher \xiion\ than the bulk of the LBG population \citep[see also][]{Reddy2022}.}
\label{fig:XiionMuv_LAEs}
\end{figure}

\begin{figure*}
    \includegraphics[width=0.85\textwidth, page=14]{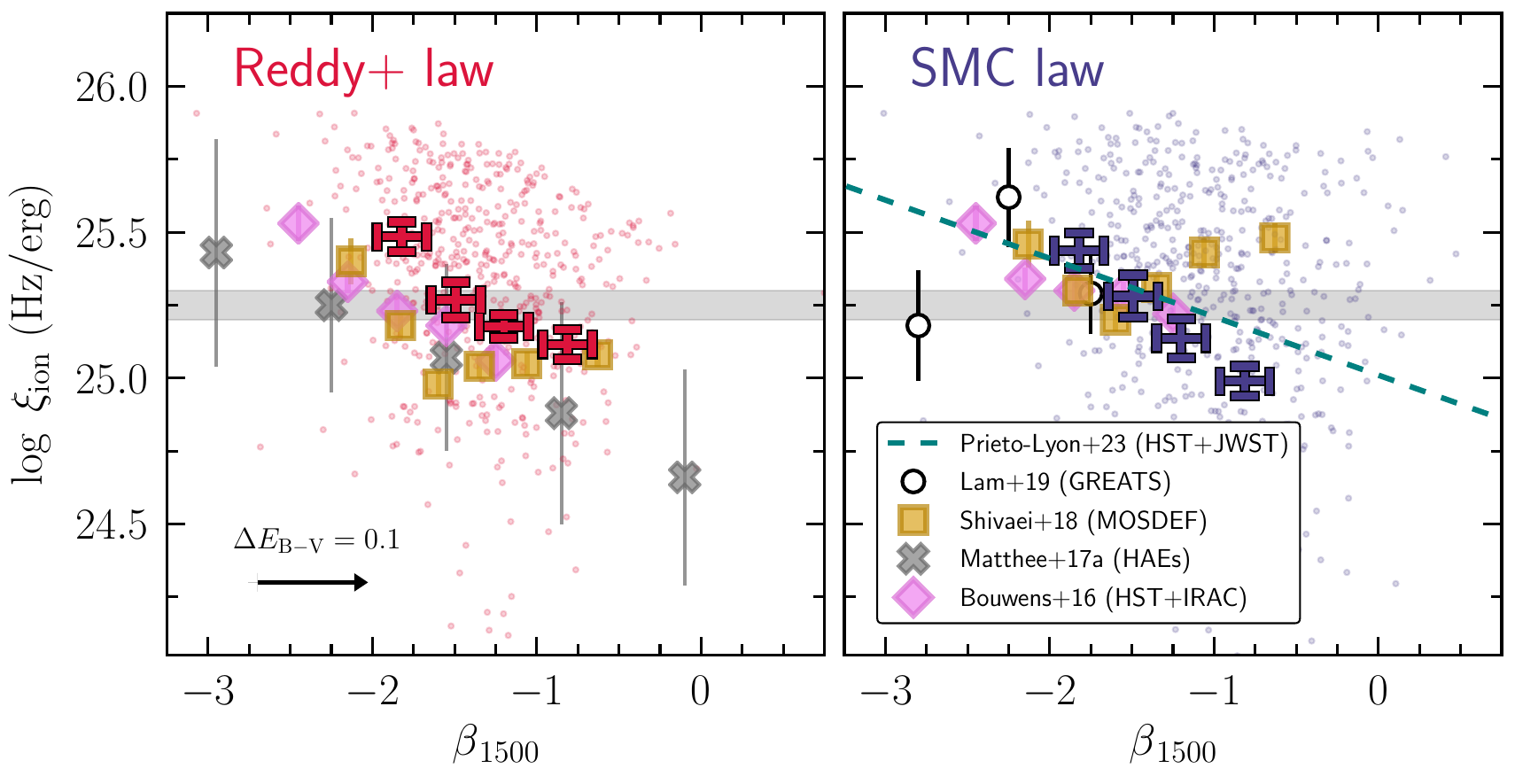}
\caption{$\log \xiion$ as function of the UV-continuum slope at 1500\AA\ ($\beta_{1500}$). The red (blue) dots show the results of our $3 \leq z \leq 5$ VANDELS individual spectra assuming \citetalias{R16} \citepalias{SMC} dust-attenuation law. Thick symbols represent the results from our composites in bins of $\beta_{1500}$. Orange squares show the MOSDEF results at $z = 2-3$ by \citet{Shivaei2018}, while blue diamonds and grey crosses correspond to the HAEs measurements in \citet{Bouwens2016} at $z = 4-5$ \citet{Matthee2017b} at $z=2-3$, respectively. Open symbols were taken from \citet{Lam2019}, and the teal dashed line corresponds to the results by \citet{PrietoLyon2022} using HST+JWST data at $z = 3-7$. 
The grey-shaded area marks the canonical $\log \xiion = 25.2-25.3$ value given by a simple stellar population at constant SFR over 100Myr \citep{Robertson2013}. $\log \xiion$ decreases with the UV slope almost unanimously for all samples.}
\label{fig:XiionBeta}
\end{figure*}

\section{The production efficiency of ionizing photons (\xiion) in high-redshift galaxies}\label{sec:discussion_xiion}
Here, we investigate the evolution of \xiion\ as a function of the UV absolute magnitude and the UV slope \citep[see][]{Bouwens2016, Chisholm2022}, quantities which will be easily measured for a significant fraction of galaxies at the EoR thanks to upcoming gound-based (\textit{ELT, GMT, TMT, ...}) and ongoing space-based facilities (\textit{JWST}). 


Fig.\ \ref{fig:XiionMuv} shows the \xiion\ parameter in stacks of the absolute UV magnitude for the VANDELS sample. As it was introduced in Sect.\ \ref{sec:results} and regardless of the attenuation law (see inset), the ionizing production efficiency monotonically, smoothly increases towards fainter UV magnitudes, so that $\log \xiion$ $\geq 25.2{~\rm Hz/erg}$ \citep{Robertson2013} at $\mobs \gtrsim -20$, i.e., the faint-end of our sample. 
Stacked results from other samples in the literature are also included in Fig.\ \ref{fig:XiionMuv}. The main comparison samples for our case of study are the MOSDEF galaxies from \citet{Shivaei2018} at $z = 2-3$, the extensive HST+IRAC searches by \citet{Bouwens2016} at $z = 4-5$, and the HAEs sample of \citet{Matthee2017b} at $z = 2-3$. Our results are qualitatively in agreement with \citet{Bouwens2016} and \citet{Matthee2017b}, both tracing photometrically-selected HAEs at similar \mobs\ than ours. Nevertheless, VANDELS samples a narrower range in \mobs\ so that the overall trend does not appear as clear as in the former works. At the faintest UV magnitude bin, our $\xiion - \mobs$ relationship disagrees with the results by \citet{Shivaei2018} who report no $\log \xiion$ evolution with \mobs, a result probably induced by the different dust-attenuation corrections between studies \citep[Balmer Decrements for MOSDEF versus global SED for the rest, see the discussion in][]{Matthee2017b}. 

Due to their lower metallicities and burstiness of their SFHs, fainter (low-mass) galaxies are expected to have high ionizing production efficiencies \citep{Ma2020}. If the LyC escape fraction and production efficiency were intertwined (see below), the fact that fainter galaxies produce more ionizing photons per UV luminosity which, at the same time, would escape more easily than in higher-mass counterparts, supports an early and slow reionization scenario in which the ionizing budget would be dominated by the same low-mass, UV-faint galaxies \citep[see][]{Finkelstein2019, Trebitsch2022}, thanks to their higher number density at the EoR \citep{Bouwens2015b, Bouwens2021}.

Recently, \citet{PrietoLyon2022} used deep HST and JWST multiband photometry to estimate the H$\alpha$ flux and thus the ionizing production efficiency extending to very faint UV magnitudes ($-23 \leq \mobs \leq -15.5$) and spanning a wide range in redshift ($z = 3-7$). Their results demonstrate that, although the general trend of increasing \xiion\ with increasing UV magnitude is very smooth, it holds at any \mobs\ \citep[see also][]{Lam2019, Emami2020}. Additionally, our thorough compilation of \xiion\ measurements (Fig.\ \ref{fig:XiionMuv}) illustrates the scatter in this relation ($\simeq 1$dex) and its dependence on the galaxy type, physical properties of the underlying galaxy-population and methodology itself (with the dust corrections among the different samples playing a critical role in the scatter). While normal high-$z$ SFGs \citep{Nakajima2018a, Lam2019} have modest production efficiencies, high-$z$ LAEs \citep{Matthee2017d,Harikane2018,Nakajima2018b} and \ion{C}{iii]} emitters \citep{Nakajima2018a} show much higher $\log \xiion$ values at similar UV magnitudes, since overall these galaxies need stronger ionizing spectra to produce such high excitation lines \citep[see e.g.,][]{Reddy2018}. 

Intending to qualitatively illustrate this behavior (Fig.\ \ref{fig:XiionMuv_LAEs}), we search for meaningful differences in \xiion\ between the bulk of the VANDELS LBG population and other sources like LCEs or LAEs, which would potentially boost \xiion\ above the canonical value. This is done by splitting the sample into LAEs and non-LAEs, where the the sources whose $\ewlya > -20$\AA\ were considered as LAEs, following the definition by \citet[][but see \citet{Stark2011,Kusakabe2020}]{Pentericci2009}. A similar analysis as we did for the LCEs/non-LCEs composites was then performed for the LAEs/non-LAEs composites, and the results are described in the following lines. 

Of particular interest is to search for \xiion\ differences within different galaxy subsamples. In Fig.\ \ref{fig:XiionMuv_LAEs} (\emph{top}), the VANDELS sample is color-coded by potential LCEs ($\fescabs \ge 0.05$, in blue) and non-LCEs ($\fescabs < 0.05$, in red), where approximately $60\%$ of the LCEs sample (39 out of 64 LCEs candidates) has $\log \xiion$ $\geq 25.2{~\rm Hz/erg}$, and they uniformly populate the whole range in UV magnitude. As said in Sect.\ \ref{sub:ion_properties}, the resulting \xiion\ for LCEs and non-LCEs are $\log \xiion~{\rm (Hz/erg)} \approx 25.38$ and $25.18$, respectively. Compared with other values in the literature, our estimated ionizing efficiency for LCEs is consistent with the average of the LCEs sample of the LzLCS \citep{Chisholm2022}. The fact that our results at $3 \leq z \leq 5$ agree with low-$z$ LCEs sample supports the argument by which the LzLCS galaxies might be analogs of higher-redshift reionizers \citep{Flury2022b}, This has been lately demonstrated by \citet{Mascia2023} through JWST spectroscopy of 29 moderately faint galaxies at $5 \leq z \leq 8$, whose optical line ratios, UV slopes, compactness, masses and predicted \fescabs\ were compatible with the strongest LCEs in LzLCS sample \citep[see also][]{Lin2023}.

Similarly, in Fig.\ \ref{fig:XiionMuv_LAEs} (\emph{bottom}) we explore the $\xiion - \mobs$ relationship for selected LAEs ($\ewlya > -20$\AA, in green) and non-LAEs ($\ewlya \leq -20$\AA, in light-green). Although the \mobs\ distribution for both galaxy types equally span along the $x-$axis of this figure, around $70\%$ of the LAEs sample (75 out of 104 identified LAEs) falls above the $\log \xiion$ canonical value in the $y-$axis. Our \xiion\ estimate for LAEs is in prefect agreement with the results by \citet{Harikane2018} at $z = 5-6$, and it matches surprisingly well the global trend found by several high-$z$ LAEs surveys at $3 \leq z \leq 7$ \citep[][see also \citet{Ning2022}]{Nakajima2018b, Harikane2018, Matthee2017d}. Similar to \lya, the ionizing efficiency also shapes the emission of other nebular lines \citep{Reddy2018} so that, for instance, higher-than-average \xiion\ values have been found among extreme high-$z$ [\ion{O}{iii}] emitters \citep{Tang2019} and local, high H$\beta$ equivalent width galaxy samples \citep{Izotov17}. 

The escape of \lya\ photons and the production of strong UV nebular emission usually requires special ISM and radiation field conditions such as low metalliticy, low dust and neutral gas content, and a vast production of (high-energy) ionizing photons. In Fig.\ \ref{fig:XiionMuv_LAEs} one can visualize a boost in the ionizing production efficiency (\xiion) for high-$z$ LAEs' samples. At the same time, a rise in the ionizing photon flux may reduce the covering fraction or column density of neutral hydrogen, which may ease the escape of LyC and \lya\ photons \citep[see][]{Erb2014}. Recently, \citet{Flury2022b} and \citet{Schaerer2022} statistically demonstrated that the detection rate of LCEs is enhanced among the LAEs and \ion{C}{iv} emitter galaxy-population \citep[see also][]{Saxena2022b, Mascia2023b}, reiterating our results. 

Moving forward with the discussion, Fig.\ \ref{fig:XiionBeta} shows \xiion\ as function of the UV-continuum slope for the VANDELS sample when using either the \citetalias{R16} (\emph{left}) or the \citetalias{SMC} (\emph{right}) attenuation law. Irrespective of the dust law, the ionizing efficiency rapidly decreases with the slope of the continuum, and the trend is more steep for the \citetalias{SMC} law. Focusing on the stacks measurements, the canonical limit in \xiion\ ($\log \xiion$ $\geq 25.2{~\rm Hz/erg}$) is only achieved for galaxies whose slopes are bluer than $\beta_{1500} \leq -1.5$. The $\xiion - \beta_{1500}$ relation of a dustless, single-burst stellar population at increasing age is not able to fully reproduce the whole range of $\beta_{1500}$ values, so that it would require ``some'' dust reddening to account for redder slopes (see horizontal arrow in Fig.\ \ref{fig:XiionBeta}, indicating the change in UV slope after adding 1~mag. of reddening). The nebular continuum would also contribute to increase the spectral slope of the youngest bursts of star formation.

Compared to the already reported relations by \citet{Bouwens2016} at $z = 4 - 5$ and by \citet{Matthee2017b} at $z = 2 - 3$, our results provide a similar slope but systematically shifted to redder \bslope. Several effects can contribute to this offset, but we mostly attribute it to the use of broad-band photometry to compute the UV slopes instead of direct spectral measurements, and the different wavelength range probed by multi-band photometry, usually broader than with spectroscopy. In contrast, the flattening of the $\xiion - \beta_{1500}$ relation proposed by \citet{Shivaei2018} at $\beta_{1500} > -1.6$ is not consistent with our findings. \citet{Izotov17} reported a similar general trend of decreasing \xiion\ with increasing UV slope for local, compact SFGs, but shifted to lower \xiion\ than ours. The new results by \citet[][teal-dashed line in the plot]{PrietoLyon2022} using combined HST+JWST photometry are broadly consistent with our estimations.

In summary, our results provide observational evidence for moderately UV-faint, low-mass and less dusty (un-obscured, blue UV colors) high-redshift galaxies most likely being able to drive the Cosmic Reionization at $6 \leq z \leq 9$ \citep{Finkelstein2019, Chisholm2022, Trebitsch2022, Lin2023}. According to recent JWST observations \citep[e.g., see ][]{Endsley2022, Cullen2023, Mascia2023}, these properties seem to be common among the galaxy population at the reionization epoch. These UV-faint galaxies would produce higher amounts of ionizing photons (through young bursts of star formation, possibly at low metallicities) and allow these photons to escape more easily due to their favoured ISM physical conditions \citep[dustless, gas-less holes in the ISM, see][\citetalias{SL22}]{Gazagnes2020}. Due to their high number density at earlier epochs \citep[$M_{\rm UV}^{\star} \approx -21$ at $z = 6$, see][]{Bouwens2021}, they might become the major contributors to the ionizing budget at the EoR (see Eq.\ \ref{eq:nion}).

\begin{figure*}
    \includegraphics[width=0.95\textwidth, page=17]{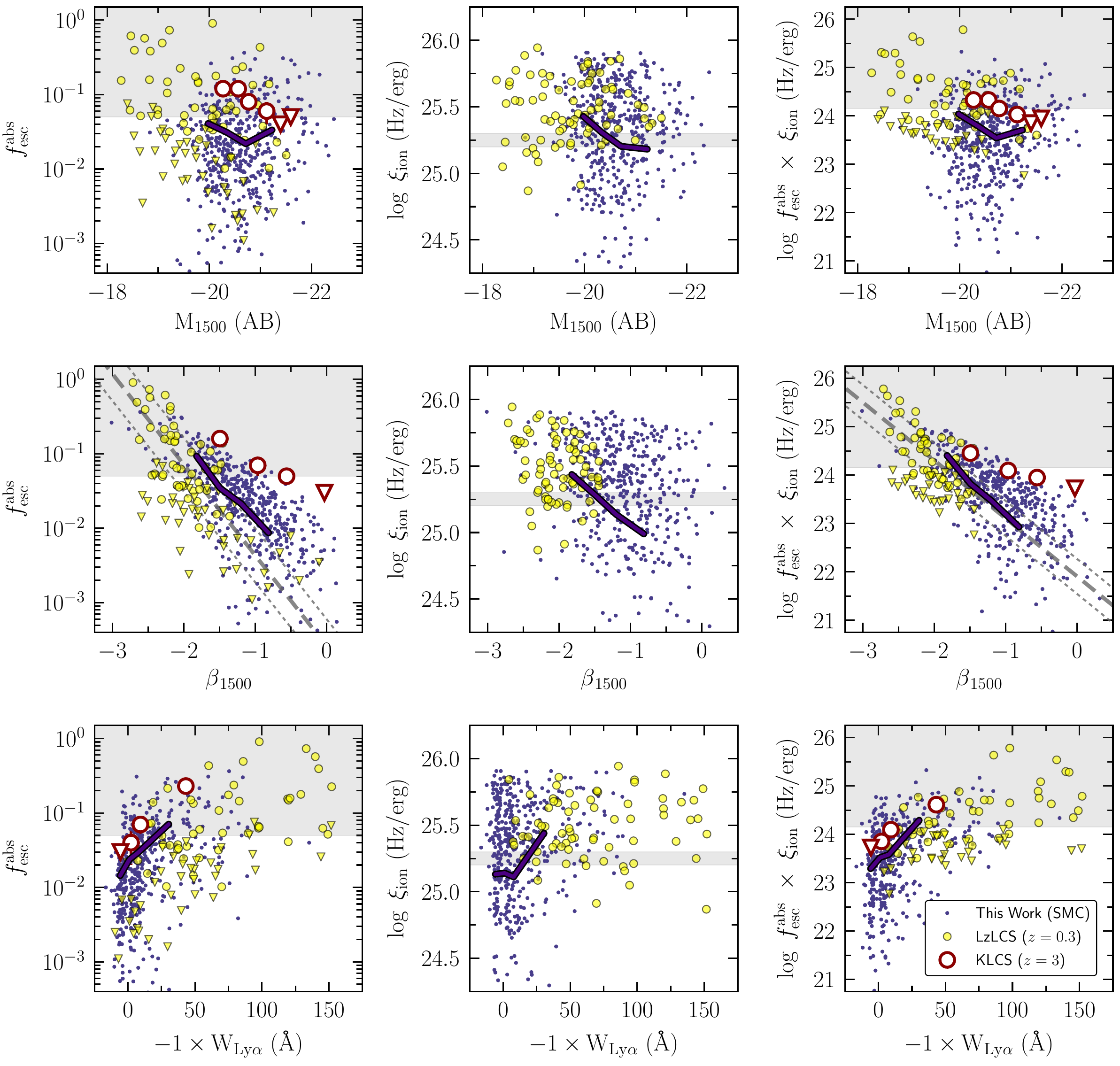}
    \caption{The absolute photon escape fraction (\fescabs), the ionizing production efficiency ($\log \xiion$) and the product of the two ($\log \fescabs \times \xiion$, see Eq.\ \ref{eq:nion}) for different low-$z$ and high-$z$ samples. Our predictions for the VANDELS galaxies at $3 \leq z \leq 5$ are shown through dark-blue points in the background (\citetalias{SMC} law), and the running locus of the stacking results is indicated via the thick blue line. For comparison, the red open symbols represent the latest KLCS \citep{Steidel2018} composite results at $z \sim 3$ by \citet{Pahl2021}, and the yellow circles show individual measurements at $z \sim 0.3$ from LzLCS \citep[][triangles are 1$\sigma$ upper limits]{Flury2022a}. The dashed grey lines draw the fit relations (median $\pm 1\sigma$ error) to the LzLCS points \citep{Chisholm2022}. Grey-shaded regions mark the canonical values of $\fescabs \geq 5\%$ and $\log \xiion = 25.2-25.3$ in classical Reionization models \citep{Robertson2013}.}
\label{fig:fesc_samples}
\end{figure*}

\section{The ionizing properties of galaxies: a detailed comparison between low- and high-redshift studies}\label{sec:discussion_lowzhighz}
\begin{figure}
    \includegraphics[width=0.95\columnwidth, page=4]{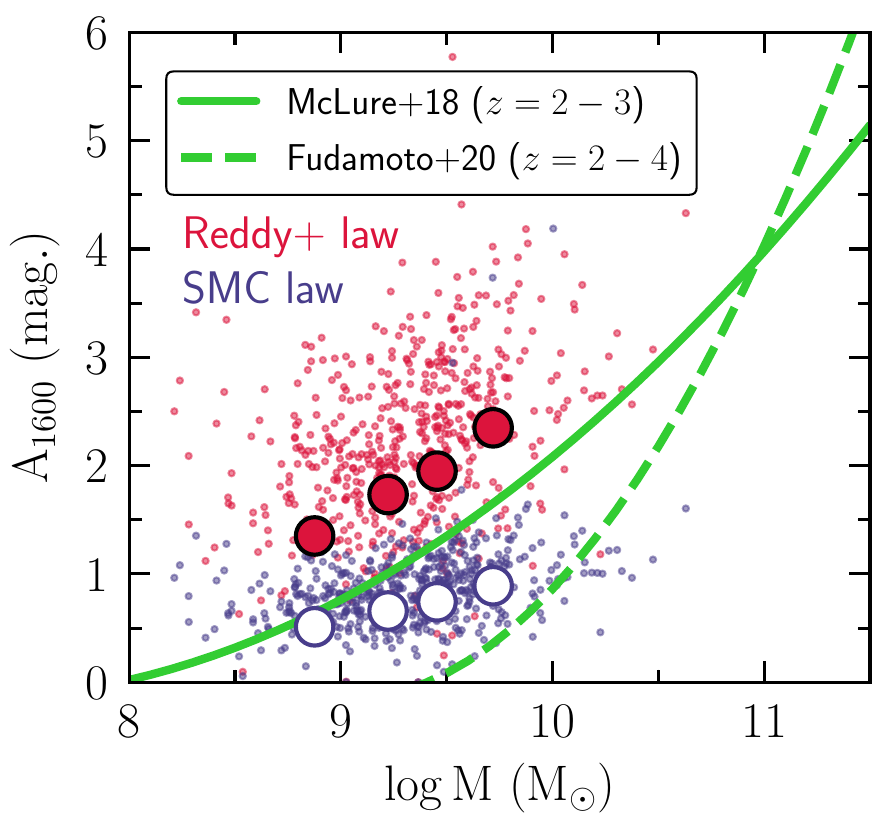}
\caption{The UV attenuation at 1600\AA\ ($A_{1600}$, in mag.) as function of the stellar mass ($M_{\star}$, in $M_{\odot}$) for individual VANDELS galaxies (dots in the background) and our stacks measurements (big circles). The solid and dashed green thick-lines represent independent estimates by \citet{McLure2018a} and \citet{Fudamoto2020} from IRX-excess measurements at $z = 2-3$ and $z = 3-4$, respectively. Red symbols assume a \citetalias{R16} dust-attenuation law while the blue symbols assume an \citetalias{SMC} attenuation curve.}
\label{fig:A1600_Mass}
\end{figure}

We now compare the ionizing properties of SFGs at $3 \leq z \leq 5$ with the estimated at other other redshifts and examine if and how they could evolve into the EoR. To do so, we put our \fescabs\ and \xiion\ trends along with observed properties of galaxies, together with other spectroscopic surveys of LCEs in the literature. On the one hand, we use the LzLCS \citep{Flury2022a}, probing emission line galaxies at $z \simeq 0.3$, as our ``EoR-analogue sample'' of galaxies. The more extreme nature of the LzLCS galaxies, characterized by high ionization parameters, high SFR surface-densities and strong nebular emission lines, makes it the perfect sample to compare with, since EoR galaxies are expected \citep{Schaerer2016, Boyett2022} and seem to hold these properties \citep[e.g.,][]{Endsley2022, Schaerer2022b, Mascia2023, Lin2023}. At $z \simeq 3$, the KLCS \citep{Steidel2018, Pahl2021, Pahl2023} is used for comparison, because it targets LBGs at similar UV magnitudes (and other properties) to the VANDELS galaxies. For both the LzLCS\footnote{We note that the \fescabs\ and \xiion\ for LzLCS galaxies were obtained by running our \textsc{FiCUS} code in \citet{Flury2022a} and \citetalias{SL22}. The \fescabs\ was calculated as the ratio between the intrinsic ionizing flux predicted by \textsc{FiCUS} and the observed flux at LyC wavelengths.} and the KLCS direct LyC measurements are available. Besides drawing conclusions on the possible redshift evolution of the ionizing properties of galaxies, we will discuss the main systematic uncertainties and caveats of our methodology. 

In Fig.\ \ref{fig:fesc_samples}, the relations between \fescabs, \xiion\ and the \fxi\ product with the UV magnitude, UV-continuum slope and \lya\ strength are compared for the three samples: VANDELS (at $z = 4$, \citetalias{SMC}), KLCS (at $z = 3$) and LzLCS (at $z = 0.3$). Particularly important is how the \fxi\ product compares between the different surveys (third column of the figure). 

Interestingly, the properties of the LzLCS and the VANDELS samples show overall fairly consistent trends, despite the redshift range difference and different methods (LzLCS represents direct \fescabs\ measurements, and VANDELS predictions are based on absorption lines). The samples are also quite complementary. Extending over a wide range of UV magnitudes ($-22 \leq \mobs \leq -18$) and stellar masses ($10^{8}~{\rm M_{\odot}} \leq M_{\star} \leq 10^{10}~{\rm M_{\odot}}$), the escape fraction of LzLCS galaxies tentatively decreases with the UV brightness and stellar mass. At variance with this, VANDELS galaxies span a much narrower range in terms of UV magnitudes, and it is unable to reveal a clear \fescabs\ correlation with \mobs. Again, the ionizing efficiencies (\xiion) of both low- and high-$z$ samples are quite comparable, except for the lack of some low \xiion\ objects and the presence of stronger \lya\ emitters in LzLCS.

LzLCS galaxies show systematically bluer UV slopes ($-2.5 \leq \beta_{1500} \leq -1$) than VANDELS galaxies, which translates into much higher \fescabs\ values for the bluest galaxies in the LzLCS (up to $\fescabs \approx 80\%$) than any of the galaxies within the VANDELS survey. However, both samples show a significant decrease in their ionizing properties (\fescabs\ and \xiion) with the UV-continuum slope. Strikingly, our VANDELS results follow the same slope provided by the fit to LzLCS data \citep[grey lines in Fig.\ \ref{fig:fesc_samples}, see][]{Chisholm2022}, but extrapolated to ($\geq 0.5$) redder UV slopes and lower values of \fescabs\ and \xiion. This is interesting as it reassures the applicability of our method for inferring the escape fraction of galaxies \citepalias[see \citet{Chisholm2018},][]{SL22}, since the \fescabs\ quoted for the LzLCS is actually the \emph{observed} escape fraction from measurements of the LyC flux. However, it is worth mentioning the shift in the intercept between the three surveys, with VANDELS and KLCS giving effectively higher \fxi\ values at a given $\beta_{1500}$ than the extrapolation by \citet{Chisholm2022}. This is interesting by itself since (1) it might hint a possible redshift evolution of the $\fxi - \beta_{1500}$ relation between $z \simeq 0.3$ (LzLCS) and $z \simeq 3$ (KLCS, VANDELS) or (2), if the high-$z$ surveys probe intrinsically lower UV slopes, they may get a better handle on the actual $\fxi - \beta_{1500}$ trend in redder galaxies. In any case, the VANDELS \fescabs\ and \fxi\ relations with $\beta_{1500}$ are compatible with \citet{Chisholm2022} within 1$\sigma$, so none of these hypothesis can be confirmed nor ruled out.

Finally, due to the more extreme nature of the LzLCS objects, their $\fescabs - \ewlya$ and $\xiion - \ewlya$ relationships are shifted to higher values of the \lya\ equivalent width respect to the VANDELS spectra. The intrinsically higher photon production efficiencies (\xiion) in LzLCS galaxies boost the strength of the \lya\ emission. Regarding the \fxi\ correlation with \ewlya\ for the LzLCS, only LCEs (yellow circles, while downward triangles represent non-LCEs) provide $\log \fxi ({\rm Hz/erg}) \geq 24.2$ \citep[see][]{Robertson2013}.

Even though our VANDELS survey populate a similar parameters space in terms of $\mobs, ~\beta_{1500}, ~\ewlya$ than the KLCS, the VANDELS \fescabs\ values fall systematically below the KLCS points, the latter acting like an upper envelope in all the correlations. In order to understand these differences, we now carefully list the main limitations and caveats of our ``picket-fence'' modeling. There are two main assumptions on the applicability of Eq.\ \ref{eq:fesc_LIS}: (1) the dust-attenuation law and (2) the conversion between the metals and \ion{H}{i} covering fractions \citepalias[see the full dedicated section in][]{SL22}.

First, the preference for a shallower \citepalias{R16} or a steeper \citepalias{SMC} attenuation curve can be tested by comparing our SED-derived attenuation versus independent measurements of the UV attenuation of galaxies. For example, one can translate IRX-excess measurements of high-$z$ galaxies into UV attenuation at 1600\AA\ ($A_{1600}$, in mag.), by using simple energy balance arguments \citep[see][]{Schaerer2013}. In Fig.\ \ref{fig:A1600_Mass} \citep[see][for details]{Cullen2018}, our VANDELS $A_{1600}$ versus stellar mass results --using both \citetalias{R16} (in red) and \citetalias{SMC} (in blue) laws-- are compared against the estimations by \citet{McLure2018a} at $z = 2-3$ and \citet{Fudamoto2020} at $z = 3-4$ from IRX-excess measurements (solid and dashed green lines, respectively). The VANDELS stacks measurements are represented through thick red and thick blue circles. If the more recent analysis of \citet{Fudamoto2020} is adopted, the \citetalias{SMC} law, and hence higher \fescabs, are preferred. However, the \citet{McLure2018a} results at similar redshift would indicate an intermediate attenuation between \citetalias{R16} and \citetalias{SMC}. The origin of these differences is difficult to understand and we currently have no way to distinguish which attenuation law is more appropriate. As discussed in Sect.\ \ref{sec:results}, assuming a most favourable attenuation law like \citetalias{SMC}, the \fescabs\ would be $\times 1.5$ higher compared to the ones derived using \citetalias{R16}.

\begin{figure}
    \includegraphics[width=0.925\columnwidth, page=16]{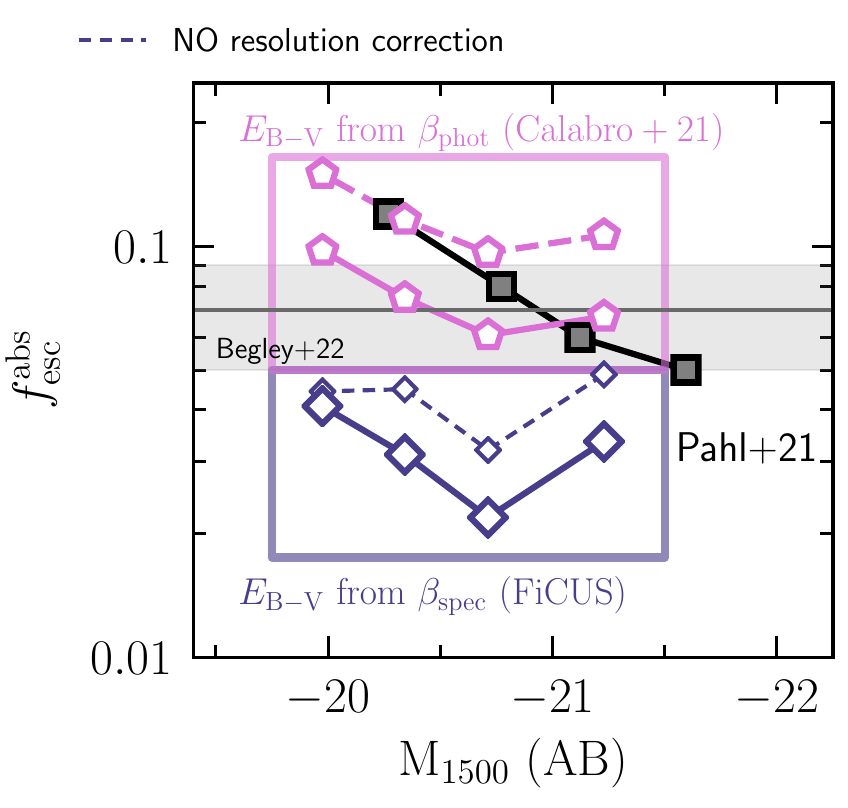}
\caption{A representation of the potential systematic bias in \fescabs\ due to the different values in the UV dust-attenuation that can be derived from dissimilar measurements of the $\beta-$slope. The blue diamonds show our default escape fractions from \textsc{FiCUS}, i.e., \ebv\ is derived from spectroscopic measurements of the UV continuum slope ($\beta_{\rm spec}$). Conversely, the pink pentagons display the \fescabs\ values when \ebv\ is inferred from the $\beta-$slope measurements by \citet{Calabro2021}, using photometry only ($\beta_{\rm phot}$). The average $\langle \fescabs \rangle = 0.07 \pm 0.02$ reported by \citet{Begley2022} at $z \simeq 3.5$ is indicated through the grey shaded-area, and the results by \citet{Pahl2021} are plotted through black squares. The effect of spectral resolution due to changes in the depths of the absorption lines is represented with the dashed line.}
\label{fig:BETAbias}
\end{figure}

Even if the attenuation law was known, \fescabs\ differences may reach depending on the way the UV dust-attenuation (\ebv) is measured. The dust-attenuation parameter at UV wavelengths is mainly constraint by the slope of the UV continuum \citep[e.g., $\beta_{1500}$, see][]{Chisholm2022}, by comparing the observed $\beta_{1500}$ against predictions of the intrinsic $\beta_{1500}^{\rm int}$ from SED fitting, either from the spectra or from multi-band photometry. In our case, \textsc{FiCUS} is able to estimate \ebv\ based on the spectral shape of the UV continuum. However, the slope of the current spectral observations might be subject to uncertainty due to different factors \citep[see][]{Garilli2021}: instrumental calibrations, reduction pipeline issues, flux losses because of the atmospheric dispersion and sky subtraction residuals. An alternative solution would be to use an independent determination of the UV slope. With this purpose, we use the $\beta_{1500}$ measurements from \citet{Calabro2021}, taking all the photometric bands whose bandwidths lie entirely inside the 1230–2750\AA\ rest-frame wavelength range, and the relation given in \citet{Chisholm2022} (Eq. 8, assuming $\beta_{1500}^{\rm int} = -2.5$) to convert this $\beta_{1500}$ to \ebv. Plugging these new dust-attenuation values into Eq.\ \ref{eq:fesc_LIS} produces higher estimations of \fescabs. The systematic deviation between the spectroscopic (\textsc{FiCUS}) and photometric \citep{Calabro2021} UV slopes by $\simeq 0.5$ on average, corresponds to $\simeq 1~$mag difference in UV attenuation at LyC wavelengths ($A_{912}$), which leads to a $\times 2.5$ higher photon escape fraction. As sketched in Fig.\ \ref{fig:BETAbias}, the new \fescabs\ values (pink pentagons) are in closer agreement with the measurements by \citet{Pahl2021} and \citet{Begley2022}, and fully consistent with the physical scenario in which the escape fraction of galaxies decreases towards brighter systems.

\begin{figure}
    \includegraphics[width=0.925\columnwidth, page=15]{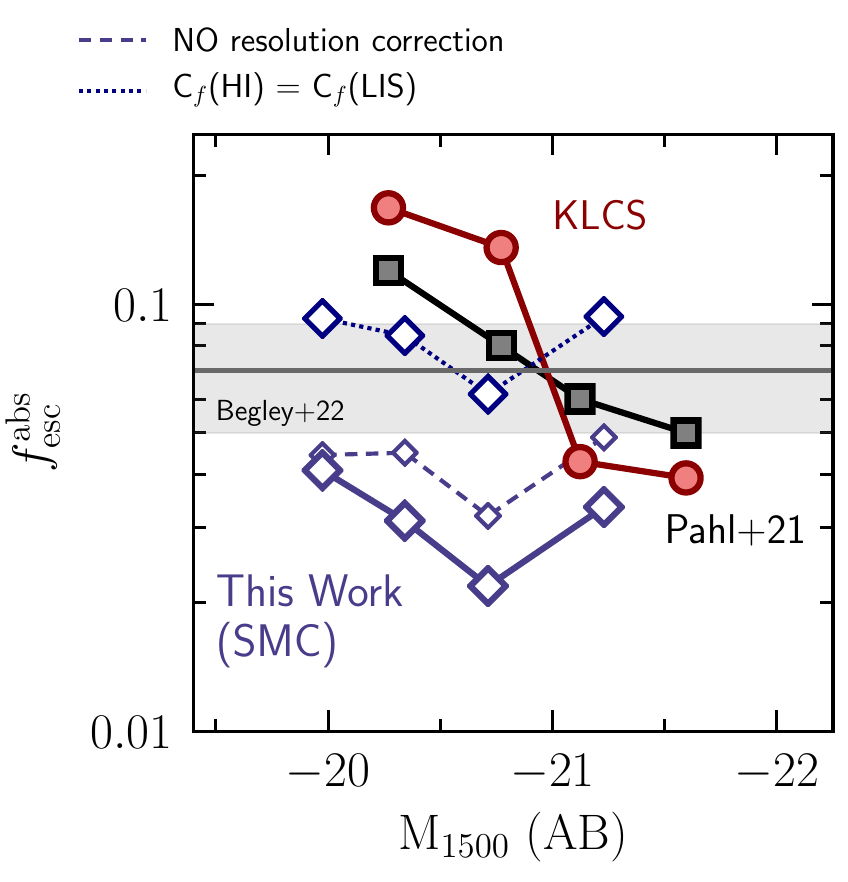}
\caption{A representation of the potential systematic bias in \fescabs\ (blue diamonds) due to the conversion between the metals and neutral gas covering fractions (blue dotted line). For comparison we also show the predicted \fescabs\ values for the KLCS stacks (red symbols), by applying our methodology to the published covering fractions and UV attenuations from \citet{Steidel2018}. The layout is the same as in Fig.\ \ref{fig:BETAbias}.}
\label{fig:LIStoHI}
\end{figure}

Second, in Fig.\ \ref{fig:LIStoHI} we study the influence that the adopted conversion between the metals and \ion{H}{i} covering fraction has on the predicted escape fractions. For comparison, we compute the predicted \fescabs\ values for the KLCS sample adopting the ``Screen Model'' \citep[see Table 9 in][]{Steidel2018}, which is identical to our default ``picket-fence'' geometry assuming a uniform slab of dust. We are not able to reconcile our escape fractions with the KLCS points, with our predictions lying a factor of $\times 2$ below them. However, if one assumes a 1:1 metals-to-\ion{H}{i} covering fraction correspondence so that $\cflis = C_f({\rm HI})$, instead of the usual linear regression by which $C_f({\rm HI} > \cflis)$ \citepalias[see \citet{Gazagnes2018} and][]{SL22}, the resulting \fescabs\ would agree with the estimate of the average $\langle \fescabs \rangle$ by \citet[][grey shaded area]{Begley2022}, which is now much more compatible with KLCS at the same time. 

Finally, the resolution correction function applied to the residual flux of the LIS lines (App.\ \ref{app:resolution}) might also affect our escape fraction estimates. In Fig.\ \ref{fig:BETAbias} and Fig.\ \ref{fig:LIStoHI}, we explicitly show the effect of our custom correction in the inferred \fescabs\ (dashed blue line). In the more conservative case in which the spectral resolution is totally dismissed and non-considered in Eq.\ \ref{eq:fesc_LIS}, \fescabs\ would increase by a factor of $\times 1-1.5$ at most: this is by far the effect with the least influence on the average \fescabs\ among the ones considered in this manuscript.

In conclusion, different effects can potentially augment our average \fescabs. Among them, the determination of the UV dust-attenuation has the strongest impact in the escape fraction estimates. In App.\ \ref{app:beta_Calabro} (Fig.\ \ref{fig:beta_Calabro}), we plot the \fescabs\ values resulting from plugging into Eq.\ \ref{eq:fesc_LIS} the dust-attenuation values derived from the $\beta_{1500}$ measurements from \citet{Calabro2021}. As we can see, all VANDELS, KLCS and LzLCS \fescabs\ correlations with physical quantities now converge to the same global trends, and they share a common behaviour by which \fescabs\ increases towards fainter UV magnitudes, bluer UV slopes and stronger \lya\ emission. In view of these results, the use of $\beta_{1500}$ from photometry as a proxy of the UV dust attenuation in combination with the depth of the absorption lines as a tracer of \hi-empty channels in the ISM of galaxies (Eq.\ \ref{eq:fesc_LIS}), seem to be a robust way of predicting the escape fraction of galaxies at any redshift. Independently, there is also the chance of (1) adopting a very steep dust-attenuation law, whose $k_{\lambda}$ extrapolation towards LyC wavelengths would be even higher than when using an \citetalias{SMC}-like law and (2) using a favourable metals-to-\ion{H}{i} covering fraction's conversion, in which the metals would spatially trace the neutral gas more closely than with the current assumed relation. 

The remaining question is whether a universal relation between the ionizing properties of galaxies (\fescabs\ and \xiion) and their physical properties ($\mobs, \beta_{1500}, \ewlya$~etc.) exists, in which the different surveys would populate different regions of the hypothetical universal trend or, contrarily, the observed differences between surveys (VANDELS, KLCS, LzLCS) indicates an evolution of these properties with redshift. Still, as far as our sample is concerned, the \fxi\ dependency along with galaxy properties is mainly due to escape fraction variations, so that the redshift evolution of this product will ultimately follow the changes in the ISM geometry, gas and dust composition across cosmic time.

In any case, and regardless the caveats that we have just described, the LzLCS, KLCS and VANDELS results suggest a scenario in which moderately UV-faint, low-mass and dustless galaxies most likely dominate the SFG ionizing photon budget at any epoch. The stellar populations of EoR-galaxies would be characterized by strong radiation fields, with high ionizing production efficiencies (boosting the nebular emission), and whose ISM conditions would be such that they facilitate the escape of a copious amount of LyC photons to the IGM. This can be speculatively attributed to the stellar mass build-up and changes in the gas and dust properties of galaxies between the Cosmic Noon ($z \simeq 3$) and the EoR ($6 \leq z \leq 9$).

%% file: 6_conclusions.tex
\section{Summary and conclusions}\label{sec:conclusions}
In this work, we fully exploit and highlight the ability of absorption line and rest-UV spectroscopy alone to decipher the ionising properties of high-z SFGs. Our novel approach make use of deep, rest-frame UV spectra from the VANDELS survey at $3 \leq z \leq 5$, to compute the ionizing absolute photon escape fraction (\fescabs) and ionizing photon production efficiency (\xiion) of galaxies. \fescabs\ has been derived by combining absorption line measurements with estimates of the UV attenuation \citep[see][]{SL22}, while the \xiion\ parameter was computed by fitting the FUV stellar continuum of the VANDELS galaxies \citep[following][]{C19}.

In particular, we have searched for correlations between \fescabs\ and \xiion\ along with different galaxy properties (UV magnitude, UV slope, \lya\ strength, etc.), and we thoroughly compared with independent literature estimates. We found that:

\begin{itemize}

    \item The predicted \fescabs\ monotonically decreases with the stellar mass, the UV-continuum slope and the \lya\ equivalent width of the VANDELS galaxies (Fig.\ \ref{fig:fescXiion_phot}, \ref{fig:lya_lis}, \ref{fig:fesc_mass}). We find a non-significant correlation between \fescabs\ and the UV magnitude, although the faintest galaxies tentatively have higher escape fractions.
    
    \item The estimated \xiion\ statistically increases towards blue UV-continuum slopes and strong \lya\ emitting galaxies (Fig. \ref{fig:XiionMuv}, \ref{fig:XiionBeta}), and it smoothly raises beyond the canonical value towards the UV-faintest galaxies in the sample.
    
    \item Potential Lyman Continuum Emitters (LCEs) and selected Lyman Alpha Emitters (LAEs) show systematically higher \xiion\ ($\log \xiion ({\rm Hz/erg}) \approx 25.38, 25.41$, respectively) than non-LCEs and non-LAEs at similar UV magnitudes ($\log \xiion ({\rm Hz/erg}) \approx 25.18, 25.14$), and our \xiion\ values are in agreement with other LAEs surveys in the literature (Fig.\ \ref{fig:XiionMuv_LAEs}).

\end{itemize}

Additionally, we constructed the average composite FUV spectrum of LCEs at $3 \leq z \leq 5$ (Fig. \ref{fig:composite_spectra}, \ref{fig:composite_fits}, \ref{fig:composite_ion}), by stacking potential, individual emitters in the VANDELS survey (selected based on our predicted $\fescabs \geq 5\%$), and explained their non-ionizing spectral properties in the framework of the ISM conditions which enable ionizing photons to escape. Our results show that the FUV spectra of typical high-$z$ LCEs would be characterized by:

\begin{itemize}

    \item Blue UV slopes ($\bslope \lesssim -2$) compared to non-LCEs, which directly translates into low UV attenuation ($\ebv \lesssim 0.1~{\rm mag.}$) and therefore low column density of dust along the line-of-sight. 
    
    \item Enhanced \lya\ emission ($\ewlya \lesssim -25$\AA) and strong UV nebular lines in contrast to the non-LCEs population, particularly high \ion{C}{iii}1908/\ion{C}{iv}1550 ratios ($\gtrsim 0.75$). Together with the intrinsically higher \xiion\ parameter for LCEs, this indicates very young underlying stellar populations ($\approx 10~{\rm Myr}$) at relatively low metallicities ($\approx 0.2~{\rm Z_{\odot}}$). 
    
    \item Weak ($\lesssim 1$\AA) ISM LIS absorption lines (e.g., \ion{Si}{ii}1260, \ion{C}{ii}1334), while the HIS absorption lines (e.g., \ion{Si}{iv}1400) are of similar strength as the bulk of non-LCEs. This, together with the low UV attenuation for LCEs, suggest the presence of dustless, low gas column-density channels in the ISM which favour the escape of ionzing photons. 
    
\end{itemize}

Finally, we have compared our findings with other LyC surveys in the literature (Fig. \ref{fig:fesc_samples}), concretely the \emph{Keck Lyman Continuum Survey} \citep[or KLCS,][]{Steidel2018, Pahl2021, Pahl2023}, targeting $z \simeq 3$ LBGs at similar UV magnitudes to VANDELS, and the \emph{Low-Redshift Lyman Continuum Survey} \citep[or LzlCS,][]{Flury2022a, Flury2022b}, which targetted emission line galaxies at $z \simeq 0.3$. VANDELS and LzLCS show overall fairly consistent trends, with LzLCS shifted to fainter UV magnitudes, bluer UV slopes and stronger \lya\ emission, and therefore their \fescabs\ and \xiion\ are enhanced with respect to VANDELS galaxies. The escape fraction of KLCS galaxies fall above our estimates at all UV magnitudes. We mainly ascribed this discrepancy to the way the amount of UV attenuation is measured, and we propose to use $\beta_{1500}$ measurements from photometry as an independent proxy of the UV dust attenuation of galaxies (Fig.\ \ref{fig:BETAbias}). The dust-attenuation law (Fig.\ \ref{fig:A1600_Mass}) and the metals-to-neutral-gas covering fraction conversion (Fig.\ \ref{fig:LIStoHI}) constitutes additional sources of uncertainty to the escape fraction.

Our joint analysis of the VANDELS, LzLCS and KLCS results shed light onto the fact that UV-faint, low-mass and dustless galaxies likely dominated the ionizing budget during the EoR. Their stellar populations would most likely be characterized by strong radiation fields, with high ionizing production efficiencies, and whose ISM conditions are such that they favour the escape of LyC photons. Recent JWST observations \citep[e.g.,][]{Endsley2022, Schaerer2022b, Cullen2023, Lin2023, Mascia2023}, have revealed that blue UV slopes and strong emission lines also characterized the less massive and moderately faint galaxies at the EoR, supporting our results.

An increasing number of high-quality, FUV spectra at $z > 6$ will be available in the mid-future thanks to upcoming ground-based (\textit{ELT, GMT, TMT, ...}) and currently ongoing space-based facilities (\textit{JWST}). This will give us the first insights into the properties of galaxies during the EoR. Nevertheless, according to our work, additional efforts are still needed in order to correctly decipher the physics underlying such vast data-sets. The two main questions that we think are urgent and clearly needed for future studies are: (1) if unique, what is the dust-attenuation law governing low-mass, low-metallicity galaxies? And (2) do metals trace the same spatial location as the neutral gas within the ISM of these galaxies?

%% file: A_appendix.tex
\section{The impact of resolution and stacking on the depth of the absorption lines}\label{app:resolution}

Here we provide two calibrations to correct the observed line-depths measurements when (1) inherently degraded by the instrumental resolution and (2) artificially smoothed during usual stacking procedures. As we discussed in the main text, the spectrograph resolution makes the absorption lines wider but less deep, conserving the flux. Together with the noise, sensitivity and other possible instrumental aberrations, this can lead to a systematically overestimated measurement of the residual flux.

To account for these effects, we simulate \ion{Si}{ii}1260 line intensities ($I_{\lambda}$) assuming a foreground dust-screen geometry of the picket-fence model, describing the line through a single gas component as \citep{Draine}:

\begin{equation}
    I_{\lambda} = C_f(\lambda) \times \exp{(-\sigma_{\lambda}N_{\lambda})} + 1-C_f(\lambda), 
\end{equation}

\noindent where $C_f(\lambda)$ represents the covering fraction of the line in question and the $\tau_{\lambda}=\sigma_{\lambda} N_{\lambda}$ product is usually known as the optical depth of the line. The line cross-section, $\sigma_{\lambda}$, which shapes the absorption profile, is given by the Voigt function so that:
\begin{equation}
    \sigma_{\lambda} = \dfrac{\sqrt{\pi} e^2 f_{\lambda}}{m_e c} \dfrac{\lambda}{b}~{\rm Voigt(\lambda, A_{\lambda}, b)}.
\end{equation}

\noindent where oscillator strengths $f_{\lambda}$ and Einstein coefficients A$_{\lambda}$ are taken from the NIST database \footnote{NIST stands for National Institute of Standards and Technology: \url{https://physics.nist.gov/PhysRefData/ASD/lines_form.html}.}. For the Voigt function, Voigt($\lambda$,A$_{\lambda}$,b), we use the numerical approximation described in \citet{Smith2015}, and assume the following Gaussian distributions ($\mathcal{N}(\mu,\,\sigma^{2})$) of the input parameters: 
\begin{enumerate}
    \item[$-$] $N \sim \mathcal{N}(16,\,1) ~{\rm cm^{-2}}$
    \item[$-$] $b \sim \mathcal{N}(125,\,25) ~{\rm km~s^{-1}}$
    \item[$-$] $v \sim \mathcal{N}(0,\,150) ~{\rm km~s^{-1}}, \lambda = \lambda_0 + v/c$
\end{enumerate}

\noindent where $\mu$ represents the median and $\sigma^2$ the standard deviation of the normal distribution. The gas column density distribution ($N, {\rm cm^{-2}}$) is chosen so that the \ion{Si}{ii}1260 equivalent width of the line falls within the optically thick limits of the curve-of-growth or, in other words, the line is \emph{always} saturated. The line velocity shift ($v, {\rm km~s^{-1}}$) randomly changes to account for inflows or outflowing gas so that the line center is blue- or red-shifted accordingly, i.e., $\lambda = \lambda_0 + v/c$, being $\lambda_0 = 1260.42$\AA\ for the \ion{Si}{ii}1260 line. Finally, the gas thermal (or Doppler) broadening ($b, {\rm km~s^{-1}}$) varies around typical values for normal star-forming galaxies \citep{Steidel2018}. The input covering fraction is fixed for every set of simulations to $C_f(\lambda) = 0.3, 0.5, 0.65, 0.75, 0.85$ and $1$, respectively. 

\begin{figure*}
    \includegraphics[width=0.80\textwidth]{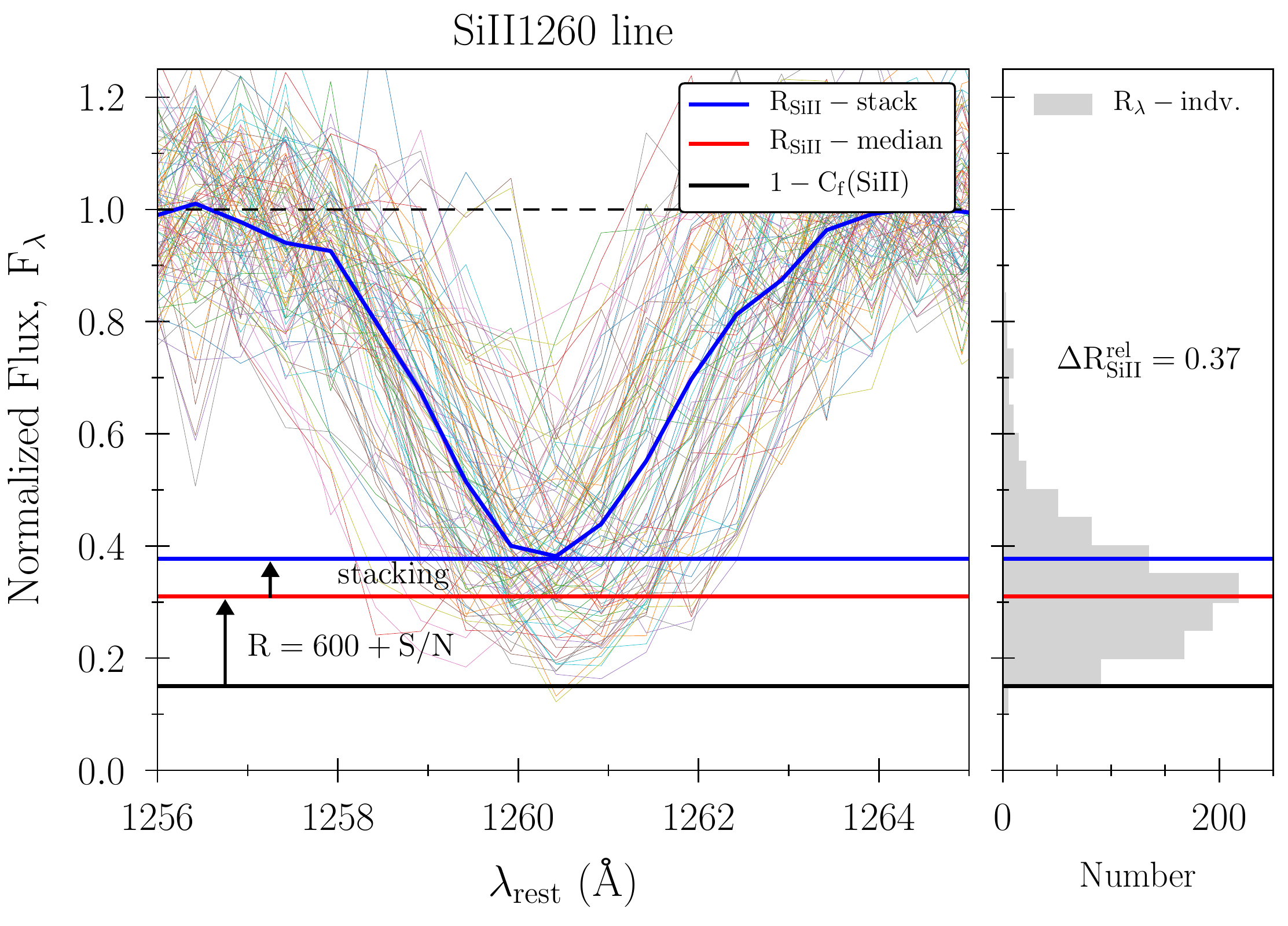}
\caption{The impact of resolution and stacking on the depth of the absorption lines. {\bf Left:} the colored lines represent a set of $\times$100 mock profiles of the SiII1260 line with an input covering fraction of $C_f = 0.85$ (equivalent to $R_{\rm SiII} = 1 - C_f = 0.15$, delimited by the horizontal black line), and where the \ion{Si}{ii} gas column density, $N_{\rm SiII}{(\rm cm^{-2})}$, the Doppler broadening parameter, $b {(\rm kms^{-1})}$, and the gas velocity, $v{(\rm kms^{-1})}$, were varied following Gaussian distributions (see text). {\bf Right:} distribution of individual $R_{\rm SiII}$ measurements for all the simulations once the lines have been degraded to a spectral resolution of $R{\rm (VIMOS)} = 600$ (red line), including a constant Gaussian noise of median ${\rm S/N} = 5$. The blue line indicates the resulting residual flux after co-adding all the spectra (stacking).}
\label{fig:sim_profile}
\end{figure*}

After performing a set of $\times 100$ ideal Voigt profiles for each covering fraction, we introduce the effect of resolution by convolving the simulated spectra with a Gaussian kernel whose FWHM matches the instrumental resolution (a.k.a., $R{\rm (VIMOS)} \approx 600$). A constant S/N value (${\rm S/N} = 5$, the median of the VANDELS sample) is then added to every single spectra. At fixed $C_f(\lambda)$, the median measured (observed) residual flux of each mock distribution is compared to the theoretical line residual flux of $R_{\lambda} = 1 - C_f(\lambda)$ according to the picket-fence configuration, producing the desired calibration. Fig.\ \ref{fig:sim_profile} presents the resulting line profiles for a set of $\times 100$ simulations of the \ion{Si}{ii}1260 line, all of them with an input covering fraction of $C_f(\lambda) = 0.85$ ($R_{\lambda} = 0.15$). The median of the distribution of individually measured $R_{\lambda}$ values is 0.3 (red line), showing that the observed line depth can statistically differ as much as 40$\%$ with respect to the real value. 

In the last step, we proceed to stack all the spectra contained on each simulation pack at fixed $C_f(\lambda)$, following the same methods described in the text. The co-added spectra for the $C_f(\lambda) = 0.85$ simulation is also shown in Fig.\ \ref{fig:sim_profile} (blue line), with a 65$\%$ relative difference in residual flux with respect to the real input value. 

\begin{figure}
    \includegraphics[width=0.90\columnwidth]{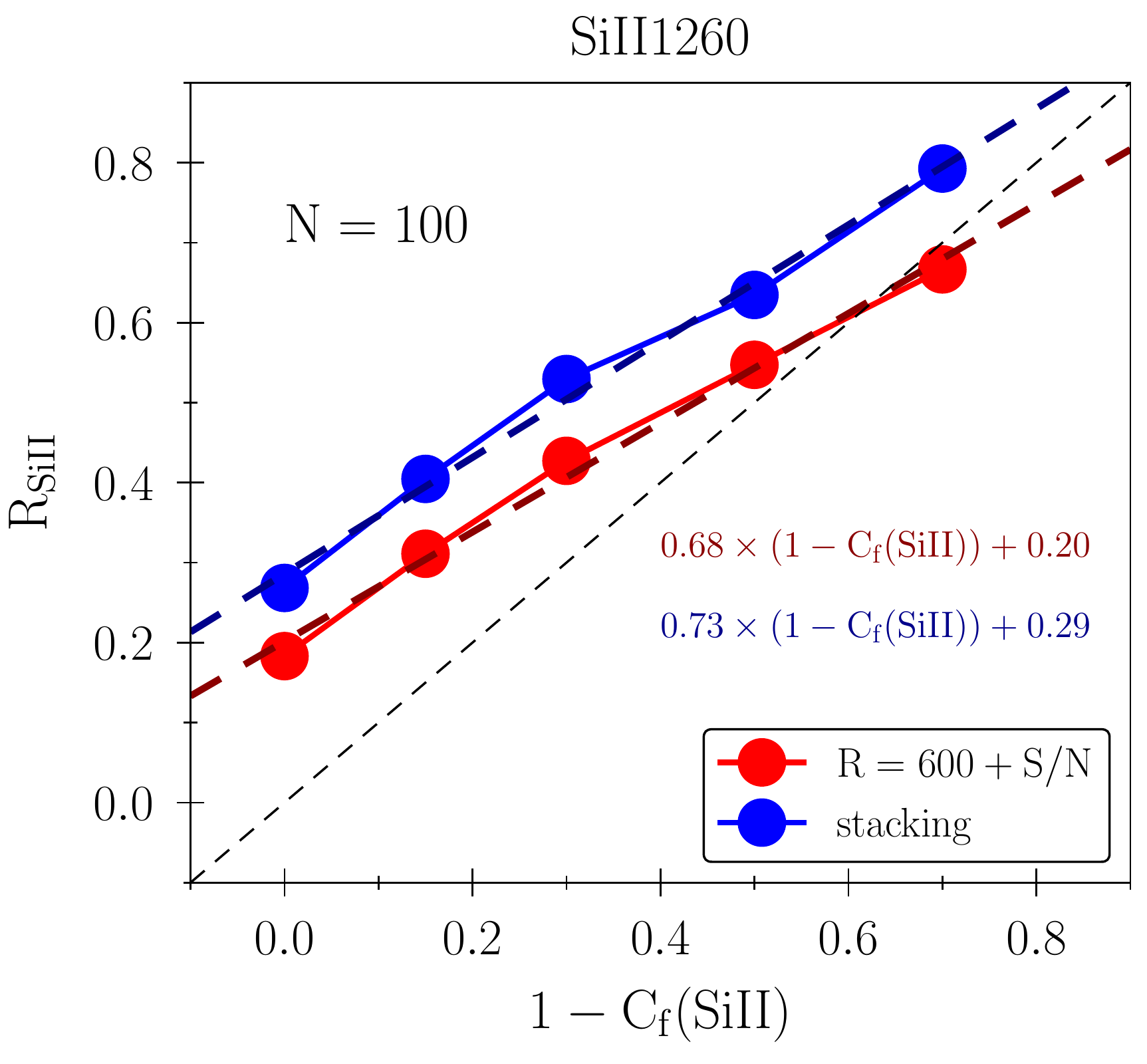}
\caption{Calibration between the theoretical residual flux of the SiII1260 line and the actual measured flux when accounting for (1, in red) the effect of a finite instrumental resolution and ${\rm S/N}$ ($R{\rm (VIMOS)} \approx 600, {\rm S/N = 5}$), and (2, in blue) the joint effect of resolution, ${\rm S/N}$ plus stacking a set of $N = 100$ simulated lines (see Fig.\ \ref{fig:sim_profile}). Dashed-black line shows the 1:1 relation.}
\label{fig:sim_diag}
\end{figure}

To sum up, the resulting calibrations --fitting a linear polynomial to the measured versus theoretical $R_{\lambda}$ values-- are:
\begin{equation}
    R_{\rm SiII}~{\rm [R + S/N]} = 0.68 \times (1 - C_f({\rm SiII})) + 0.20, 
    \label{eq:cal_R}
\end{equation}

\noindent and 
\begin{equation}
    R_{\rm SiII}~{\rm [stacking]} = 0.73 \times (1 - C_f({\rm SiII})) + 0.29, 
    \label{eq:cal_stack}
\end{equation}

\noindent when accounting for instrumental resolution plus S/N (Eq.\ \ref{eq:cal_R}), and spectral stacking (Eq.\ \ref{eq:cal_stack}), respectively. Both relations are plotted in Fig.\ \ref{fig:sim_diag} through the dashed red and dashed blue lines. These calibrations were applied to individual residual flux measurements as well as when measuring the residual flux over composite spectra.

\section{Stack measurements and \textsc{FiCUS} SED results}\label{app:stack_results}
As stated in Sect.\ \ref{sub:stacks}, the VANDELS sample was divided according to the 25$^{th}$, 50$^{th}$ and 75$^{th}$ ditribution percentiles (quartiles) of the different targeted quantities, resulting in four sub-samples for each quantity which were quoted as Q1, Q2, Q3 and Q4. Then, stacked spectra in bins of UV magnitude (\mobs), UV intrinsic luminosity (\mint), UV continuum slope (\bslope) and \lya\ equivalent width (\ewlya) were built following the methodology described in the same section. 

Table \ref{tab:stack_table} includes the ionizing escape fractions (\fescabs) and production efficiencies (\xiion) derived for every composite spectra. It also contains the same results when using either the \citetalias{R16} or \citetalias{SMC} dust-attenuation laws, for comparison. Table \ref{tab:lces_table} summarizes the outputs of our \textsc{FiCUS} SED fits and the main UV spectral measurements for the LCEs, non-LCEs, LAEs and non-LAEs stacks. 

\begin{table*}
\begin{center}
\caption{Inferred ionizing absolute escape fractions (\fescabs) and production efficiencies (\xiion) for the VANDELS composites in this work.}
\begin{tabular}{cccccccc}
\toprule
Composite ID & Bin range & Median & ${\rm N_{gal}}$ & $f_{\rm esc}^{\rm abs}~{\rm (R16,~\%)}$ & $\xi_{\rm ion}~{\rm (R16, ~Hz/erg)}$ & $f_{\rm esc}^{\rm abs}~{\rm (SMC,~\%)}$ & $\xi_{\rm ion}~{\rm (SMC, ~Hz/erg)}$ \\
\midrule
 & \multicolumn{6}{c}{Stacks in bins of observed absolute UV magnitude (AB)} & \\
\midrule
${\rm M_{1500}~(Q1)}$ & $\leq -20.94$ & -21.22 & 63 & 1.70$~\pm~$0.30 & 25.21$~\pm~$0.04 & 3.35$~\pm~$0.47 & 25.18$~\pm~$0.04 \\
${\rm M_{1500}~(Q2)}$ & $-20.94, -20.53$ & -20.70 & 71 & 1.01$~\pm~$0.23 & 25.26$~\pm~$0.05 & 2.19$~\pm~$0.40 & 25.20$~\pm~$0.06 \\
${\rm M_{1500}~(Q3)}$ & $-20.53, -20.16$ & -20.34 & 73 & 1.55$~\pm~$0.34 & 25.30$~\pm~$0.09 & 3.11$~\pm~$0.62 & 25.31$~\pm~$0.08 \\
${\rm M_{1500}~(Q4)}$ & $> -20.16$ & -19.95 & 90 & 2.38$~\pm~$0.33 & 25.42$~\pm~$0.09 & 4.09$~\pm~$0.56 & 25.43$~\pm~$0.08 \\
\midrule
 & \multicolumn{6}{c}{Stacks in bins of intrinsic UV luminosity (AB)} & \\
\midrule
${\rm M_{1500}^{int}~(Q1)}$ & $\leq -23.49$ & -23.91 & 74 & 0.43$~\pm~$0.10 & 25.20$~\pm~$0.03 & 1.39$~\pm~$0.28 & 25.09$~\pm~$0.04 \\
${\rm M_{1500}^{int}~(Q2)}$ & $-23.49, -22.86$ & -23.16 & 76 & 1.02$~\pm~$0.22 & 25.26$~\pm~$0.08 & 2.45$~\pm~$0.52 & 25.16$~\pm~$0.08 \\
${\rm M_{1500}^{int}~(Q3)}$ & $-22.86, -22.29$ & -22.61 & 71 & 2.20$~\pm~$0.57 & 25.29$~\pm~$0.09 & 3.78$~\pm~$0.67 & 25.40$~\pm~$0.08 \\
${\rm M_{1500}^{int}~(Q4)}$ & $> -22.29$ & -21.85 & 71 & 5.64$~\pm~$1.19 & 25.28$~\pm~$0.05 & 7.21$~\pm~$1.28 & 25.28$~\pm~$0.05 \\
\midrule
 & \multicolumn{6}{c}{Stacks in bins of UV continuum slope} & \\
\midrule
$\beta{\rm _{spec}^{1500}}~{\rm (Q1)}$ & $\leq -1.65$ & -1.82 & 69 & 6.37$~\pm~$1.21 & 25.48$~\pm~$0.05 & 9.31$~\pm~$1.70 & 25.44$~\pm~$0.06 \\
$\beta{\rm _{spec}^{1500}}~{\rm (Q2)}$ & $-1.65, -1.34$ & -1.49 & 78 & 1.94$~\pm~$0.46 & 25.27$~\pm~$0.06 & 3.45$~\pm~$0.56 & 25.28$~\pm~$0.07 \\
$\beta{\rm _{spec}^{1500}}~{\rm (Q3)}$ & $-1.34, -1.03$ & -1.20 & 74 & 0.97$~\pm~$0.24 & 25.18$~\pm~$0.04 & 2.16$~\pm~$0.40 & 25.14$~\pm~$0.07 \\
$\beta{\rm _{spec}^{1500}}~{\rm (Q4)}$ & $> -1.03$ & -0.81 & 76 & 0.27$~\pm~$0.09 & 25.12$~\pm~$0.05 & 0.87$~\pm~$0.20 & 24.99$~\pm~$0.05 \\
\midrule
 & \multicolumn{6}{c}{Stacks in bins of \lya\ equivalent width (\AA)} & \\
\midrule
$-1 \times {\rm W_{Ly \alpha}~(Q1)}$ & $\leq -3.61$ & -5.86 & 71 & 0.60$~\pm~$0.17 & 25.18$~\pm~$0.02 & 1.44$~\pm~$0.31 & 25.13$~\pm~$0.04 \\
$-1 \times {\rm W_{Ly \alpha}~(Q2)}$ & $-3.61, 4.58$ & 1.13 & 67 & 1.14$~\pm~$0.23 & 25.14$~\pm~$0.06 & 2.38$~\pm~$0.40 & 25.14$~\pm~$0.08 \\
$-1 \times {\rm W_{Ly \alpha}~(Q3)}$ & $4.58, 15.60$ & 7.75 & 79 & 1.44$~\pm~$0.37 & 25.13$~\pm~$0.07 & 2.95$~\pm~$0.55 & 25.11$~\pm~$0.07 \\
$-1 \times {\rm W_{Ly \alpha}~(Q4)}$ & $> 15.60$ & 30.42 & 80 & 4.24$~\pm~$0.92 & 25.41$~\pm~$0.07 & 7.09$~\pm~$1.27 & 25.44$~\pm~$0.07 \\
\bottomrule
\end{tabular}
\label{tab:stack_table}
\end{center}
{\bf Notes.} Column 1: composite identifier. Column 2: interquartile range for each magnitude bin (units are in the spanning titles). Column 3: median of the interquartile range. Column 4: number of objects included in the stacks. Columns 5 and 6: estimated escape fractions (in $\%$) and production efficiencies (in Hz/erg) when using \citetalias{R16} dust-attenuation law. Columns 7 and 8: estimated escape fractions (in $\%$) and production efficiencies (in Hz/erg) when using \citetalias{SMC} dust-attenuation law. 
\end{table*}

\begin{table*}
\begin{center}
\caption{Summary chart containing the main spectral measurements and \textsc{FiCUS} SED fitting results for LCEs, non-LCEs, LAEs and non-LAEs composites.}
\begin{tabular}{cccccc}
\toprule
Parameter$^{\rm ~a}$ & Units & LCEs ($\fescabs \geq 5\%$) & non-LCEs ($\fescabs < 5\%$) & LAEs ($\ewlya \leq -20$\AA) & non-LAEs ($\ewlya > -20$\AA) \\
\midrule
$W_{\rm Ly\alpha}$ & \AA & $-29.71~\pm~2.46$ & $-3.62~\pm~0.41$ & $-33.45~\pm~1.10$ & $-1.63~\pm~0.22$ \\
$\beta _{\rm spec}^{1500}$ & -- & $-2.17~\pm~0.03$ & $-1.23~\pm~0.01$ & $-1.68~\pm~0.03$ & $-1.22~\pm~0.01$ \\
$W_{\rm LIS}$ & \AA & $0.86~\pm~0.07$ & $1.90~\pm~0.02$ & $1.04~\pm~0.05$ & $2.01~\pm~0.03$ \\
$R_{\rm LIS}^{\rm ~b}$ & -- & $0.73~\pm~0.06$ & $0.47~\pm~0.02$ & $0.64~\pm~0.05$ & $0.43~\pm~0.03$ \\
$W_{\rm CIV1550}$ & \AA & $-0.72~\pm~0.18$ & $-0.26~\pm~0.05$ & $-0.55~\pm~0.10$ & $-0.25~\pm~0.05$ \\
$W_{\rm HeII1640}$ & \AA & $-1.11~\pm~0.25$ & $-0.81~\pm~0.06$ & $-1.16~\pm~0.15$ & $-0.76~\pm~0.06$ \\
$W_{\rm CIII]1908}$ & \AA & $-2.15~\pm~0.67$ & $-0.73~\pm~0.13$ & $-2.28~\pm~0.31$ & $-0.46~\pm~0.14$ \\
$\ion{C}{iv}/\ion{C}{iii}$ & -- & $0.79~\pm~0.21$ & $0.42~\pm~0.20$ & -- & -- \\
${\rm Age}$ & Myr & $13.60~\pm~2.73$ & $19.14~\pm~0.92$ & $13.63~\pm~1.97$ & $20.26~\pm~0.90$ \\
${\rm Z_{\star}}$ & ${Z_{\odot}}$ & $0.18~\pm~0.02$ & $0.21~\pm~0.01$ & $0.19~\pm~0.02$ & $0.21~\pm~0.01$ \\
$E_{\rm B-V}$ & mag. & $0.07~\pm~0.01$ & $0.23~\pm~0.00$ & $0.16~\pm~0.00$ & $0.24~\pm~0.00$ \\
${\rm (F_{\lambda900}/F_{\lambda1500})_{int}}$ & -- & $0.64~\pm~0.09$ & $0.47~\pm~0.02$ & $0.69~\pm~0.07$ & $0.43~\pm~0.02$ \\
$\xi_{\rm ion}$ & ${\rm Hz/erg}$ & $25.38~\pm~0.08$ & $25.18~\pm~0.03$ & $25.41~\pm~0.07$ & $25.14~\pm~0.03$ \\
\bottomrule
\end{tabular}
\label{tab:lces_table}
\end{center}
{\bf Notes.} Column 1: galaxy properties. Column 2: physical units.\\
Columns 3 and 4: list of measurements for the LCEs and non-LCEs stacks.\\
$^{\rm a}$All tabulated results use the \citetalias{R16} dust-attenuation law.\\
$^{\rm b}$ Measured residual fluxes not corrected by the spectral resolution (see App.\ \ref{app:resolution}).
\end{table*}

\section{Alternative escape fraction estimates}\label{app:beta_Calabro}
\begin{figure}
    \includegraphics[width=0.99\columnwidth, page=18]{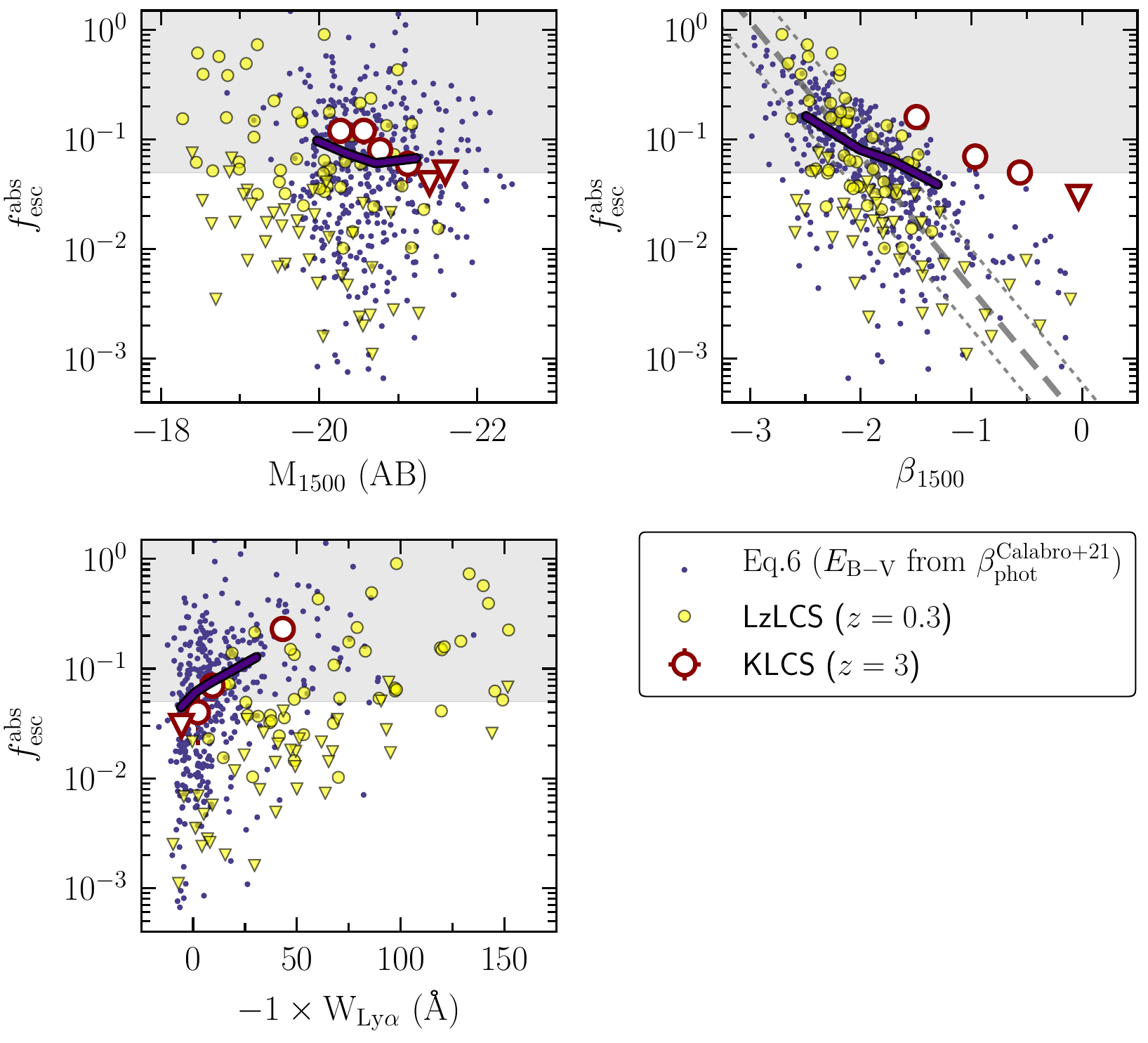}
\caption{The ionizing photon escape fraction (\fescabs) versus the UV magnitude (\mobs), the UV slope at 1500\AA\ ($\beta_{1500}$) and the \lya\ equivalent width (\ewlya). Blue dots represent the predicted \fescabs\ values for VANDELS galaxies (Eq.\ \ref{eq:fesc_LIS}, using \citetalias{SMC}) when the dust-attenuation (\ebv) is inferred from the $\beta_{1500}$ measurement by \citet{Calabro2021}. Layout is the same as Fig.\ \ref{fig:fesc_samples}.}
\label{fig:beta_Calabro}
\end{figure}

Here we offer complementary predictions for the ionizing photon escape fraction. \fescabs\ is derived by combining absorption line measurements with estimates of the UV attenuation, in a similar manner as in the main text (Eq.\ \ref{eq:fesc_LIS}). However, the dust-attenuation parameter (\ebv) has now been estimated from the UV slope measurements of \citet{Calabro2021}, by fitting a power law to the VANDELS photometry, requiring the whole bandwidths to reside between 1230 and 2750\AA\ rest-frame. Eq. 8 in \citet{Chisholm2022} was then used to convert $\beta_{1500}$ to \ebv\ assuming an \citetalias{SMC} attenuation law. VANDELS, KLCS and LzLCS \fescabs\ correlations with physical quantities converge to the same global trends (see Fig.\ \ref{fig:beta_Calabro}), with \fescabs\ increasing towards fainter UV magnitudes, bluer UV slopes and stronger \lya\ emission. We refer to Sect.\ \ref{sec:discussion_lowzhighz} for more details.

